# *Ab initio* quantum-statistical approach to kinetic theory of low-temperature dilute gases of hydrogen-like atoms

Yu.V. Slyusarenko[1,2,a)], O.Yu. Sliusarenko[1]

[1] *Akhiezer Institute for Theoretical Physics, NSC KIPT, 1 Akademichna Str., 61108 Kharkiv, Ukraine*

[2] *Karazin Kharkiv National University, 4 Svobody Sq., 61077 Kharkiv, Ukraine*

We develop a microscopic approach to the consistent construction of the kinetic theory of dilute weakly ionized gases of hydrogen-like atoms. The approach is based on the framework of the second quantization method in the presence of bound states of particles. It is assumed that a bound state (hydrogen-like atom of an alkali metal) is formed by two different kinds of charged fermions - valence electron and the core. The basis of the derivation of kinetic equations is the method of reduced description of relaxation processes. Within the approach we developed the first-order perturbation theory over the weak interaction for a system of kinetic equations for the Wigner distribution functions of free fermions of both kinds and their bound states, the hydrogen-like atoms. Kinetic equations are used to study the spectra of elementary excitations in the system. To this end, they are linearized near spatially-homogeneous equilibrium state, characterized by the fact that all components of the system are far from degeneration. It is shown that the conditions of low-temperature approximation, of the gas non-degeneracy and the approximation of weak interaction are realistic and can be met in a wide range of temperatures and the densities of the studied system. We obtain dispersion equations for determining the frequency and wave attenuation coefficients in dilute weakly ionized gas of hydrogen-like atoms near the described equilibrium state. In the two-level atom approximation it is shown that in the system there are longitudinal waves of matter polarization and transverse waves with the behavior characteristic of plasmon polaritons. The expressions for the dependence of the frequency and the Landau damping coefficients on the wave vector for all branches of the oscillations detected, are obtained. Quantitative estimations of the characteristics of the elementary perturbations in the system on an example of a weakly ionized dilute gas of $^{23}$Na atoms are presented. The possibility of using the results of the theory developed to describe the properties of a Bose condensate of photons in dilute weakly ionized gas of hydrogen-like atoms is noted and the directions of its generalizations are discussed.

---

a) Electronic mail: slusarenko@kipt.kharkov.ua

**I. INTRODUCTION**

The topic of present investigations dates back to a second half of the last century when a consensus in the classification of plasma by its characteristics and properties was reached, and there was an understanding of quantum plasma (Refs. 1-5) and weakly ionized gases (Refs. 6, 7) features. We should note, however, that a number of literature devoted to plasma physics including monographs and students books, is so huge that we here mention only several citations serving something like rappers that would define a chronology of the investigations (altogether with the citations in them).

The interest to investigations on quantum plasma and excited gases was revoked due to several reasons. Firstly, it was a substantial technical progress that allowed to study and even to use the quantum effects in plasma (see in this regard methodological notes Ref. 8 and the references therein). The one itself was stimulated by a transition of metallic and semiconducting structures to nano-sized scales (thin metallic films, nanowires, quantum points) that does not allow to neglect the quantum effects in the collective plasma processes in such systems.

An another important circumstance that favored an interest to understanding the kinetics of quantum gases is the intense investigation of systems with Bose-Einstein condensate (BEC), see Refs. 9, 10. Indeed, the BEC is an excellent example of a quantum-mechanical origin of matter demonstrated at microlevel. Besides, the BEC phenomenon was primarily observed in alkali atoms vapor at temperatures of some hundreds nano-Kelvin (quantum gases)[11, 12]. The most substantial argument for studying the kinetics of weakly ionized and excited gases are the unique effects of their interaction with electromagnetic field, and first of all, the phenomenon of a strong slowdown or even a full stop of light in gases with BEC. An experimental evidence of these may be found in Refs. 13, 14. A consistent theoretical description of the interaction of electromagnetic fields with gasses with BEC was suggested in Refs. 15-17. Being based on a new formulation of the secondary quantization method for the presence of bound particles states[18], it allowed in particular, to prove a principled possibility of a strong light slowdown in ultracold dilute Bose-gases without using an artificially stimulated medium transparency near resonances (see in this respect Ref. 14). Moreover, many other intriguing effects of ultracold BEC gases response to excitations by electromagnetic fields were predicted within this method's framework. It illustrated a possibility of microwaves slowdown in such systems, to group velocity values of $10^{-2} \, cm/s$ order[19], predicted the ability to govern the group velocity of light by means of an external magnetic field[20], to filter electromagnetic signals by using these systems[21] and even a hard-to-believe phenomenon of charged particles acceleration in ultracold gases with BEC, see Ref. 22. We underline that in the cases mentioned, the systems of many identical particles are under extremely low temperatures, that is a requirement for atomic or molecular BEC implementation at contemporary experimental possibilities. Since in these conditions the charged components densities of quantum plasma are exponentially (by temperature) small (see in this regard Ref. 23), the systems under consideration may be treated as weakly excited ultracold gases. In other words, a contribution of charged components of quantum plasma to the stated effects can be neglected or taken into account in a perturbation theory.

However, such a situation is not typical for all BEC systems. It was reported recently of a photons BEC observation in a real experiment with a special colorant, at room temperatures[24, 25]. Soon came a number of theoretical papers devoted to the description of this phenomenon (see, e.g., Refs. 26-30), some of them predicted a possibility to realize a photons BEC in excited gases, or even in quantum plasma[28-30]. In the latter cases, kinetic processes play an essential role in the photons BEC formation. Indeed, the possibility of the photons BEC is connected with photons effective mass acquisition, which is only possible in a medium[24-30]. The photon's effective mass is defined by a so-called "cut-off frequency". This frequency is a (finite!) frequency value which is obtained when tending to zero the wave vector from a photon dispersion law in medium. The photon dispersion law in medium itself (the dispersion law of electromagnetic waves in the systems) should be determined by kinetic processes connected with an interaction between radiation and the structural components of this substance (atoms, gas molecules or neutrals, ions and electrons, if we consider quantum plasma). In other words, kinetic processes in the systems form photon's effective mass ("rest mass"). Moreover, the same processes should determine also a thermalization of photons in medium that allows to decrease the photons subsystem's temperature and, therefore, to achieve the state of BEC in it. We note that in a real experiment, the effective mass of a photon may be formed also by a standing wave along some direction in the system due to mirrors that do not allow the photons to leave the system, see Refs. 24,25. In this case we speak about effectively two-dimensional systems.

Thus, the description of phenomena and effects of photons BEC formation in excited gases and weakly ionized plasma implies a problem of a construction of a kinetic theory for such systems, that is, of a coupled system of kinetic equations for all possible system components, including radiation (photons). This theory should be microscopic, based on the primary principles of quantum statistics. We shall especially note that a microscopic method developed and used in Refs. 15-23 cannot be used. The latter becomes inapplicable at small frequencies and wave vectors domain and needs a substantial modification, which, in its turn requires the usage of a kinetic theory of matter with quantum origin[31]. This circumstance leads us back to the construction of a kinetic theory of such systems from primary principles.

The creation of the microscopic approach to the kinetic theory development is quite a complex problem (see in this regard, e.g., Ref. 8). However, as we are going to demonstrate in the current paper, lots of hardships may be overcome if for the sake of construction of a kinetic theory of excited gases or weakly ionized plasma one uses the reduced description method for relaxation processes in multiparticle systems suggested by N.N. Bogolyubov[32]. This method was developed for classical (non-quantum) systems of identical particles and allows constructing a systematic procedure for obtaining coupled dissipative kinetic equations based on the Bogolyubov-Born-Green-Kirkwood-Yvon (BBGKY) chain of reversible equations for multiparticle distribution functions. For quantum systems, a reduced description method was developed by Peletminskii and co-authors[31]. This method proved its effectiveness at description of irreversible processes in systems with spontaneously broken symmetry (solid matter, magnetics, superfluid and superconducting[31]). It demonstrated its prospects at studying the peculiarities of relaxation processes in systems with long non-equilibrium fluctuations[33], systems of neutrons interacting with hydrodynamic media[34] or even prototypes of active matter[35]. Due to these, to achieve our goals we shall use the very methods of the reduced description method stated in Ref. 31, especially given a quantum nature of the objects under study.

The reduced description method is of the most effectiveness when the system Hamiltonian may be separated into two summands $\hat{H}_0$ and $\hat{V}$, $\hat{H} = \hat{H}_0 + \hat{V}$, where $\hat{H}_0$ includes basic interactions, while $\hat{V}$ describes relatively small ones. Thus, the development of the kinetic theory of weakly ionized plasma (excited gas may be considered as a particular case of weakly ionized plasma) needs to be started with a statement of the system Hamiltonian's explicit form.

## II. HAMILTONIAN OF HYDROGEN-LIKE PLASMA IN EXTERNAL ELECTROMAGNETIC FIELD

The problem of construction of the Hamiltonian of weakly ionized plasma in external electromagnetic field is, in fact, solved. In Ref. 18 an approximate secondary quantization method for multiparticle systems with bound particles states is developed. It considers a system of three different gas components: the subsystems of two different oppositely charged fermions and their bound states. A case when the bound states of these two oppositely charged fermions are hydrogen-like atoms is studied in detail. One kind of the fermions is identified as electrons, while the other one as cores. This system, therefore, may be regarded as hydrogen-like plasma. The choice of such model is not associated with principle difficulties and makes it possible to simplify calculations and to obtain the visual results. Secondary quantization method for the chosen system may be applied only when particles mean kinetic energy is lower than the binding energy of compound particles. In this approximation, both compound and "elementary" particles may be equally considered. This means that creation and annihilation operators of compound particles can be introduced. In terms of introduced particles creation and annihilation operators, the main system's physical characteristics operators are written, in particular, the Hamiltonians. The Hamiltonians, which specify the interactions between compound and elementary particles and between compound particles themselves, are found in terms of the interaction amplitudes for elementary particles. The Hamiltonians that characterize an interaction between charged fermions with their bound states (excited atoms) and electromagnetic field are obtained. The nonrelativistic quantum electrodynamics is developed for systems containing both elementary and compound particles. As applications of the developed theory, Ref. 18 studies a spontaneous radiation of two-component excited atom and obtains the expression for its probability. It also investigates the attraction forces acting between neutral atoms at long distances (the van der Waals forces) and derives the expression for the potential of the van der Waals forces. Finally, it examines a question concerning the scattering of photons and elementary particles by bound states. Moreover, the developed in Ref. 18 secondary quantization method creates a basic of a microscopic approach of Refs. 15-23 to studying various aspects of gases with BEC with electromagnetic field interaction. Unfortunately, as we already have noted, the method becomes inapplicable in the long-wave domain of electromagnetic waves spectrum. This remark, however, does not deal with Hamiltonians expressions received in Ref. 18, which are valid in the long-wave domain of spectrum. For this reason, the use of these Hamiltonians together with the modified reduced description method for weakly ionized gas of hydrogen-like atoms at low temperatures allows to develop a controlled kinetic theory for the system.

Before we write an expression for the Hamiltonian of weakly ionized gas of hydrogen-like atoms at low temperatures we briefly present the basic postulates of the theory developed in Ref. 18. First of all we ought to explain why we constantly use the terms "low temperatures" and "weakly ionized gas". As we mentioned above, the secondary quantization method is valid only in the case when mean kinetic energy of system's particles is small compared to the energies of bound states (atoms). In close to equilibrium systems this condition is fulfilled at low temperatures. The approximate (and new at that time) form of the secondary quantization method for systems with bound states obtained in Ref. 18 is as follows. A low-temperature hydrogen-like plasma is a multiparticle multicomponent system with subsystems of oppositely charged free fermions (electrons and positively charged cores) and bound states of these fermions – neutral hydrogen-like atoms (bosons) that may be in excited states. Creation and annihilation operators of fermions of the first and second kinds in the momentum representation

$$\hat{a}_l(\mathbf{p}), \ \hat{a}_l^+(\mathbf{p}), l = 1, 2 \tag{1}$$

obey the usual (Fermi) commutation

$$\{\hat{a}_l(\mathbf{p}), \hat{a}_{l'}^+(\mathbf{p'})\} \equiv \hat{a}_l(\mathbf{p})\hat{a}_{l'}^+(\mathbf{p'}) + \hat{a}_{l'}^+(\mathbf{p'})\hat{a}_l(\mathbf{p}) = \Delta(\mathbf{p}-\mathbf{p'})\Delta(l-l'),$$
$$\{\hat{a}_l(\mathbf{p}), \hat{a}_{l'}(\mathbf{p'})\} = 0, \quad \{\hat{a}_l^+(\mathbf{p}), \hat{a}_{l'}^+(\mathbf{p'})\} = 0, \tag{2}$$

where the quantities $\Delta(\mathbf{p}-\mathbf{p'})$ и $\Delta(l-l')$ are the Kronecker symbols. For definiteness, in what follows we assume that the index $l = 1$ corresponds to subsystem of electrons, and $l = 2$ -- that of skeletons. Note that for simplicity, the calculations in Ref. 18 do not take into account the presence of the spin variables as the individual characteristics of the quantum particles that make up the subsystem. Nor will it be considered in the present work for the same reason.

It is possible to introduce creation $\hat{\eta}_\alpha^+(\mathbf{p})$ and annihilation $\hat{\eta}_\alpha(\mathbf{p})$ operators for bound states (hydrogen-like atoms) in low-energy domain that also obey the regular Bose commutation relations:

$$\left[\hat{\eta}_\alpha(\mathbf{p}), \hat{\eta}_\beta^+(\mathbf{p}')\right] \equiv \hat{\eta}_\alpha(\mathbf{p})\hat{\eta}_\beta^+(\mathbf{p}') - \hat{\eta}_\beta^+(\mathbf{p}')\hat{\eta}_\alpha(\mathbf{p}) = \Delta(\mathbf{p}-\mathbf{p}')\Delta(\alpha-\beta), \qquad (3)$$

where index "$\alpha$" (or "$\beta$") denotes a set of quantum numbers characterizing a quantum state of the hydrogen-like atom.

Besides, an impact of an external electromagnetic field characterized by a scalar $\varphi^{(e)}(\mathbf{x},t)$ and a vector $\mathbf{A}^{(e)}(\mathbf{x},t)$ potentials is possible. The authors also take into account the existence of photons with the dispersion law $\omega(k)$ ($\omega$ is frequency, $\mathbf{k}$ is wave vector) by using creation operators $\hat{C}_\lambda^+(\mathbf{k})$ for a photon with the wave vector $\mathbf{k}$ and polarization $\lambda = 1, 2$ and annihilation operators $\hat{C}_\lambda(\mathbf{k})$,

$$\left[\hat{C}_\lambda(\mathbf{k}), \hat{C}_{\lambda'}^+(\mathbf{k}')\right] = \Delta(\mathbf{k}-\mathbf{k}'), \quad \omega(k) = ck, \qquad (4)$$

In terms of the introduced particles creation and annihilation operators (1) – (4) the Hamiltonian of low-temperature hydrogen-like plasma, according to Ref. 18 reads as:

$$\hat{H} = \hat{H}_0 + \hat{V}(t) + \hat{V}, \qquad (5)$$

where $\hat{H}_0$ is the free particles Hamiltonian:

$$\hat{H}_0 = \sum_{l=1}^{2}\sum_{\mathbf{p}} \varepsilon_l(\mathbf{p})\hat{a}_l^+(\mathbf{p})\hat{a}_l(\mathbf{p}) + \sum_{\alpha}\sum_{\mathbf{p}} \varepsilon_\alpha(\mathbf{p})\hat{\eta}_{\alpha\mathbf{p}}^+\hat{\eta}_{\alpha\mathbf{p}} + \sum_{\lambda,\mathbf{k}} \omega(\mathbf{k})\hat{C}_\lambda^+(\mathbf{k})\hat{C}_\lambda(\mathbf{k}),$$

$$\varepsilon_l(\mathbf{p}) = \frac{\mathbf{p}^2}{2m_l}, \quad l \equiv \{1,2\}, \qquad \varepsilon_\alpha(\mathbf{p}) = \varepsilon_\alpha + \frac{\mathbf{p}^2}{2M}, \qquad (6)$$

while the quantity $\varepsilon_\alpha < 0$ is the bound state (hydrogen-like atom) energy in the state of quantum numbers set $\alpha$. In (6) and in what follows, as usual, we formally equal the Planck constant $\hbar$ to unity, $\hbar \equiv 1$; if necessary, the dependence of the results on $\hbar$ is easily recovered.

Hamiltonian $\hat{V}(t)$ of interaction of the particles system with electromagnetic field can be written as follows:

$$\hat{V}(t) = -\frac{1}{c}\int d\mathbf{x}\,\hat{\mathbf{A}}(\mathbf{x},t)\hat{\mathbf{j}}(\mathbf{x}) + \frac{1}{2c^2}\int d\mathbf{x}\,\hat{\mathbf{A}}^2(\mathbf{x},t)\hat{I}(\mathbf{x}) + \int d\mathbf{x}\,\varphi^{(e)}(\mathbf{x},t)\hat{\sigma}(\mathbf{x}), \tag{7}$$

where the operator $\hat{\mathbf{A}}(\mathbf{x},t)$,

$$\hat{\mathbf{A}}(\mathbf{x},t) \equiv \mathbf{A}^{(e)}(\mathbf{x},t) + \hat{\mathbf{a}}(\mathbf{x}) \tag{8}$$

is a superposition of the external electromagnetic field vector potential $\mathbf{A}^{(e)}(\mathbf{x},t)$ and the radiation field vector potential $\hat{\mathbf{a}}(\mathbf{x})$, the latter in terms of photons creation and annihilation operators reads (see, e.g., Ref. 31):

$$\hat{\mathbf{a}}(\mathbf{x}) = \left(\frac{2\pi}{V}\right)^{1/2} c \sum_{\mathbf{k}} \sum_{\lambda=1}^{2} \omega^{-1/2}(k)\mathbf{e}_\lambda(\mathbf{k})\left\{\hat{C}_\lambda^+(\mathbf{k})e^{-i\mathbf{k}\mathbf{x}} + \hat{C}_\lambda(\mathbf{k})e^{i\mathbf{k}\mathbf{x}}\right\} \tag{9}$$

($\mathbf{e}_\lambda(\mathbf{k})$ is the polarization vector of a photon in the state $\mathbf{k}$ and $\lambda = 1, 2$). We note, that for the field we use Coulomb calibration.

The current density operator $\hat{\mathbf{j}}(\mathbf{x})$ in (7) is defined by the next formulas:

$$\begin{aligned}
\hat{\mathbf{j}}(\mathbf{x}) &= \sum_{a=0}^{2} \hat{\mathbf{j}}_a(\mathbf{x}), \qquad a = \{0, l\}; \\
\hat{\mathbf{j}}_l(\mathbf{x}) &= -\frac{ie_l}{2m_l V}\sum_{\mathbf{p},\mathbf{p}'} e^{i\mathbf{x}(\mathbf{p}-\mathbf{p}')}(\mathbf{p}+\mathbf{p}')\hat{a}_l^+(\mathbf{p})\hat{a}_l(\mathbf{p}'), \qquad e_1 = -e_2 = e; \\
\hat{\mathbf{j}}_0(\mathbf{x}) &= \frac{1}{V}\sum_{\mathbf{p},\mathbf{p}'}\sum_{\alpha,\beta} e^{i\mathbf{x}(\mathbf{p}'-\mathbf{p})}\left[\frac{(\mathbf{p}+\mathbf{p}')}{2M}\sigma_{\alpha\beta}(\mathbf{p}-\mathbf{p}') + \mathbf{j}_{\alpha\beta}(\mathbf{p}-\mathbf{p}')\right]\hat{\eta}_\alpha^+(\mathbf{p})\hat{\eta}_\beta(\mathbf{p}'),
\end{aligned} \tag{10}$$

where $V$ is the system's volume, $e$ is the elementary charge, $m_1$ and $m_2$ are the masses of an electron and a core, respectively, and the quantities $\sigma_{\alpha\beta}(\mathbf{k})$ и $\mathbf{j}_{\alpha\beta}(\mathbf{k})$ are given by the expressions

$$\begin{aligned}
\sigma_{\alpha\beta}(\mathbf{k}) &\equiv e\int d\mathbf{y}\,\varphi_\alpha^*(\mathbf{y})\varphi_\beta(\mathbf{y})\left[\exp\left(-i\frac{m_1}{M}\mathbf{k}\mathbf{y}\right) - \exp\left(i\frac{m_2}{M}\mathbf{k}\mathbf{y}\right)\right], \qquad M \equiv m_1 + m_2; \\
\mathbf{j}_{\alpha\beta}(\mathbf{k}) &\equiv e\frac{i}{2}\int d\mathbf{y}\left(\varphi_\alpha^*(\mathbf{y})\frac{\partial\varphi_\beta(\mathbf{y})}{\partial\mathbf{y}} - \frac{\partial\varphi_\alpha^*(\mathbf{y})}{\partial\mathbf{y}}\varphi_\beta(\mathbf{y})\right)\left[\frac{1}{m_1}\exp\left(i\frac{m_2}{M}\mathbf{k}\mathbf{y}\right) + \frac{1}{m_2}\exp\left(-i\frac{m_1}{M}\mathbf{k}\mathbf{y}\right)\right],
\end{aligned} \tag{11}$$

where $M$ is atom mass and $\varphi_\alpha(\mathbf{y})$ is the wave function of a hydrogen-like atom in state $\alpha$, and in considered to be known.

The quantity $\hat{\sigma}(\mathbf{x})$ in (7), is the density operator of system's charges:

$$\hat{\sigma}(\mathbf{x}) = \sum_{a=0}^{2} \hat{\sigma}_a(\mathbf{x}), \tag{12}$$

and,

$$\hat{\sigma}_l(\mathbf{x}) = \frac{e_l}{V} \sum_{\mathbf{p},\mathbf{p}'} e^{i\mathbf{x}(\mathbf{p}'-\mathbf{p})} \hat{a}_l^+(\mathbf{p}) \hat{a}_l(\mathbf{p}'), \quad \hat{\sigma}_0(\mathbf{x}) = \frac{1}{V} \sum_{\mathbf{p},\mathbf{p}'} \sum_{\alpha,\beta} e^{i\mathbf{x}(\mathbf{p}'-\mathbf{p})} \sigma_{\alpha\beta}(\mathbf{p}-\mathbf{p}') \hat{\eta}_\alpha^+(\mathbf{p}) \hat{\eta}_\beta(\mathbf{p}'). \tag{13}$$

Finally, the operator $\hat{I}(\mathbf{x})$ from (7), according to Ref. 31 can be written as

$$\hat{I}(\mathbf{x}) = \sum_{a=0}^{2} \hat{I}_a(\mathbf{x}), \tag{14}$$

where

$$\hat{I}_l(\mathbf{x}) \equiv \frac{e^2}{Vm_l} \sum_{\mathbf{p},\mathbf{p}'} e^{i\mathbf{x}(\mathbf{p}'-\mathbf{p})} \hat{a}_l^+(\mathbf{p}) \hat{a}_l(\mathbf{p}'), \quad \hat{I}_0(\mathbf{x}) \equiv \frac{1}{V} \sum_{\mathbf{p},\mathbf{p}'} \sum_{\alpha,\beta} e^{i\mathbf{x}(\mathbf{p}'-\mathbf{p})} I_{\alpha\beta}(\mathbf{p}-\mathbf{p}') \hat{\eta}_\alpha^+(\mathbf{p}) \hat{\eta}_\beta(\mathbf{p}'), \tag{15}$$

and the tensor $I_{\alpha\beta}(\mathbf{k})$ is defined by the following expression:

$$I_{\alpha\beta}(\mathbf{k}) \equiv e^2 \int d\mathbf{y}\, \varphi_\alpha^*(\mathbf{y}) \varphi_\beta(\mathbf{y}) \left[ \frac{1}{m_1} \exp\left(i\frac{m_2}{M}\mathbf{k}\mathbf{y}\right) + \frac{1}{m_2} \exp\left(-i\frac{m_1}{M}\mathbf{k}\mathbf{y}\right) \right]. \tag{16}$$

Thus, the equations (7) – (16) entirely define the Hamiltonian of a hydrogen-like low-temperature plasma interaction with electromagnetic field.

We note, that under the assumption of an external electromagnetic field weakness and in the principal approximation over fine structure's constant $e^2/\hbar c$ the interaction Hamiltonian $\hat{V}(t)$ of electromagnetic field with matter simplifies to a sum of the two Hamiltonians $\hat{V}_{ext}(t)$ and $\hat{V}_{int}(t)$:

$$\hat{V}(t) = \hat{V}_{ext}(t) + \hat{V}_{int}, \tag{17}$$

where the interaction Hamiltonian of matter with external electromagnetic field $\hat{V}_{ext}(t)$ obeys the following:

$$\hat{V}_{ext}(t) = \hat{V}_{ext}^{(1)}(t) + \hat{V}_{ext}^{(2)}(t) + \hat{V}_{ext}^{(0)}(t), \tag{18}$$

$$\hat{V}_{ext}^{(1)}(t) = \frac{e}{V}\sum_{\mathbf{p}_1\mathbf{p}_2}\left\{\frac{i}{2m_1 c}\mathbf{A}^{(e)}(\mathbf{p}_1-\mathbf{p}_2,t)(\mathbf{p}_1+\mathbf{p}_2)+\varphi^{(e)}(\mathbf{p}_1-\mathbf{p}_2,t)\right\}\hat{a}_1^+(\mathbf{p}_1)\hat{a}_1(\mathbf{p}_2),$$

$$\hat{V}_{ext}^{(2)}(t) = \frac{e}{V}\sum_{\mathbf{p}_1\mathbf{p}_2}\left\{\frac{i}{2m_2 c}\mathbf{A}^{(e)}(\mathbf{p}_1-\mathbf{p}_2,t)(\mathbf{p}_1+\mathbf{p}_2)+\varphi^{(e)}(\mathbf{p}_1-\mathbf{p}_2,t)\right\}\hat{a}_2^+(\mathbf{p}_1)\hat{a}_2(\mathbf{p}_2)$$

$$\hat{V}_{ext}^{(0)}(t) = -\frac{1}{Vc}\sum_{\alpha,\beta}\sum_{\mathbf{p}_1\mathbf{p}_2}\mathbf{A}^{(e)}(\mathbf{p}_1-\mathbf{p}_2,t)\left[\frac{(\mathbf{p}_1+\mathbf{p}_2)}{2M}\sigma_{\alpha\beta}(\mathbf{p}_1-\mathbf{p}_2)+\mathbf{j}_{\alpha\beta}(\mathbf{p}_1-\mathbf{p}_2)\right]\hat{\eta}_\alpha^+(\mathbf{p}_1)\hat{\eta}_\beta(\mathbf{p}_2)$$
$$+\frac{1}{V}\sum_{\mathbf{p}_1,\mathbf{p}_2}\sum_{\alpha,\beta}\varphi^{(e)}(\mathbf{p}_1-\mathbf{p}_2,t)\sigma_{\alpha\beta}(\mathbf{p}_1-\mathbf{p}_2)\hat{\eta}_\alpha^+(\mathbf{p}_1)\hat{\eta}_\beta(\mathbf{p}_2)$$

where the functions $\mathbf{A}^{(e)}(\mathbf{p},t)$, $\varphi^{(e)}(\mathbf{p},t)$ are Fourier-images of the external electromagnetic field potentials $\mathbf{A}^{(e)}(\mathbf{x},t)$ and $\varphi^{(e)}(\mathbf{x},t)$:

$$\mathbf{A}^{(e)}(\mathbf{p},t) = \int d\mathbf{x} e^{-i\mathbf{p}\mathbf{x}}\mathbf{A}^{(e)}(\mathbf{x},t),\; \varphi^{(e)}(\mathbf{p},t) = \int d\mathbf{x} e^{-i\mathbf{p}\mathbf{x}}\varphi^{(e)}(\mathbf{x},t). \tag{19}$$

The Hamiltonian of interaction of matter with radiation is given by these relations:

$$\hat{V}_{int} = \hat{V}_{int}^{(1)} + \hat{V}_{int}^{(2)} + \hat{V}_{int}^{(0)}, \tag{20}$$

$$\hat{V}_{int}^{(0)} \equiv -\sum_{\lambda=1}^{2}\sum_{\mathbf{k},\mathbf{p}}\mathbf{e}_\lambda(\mathbf{k})\left(\frac{2\pi}{V\omega_\mathbf{k}}\right)^{1/2}\sum_{\alpha,\beta}\left[\frac{(2\mathbf{p}-\mathbf{k})}{2M}\sigma_{\alpha\beta}(-\mathbf{k})+\mathbf{j}_{\alpha\beta}(-\mathbf{k})\right]\hat{\eta}_{\alpha\mathbf{p}}^+\hat{\eta}_{\beta\mathbf{p}-\mathbf{k}}\hat{C}_{\mathbf{k}\lambda}$$
$$-\sum_{\lambda=1}^{2}\sum_{\mathbf{k},\mathbf{p}}\mathbf{e}_\lambda(\mathbf{k})\left(\frac{2\pi}{V\omega_\mathbf{k}}\right)^{1/2}\sum_{\alpha,\beta}\left[\frac{(2\mathbf{p}+\mathbf{k})}{2M}\sigma_{\alpha\beta}(\mathbf{k})+\mathbf{j}_{\alpha\beta}(\mathbf{k})\right]\hat{\eta}_{\alpha\mathbf{p}}^+\hat{\eta}_{\beta\mathbf{p}+\mathbf{k}}\hat{C}_{\mathbf{k}\lambda}^+$$

$$\hat{V}_{int}^{(1)} = \frac{e}{2m_1}\sum_{\mathbf{p},\mathbf{p}',\mathbf{k}}\sum_{\lambda=1}^{2}\left(\frac{2\pi}{V\omega_\mathbf{k}}\right)^{1/2}\mathbf{e}_\lambda(\mathbf{k})(\mathbf{p}'+\mathbf{p})\hat{a}_1^+(\mathbf{p})\hat{a}_1(\mathbf{p}')\left(\Delta(\mathbf{p}'-\mathbf{p}+\mathbf{k})\hat{C}_{\mathbf{k}\lambda}+\Delta(\mathbf{p}'-\mathbf{p}-\mathbf{k})\hat{C}_{\mathbf{k}\lambda}^+\right)$$

$$\hat{V}_{\text{int}}^{(2)} = -\frac{e}{2m_2} \sum_{\mathbf{p},\mathbf{p}',\mathbf{k}} \sum_{\lambda=1}^{2} \left(\frac{2\pi}{V\omega_{\mathbf{k}}}\right)^{1/2} \mathbf{e}_\lambda(\mathbf{k})(\mathbf{p}'+\mathbf{p})\hat{a}_2^+(\mathbf{p})\hat{a}_2(\mathbf{p}')\left[\Delta(\mathbf{p}'-\mathbf{p}+\mathbf{k})\hat{C}_{\mathbf{k}\lambda} + \Delta(\mathbf{p}'-\mathbf{p}-\mathbf{k})\hat{C}_{\mathbf{k}\lambda}^+\right].$$

We note, that the Hamiltonian (18) plays the major role in describing the system's response to an external perturbation with a weak electromagnetic field (see Refs. 15-23). The Hamiltonian (20) defines relaxation processes in the photonic subsystem. Actually, this Hamiltonian correctly takes into account the photon emission and absorption processes, but leaves behind the scattering of photons by atoms. The latter is under the responsibility of the unaccounted in Hamiltonians (17) – (20) summands, which are quadratic over the fine structure constant $e^2/\hbar c$ (the full Hamiltonian (7) does contain them). However, the same contribution to relaxation processes in the system (e.g. to collision integral) is given also by a quadratic approximation over $\hat{V}_{\text{int}}$, see Ref. 31.

Finally, we obtain the last summand in (5), the interaction Hamiltonian $\hat{V}$ of particles of all system components. According to Ref. 18 it can be defined as the three summands:

$$\hat{V} = \hat{V}^{(1)} + \hat{V}^{(2)} + \hat{V}^{(3)}, \tag{21}$$

where $\hat{V}^{(1)}$ is the interaction Hamiltonian between free fermions of both kinds and hydrogen-like atoms:

$$\hat{V}^{(1)} = \frac{e}{V} \sum_{\mathbf{p}_1\mathbf{p}_2\mathbf{p}_3\mathbf{p}_4} \Phi_{\alpha\beta}(\mathbf{p}_1,\mathbf{p}_2;\mathbf{p}_3,\mathbf{p}_4)\hat{\eta}_\alpha^+(\mathbf{p}_3)\hat{\eta}_\beta(\mathbf{p}_4)\left\{a_2^+(\mathbf{p}_1)a_2(\mathbf{p}_2) - a_1^+(\mathbf{p}_1)a_1(\mathbf{p}_2)\right\}, \tag{22}$$

$$\Phi_{\alpha\beta}(\mathbf{p}_1,\mathbf{p}_2;\mathbf{p}_3,\mathbf{p}_4) \equiv \Delta(\mathbf{p}_4 - \mathbf{p}_3 - \mathbf{p}_1 + \mathbf{p}_2)\nu(\mathbf{p}_1 - \mathbf{p}_2)\sigma_{\alpha\beta}(\mathbf{p}_2 - \mathbf{p}_1).$$

The Hamiltonian $\hat{V}^{(2)}$ in (21) describes interaction between atoms in different quantum-mechanical states:

$$\hat{V}^{(2)} = \frac{1}{4V} \sum_{\mathbf{p}_1\mathbf{p}_2\mathbf{p}_3\mathbf{p}_4} \Phi_{\alpha_1\alpha_2;\alpha_3\alpha_4}(\mathbf{p}_1,\mathbf{p}_2;\mathbf{p}_3,\mathbf{p}_4)\hat{\eta}_{\alpha_1}^+(\mathbf{p}_1)\hat{\eta}_{\alpha_2}^+(\mathbf{p}_2)\hat{\eta}_{\alpha_3}(\mathbf{p}_3)\hat{\eta}_{\alpha_4}(\mathbf{p}_4), \tag{23}$$

$$\Phi_{\alpha_1\alpha_2;\alpha_3\alpha_4}(\mathbf{p}_1,\mathbf{p}_2;\mathbf{p}_3,\mathbf{p}_4) \equiv \frac{1}{2V}\Delta(\mathbf{p}_4 + \mathbf{p}_3 - \mathbf{p}_1 - \mathbf{p}_2)$$
$$\times\{\nu(\mathbf{p}_3 - \mathbf{p}_2)\sigma_{\alpha_1\alpha_4}(\mathbf{p}_3 - \mathbf{p}_2)\sigma_{\alpha_2\alpha_3}(\mathbf{p}_2 - \mathbf{p}_3) + \nu(\mathbf{p}_3 - \mathbf{p}_1)\sigma_{\alpha_2\alpha_4}(\mathbf{p}_3 - \mathbf{p}_1)\sigma_{\alpha_1\alpha_3}(\mathbf{p}_1 - \mathbf{p}_3)$$
$$+\nu(\mathbf{p}_4 - \mathbf{p}_2)\sigma_{\alpha_1\alpha_3}(\mathbf{p}_4 - \mathbf{p}_2)\sigma_{\alpha_2\alpha_4}(\mathbf{p}_2 - \mathbf{p}_4) + \nu(\mathbf{p}_4 - \mathbf{p}_1)\sigma_{\alpha_2\alpha_3}(\mathbf{p}_4 - \mathbf{p}_1)\sigma_{\alpha_1\alpha_4}(\mathbf{p}_1 - \mathbf{p}_4)\},$$

and the Hamiltonian $\hat{V}^{(3)}$ defines the interaction of free fermions with each other:

$$\hat{H}_{int}^{(3)} = -\frac{e^2}{V} \sum_{\mathbf{p}_1\mathbf{p}_2\mathbf{p}_3\mathbf{p}_4} \Phi_1^{(3)}(\mathbf{p}_1,\mathbf{p}_2;\mathbf{p}_3,\mathbf{p}_4) \hat{a}_1^+(\mathbf{p}_2)\hat{a}_1(\mathbf{p}_3)\hat{a}_2^+(\mathbf{p}_1)\hat{a}_2(\mathbf{p}_4)$$
$$+\frac{1}{4}\sum_{\mathbf{p}_1\mathbf{p}_2\mathbf{p}_3\mathbf{p}_4}\Phi_2^{(3)}(\mathbf{p}_1,\mathbf{p}_2;\mathbf{p}_3,\mathbf{p}_4)\left[\hat{a}_{1\mathbf{p}_1}^+\hat{a}_{1\mathbf{p}_2}^+\hat{a}_{1\mathbf{p}_3}\hat{a}_{1\mathbf{p}_4} + \hat{a}_{2\mathbf{p}_1}^+\hat{a}_{2\mathbf{p}_2}^+\hat{a}_{2\mathbf{p}_3}\hat{a}_{2\mathbf{p}_4}\right],$$

(24)

$$\Phi_1^{(3)}(\mathbf{p}_1,\mathbf{p}_2;\mathbf{p}_3,\mathbf{p}_4) \equiv \Delta(\mathbf{p}_4-\mathbf{p}_1+\mathbf{p}_3-\mathbf{p}_2)\nu(\mathbf{p}_2-\mathbf{p}_3)$$

$$\Phi_2^{(3)}(\mathbf{p}_1,\mathbf{p}_2;\mathbf{p}_3,\mathbf{p}_4) \equiv \frac{e^2}{2V}\Delta(\mathbf{p}_4-\mathbf{p}_1+\mathbf{p}_3-\mathbf{p}_2)$$
$$\times\left(\nu(\mathbf{p}_2-\mathbf{p}_3)+\nu(\mathbf{p}_1-\mathbf{p}_4)-\nu(\mathbf{p}_1-\mathbf{p}_3)-\nu(\mathbf{p}_2-\mathbf{p}_4)\right).$$

The quantity $\nu(\mathbf{p})$ in equations (22) – (24) is a Fourier-image of Coulomb potential divided by $e^2$, elementary charge squared:

$$\nu(\mathbf{p}) = \frac{4\pi}{\mathbf{p}^2}. \tag{25}$$

Thus, the expressions (1) – (25) define all types of interactions between the components of weakly ionized hydrogen-like gas at low temperatures and the interaction of the system with an external random field. As stated above, the system Hamiltonian in form of (5) will be used to describe the system within the specifically modified reduced description method framework. However, we first set out the main ideas of this method closely following the wordings of Ref. 31.

### III. MAIN REDUCED DESCRIPTION METHOD STATEMENTS FOR SPATIALLY INHOMOGENEOUS STATES OF QUANTUM MULTIPARTICLE SYSTEMS

Let us start from the most general formulations of the reduced description method presented in Ref. 31 for spatially inhomogeneous quantum multiparticle systems. As mentioned above, the method describes the systems with the Hamiltonian

$$\hat{H} = \hat{H}_0 + \hat{V}, \tag{26}$$

where $\hat{H}_0$ includes the main interactions, and $\hat{V}$ accounts for relatively weak. In case the Hamiltonian $\hat{H}$ does not depend on time (conservative systems), the system is described by a statistical operator (or density matrix) $\rho(t)$ that obeys the Liouville equation (we remind, that in our expressions we assume $\hbar=1$):

$$i\frac{\partial \rho(t)}{\partial t} = \left[\hat{H}, \rho(t)\right]. \tag{27}$$

Ergodic hypothesis, which is one of the postulates of the reduced description method of systems, claims that in a conservative system at sufficiently long times ($t \to \infty$) as a result of evolution, an equilibrium distribution is established. The latter is described with the Gibbs statistical operator $w$:

$$\rho(t) = e^{-i\hat{H}t} \rho e^{i\hat{H}t} \underset{t \to \infty}{\to} w, \tag{28}$$

where $\rho$ is the initial value of the statistical operator, and the Gibbs operator takes the form of

$$w = \exp\{\Omega - Y_a \hat{\gamma}_a\}, \tag{29}$$

in which operators $\hat{\gamma}_a$ are the operators of additive motion integrals with respect to the Hamiltonian $\hat{H}$:

$$\left[\hat{H}, \hat{\gamma}_a\right] = 0. \tag{30}$$

Index $a$ in formulas (29), (30) enumerates all the set of these additive motion integrals; the repeated index implies summation over it (see (29)). Thermodynamic potential $\Omega$ and generalized thermodynamic forces $Y_a(\gamma)$ are defined from equations

$$\text{Sp } w = 1, \quad \text{Sp } w\hat{\gamma}_a = \gamma_a, \tag{31}$$

in which the quantities $\gamma_a$ are known mean values of the additive integrals in the system.

However, if we study the system's evolution with the "truncated" or incomplete Hamiltonian $\hat{H}_0$, then, after sufficiently large time, the statistical operator $\rho(t)$ in general will not converge to the Gibbs equilibrium distribution (find this in more details in Ref. 31). And, according to the ergodic relation (28) we may prove that at large times $t \gg \tau_0$ (where $\tau_0$ is a so-called chaotization time), a universal state different from the equilibrium establishes. This universal state is characterized by some set of operators $\hat{\gamma}_a$ (in general case other than the ones from (28) – (31)), defined by the structure of the Hamiltonian $\hat{H}_0$ and its symmetrical properties[31]. The latter statement can be presented as follows:

$$e^{-i\hat{H}_0 t}\rho e^{i\hat{H}_0 t} \underset{t\to\infty}{\to} \rho^{(0)}\left(e^{iat}\operatorname{Sp}\rho\hat{\gamma}_a\right), \qquad (32)$$

where $\rho$ is the initial value of the system statistical operator, and statistical operator $\rho^{(0)}$ obeys the formula

$$\rho^{(0)}(\gamma) = \exp\{\Omega(\gamma) - \hat{\gamma}_a Y_a(\gamma)\}, \qquad (33)$$

in which the quantities $\Omega(\gamma)$ and $Y_a(\gamma)$ are retrieved from equations

$$\operatorname{Sp}\rho^{(0)} = 1, \quad \operatorname{Sp}\rho^{(0)}\hat{\gamma}_a = \gamma_a. \qquad (34)$$

The quantities $\gamma_a$, the operators of which are present in (32) – (34), represent the sought-for system's reduced description parameters; the motion equations for them will serve the system's evolution equations at times $t \gg \tau_0$. These equations need to be derived from the Liouville equation (27). Index "$a$" enumerates the whole set of the reduced description parameters $\gamma_a$. As mentioned above, the operators $\hat{\gamma}_a$ depend on the symmetry properties of the Hamiltonian $\hat{H}_0$, that is reflected in (32), where the matrix $\|a\|$, is defined by a structure and symmetry of the Hamiltonian $\hat{H}_0$:

$$\left[\hat{H}_0, \hat{\gamma}_a\right] = a_{ab}\hat{\gamma}_b. \qquad (35)$$

In equations (33), (35) the repeating indexes $a, b$ imply summation over them. We note, that the derivation of the set of operators $\hat{\gamma}_a$ for known Hamiltonian $\hat{H}_0$ may be quite a complex task. In the general formulations of the reduced description method it is supposed to be already solved, and the main attention is paid to the derivation of the evolutional equations for these reduced description parameters.

The circumstance that the quantities $\gamma_a(t) = Sp\rho(t)\hat{\gamma}_a$ are the reduced description parameters implies that taking into account all the interactions in the system at times $t \gg \tau_0$ the dependence on time and the initial statistical operator $\rho = \rho(0)$ is expressed only through the parameters $\gamma_a(t;\rho)$:

$$e^{-i\hat{H}t}\rho e^{i\hat{H}t} \underset{t\gg\tau_0}{\to} \sigma(\gamma(t;\rho)). \qquad (35)$$

If the system evolution has been described with only Hamiltonian $\hat{H}_0$, then the parameters $\gamma_a(t)$ would have evolve in time as

$$\gamma_a(t) = \left[\exp(iat)\gamma(0)\right]_a. \tag{36}$$

When taking into account also the weak interactions $\hat{V}$ (see (26)) the quantities $\gamma_a(t)$ experience additional change that should be weak compared to that given by (36) and caused by the Hamiltonian $\hat{H}_0$, that leads to an establishment of a universal distribution $\rho^{(0)}(\gamma)$, see (33). I.e., the system will manage to "tune up" for moment nonequilibrium values of the parameters $\gamma_a(t)$. This should take place since the relaxation time $\tau_r$ of the parameters $\gamma_a(t)$ to their equilibrium values is much larger than the chaotization time $\tau_0$, $\tau_r \gg \tau_0$. This inequality is always fulfilled in case of a weak interaction, because the chaotization time does not depend on the interaction intensity, but the relaxation time $\tau_r$ is defined by the interaction intensity $\hat{V}$ itself as it tends to infinity $\tau_r \to \infty$ at $\hat{V} \to 0$, see Ref. 31.

The statistical operator $\sigma(\gamma(t;\rho))$ is called the coarse (crude) statistical operator. The forthcoming problem is while not considering the processes that lead to the establishment of the rude statistical operator $\sigma(\gamma(t;\rho))$, to find the structure of this operator and the dependence of the reduced description parameters $\gamma_a(t;\rho)$ on time and the initial statistical operator $\rho$. Since $\sigma(\gamma(t;\rho))$ is a statistical operator at moment of time $t$, when the parameters $\gamma_a$ have the values of $\gamma_a(t;\rho)$, then the following should fulfill:

$$\mathrm{Sp}\,\sigma(\gamma(t;\rho))\hat{\gamma}_a = \gamma_a(t;\rho), \tag{37}$$

while the rude statistical operator $\sigma(\gamma(t;\rho))$ is also to obey the Liouville equation (see (27)):

$$i\frac{\partial \sigma(\gamma(t;\rho))}{\partial t} = \left[\hat{H}, \sigma(\gamma(t;\rho))\right], \tag{38}$$

or, considering that this operator depends on time only through the description parameters $\gamma_a(t;\rho)$:

$$i\frac{\partial \sigma(\gamma)}{\partial \gamma_a}\dot{\gamma}_a = \left[\hat{H},\sigma(\gamma)\right], \tag{39}$$

The "boundary" condition for the Liouville equation (39) arises from the ergodic relation (32) – (33) and may be presented as follows:

$$e^{-i\hat{H}_0\tau}\sigma(\gamma)e^{i\hat{H}_0\tau} \underset{\tau\to\infty}{\to} \rho^{(0)}\left(e^{ia\tau}\gamma\right), \tag{40}$$

or, in an equivalent form (see Ref. 31):

$$\lim_{\tau\to\infty} e^{-i\hat{H}_0\tau}\sigma\left(e^{-ia\tau}\gamma\right)e^{i\hat{H}_0\tau} \underset{\tau\to\infty}{\to} \rho^{(0)}(\gamma). \tag{41}$$

Multiplying the Liouville equation (38) by $\hat{\gamma}_a$ and taking a trace, remembering (26), (37) we get

$$\dot{\gamma}_a = i\operatorname{Sp}\sigma(\gamma)\left[\hat{H},\hat{\gamma}_a\right] \equiv L_a(\gamma), \quad L_a(\gamma) = i\operatorname{Sp}\sigma(\gamma)\left[\hat{H}_0+\hat{V},\hat{\gamma}_a\right]. \tag{42}$$

Now equation (39) is

$$i\frac{\partial \sigma(\gamma)}{\partial \gamma_a}L_a(\gamma) = \left[\hat{H},\sigma(\gamma)\right]. \tag{39}$$

Ref. 31 clarifies in detail the derivation of an integral equation for $\sigma(\gamma)$ by using Eqs. (32) – (39), which will be convenient to develop the perturbation theory over the weak interaction $\hat{V}$:

$$\sigma(\gamma) = \rho^{(0)}(\gamma) + i\int_{-\infty}^{0} d\tau\, e^{\eta\tau}e^{i\hat{H}_0\tau}\left\{\left[\sigma(\gamma),\hat{V}\right] + i\frac{\partial\sigma(\gamma)}{\partial\gamma_a}L_a(\gamma)\right\}_{\gamma\to e^{ia\tau}\gamma} e^{-i\hat{H}_0\tau}, \tag{40}$$

and the development of the motion equations for the parameters $\gamma_a$ within the framework of the same perturbation theory:

$$\dot{\gamma}_a = L_a^{(0)}(\gamma) + L_a(\gamma), \tag{41}$$

where we introduce

$$L_a^{(0)}(\gamma) \equiv iSp\sigma(\gamma)\left[\hat{H}_0, \hat{\gamma}_a\right] = a_{ab}\gamma_b, \quad L_a(\gamma) \equiv iSp\sigma(\gamma)\left[\hat{V}, \hat{\gamma}_a\right]. \tag{42}$$

Eqs. (40) – (42) show that the statistical operator $\sigma(\gamma)$ may be obtained in an arbitrary order of the perturbation theory over the weak interaction. This statement refers also to the evolution equations for the reduced description parameters, see (41), (42). For example, in the second order of the perturbation theory $\hat{V}$ these equations look like[31]:

$$\begin{aligned}
\dot{\gamma}_a &= L_a^{(0)}(\gamma) + L_a^{(1)}(\gamma) + L_a^{(2)}(\gamma), \\
L_a^{(0)}(\gamma) &= a_{ab}\gamma_b, \quad L_a^{(1)}(\gamma) = i\operatorname{Sp}\rho^{(0)}(\gamma)\left[\hat{V}, \hat{\gamma}_a\right], \\
L_a^{(2)}(\gamma) &= -i\int_{-\infty}^{0} d\tau e^{\eta\tau} \operatorname{Sp}\rho^{(0)}(\gamma)\left[\hat{V}(\tau), \left[\hat{V}, \hat{\gamma}_a\right] + i\hat{\gamma}_b \frac{\partial L_a^{(1)}(\gamma)}{\partial \gamma_b}\right], \\
\hat{V}(\tau) &\equiv e^{i\hat{H}_0\tau}\hat{V}e^{-i\hat{H}_0\tau}.
\end{aligned} \tag{43}$$

We note, that the parameter $\eta$ in (40), (43) should be made vanished after calculating of averages with operator $\sigma(\gamma)$ (in any order of the perturbation theory $\hat{V}$) and calculating the integrals over $\tau$.

These obtained expressions (40), (43) will be used to derive the kinetic equations for weakly ionized low-temperature gas of hydrogen-like atoms.

## IV. KINETIC EQUATIONS FOR WEAKLY IONIZED GAS OF HYDROGEN-LIKE ATOMS.

Ref. 31 explains, that for quantum gases with a Hamiltonian

$$\hat{H} = \hat{H}_0 + \hat{V}, \tag{44}$$

where

$$\hat{H}_0 = \sum_i \varepsilon_i \hat{a}_i^+ \hat{a}_i, \quad \hat{V} = \frac{1}{4V}\sum_{i_1 i_2 i_3 i_4} \Phi(i_1 i_2; i_3 i_4) \hat{a}_{i_1}^+ \hat{a}_{i_2}^+ \hat{a}_{i_3} \hat{a}_{i_4}, \tag{45}$$

as examples of the reduced description parameters $\gamma_a$ introduced in the latter Section formally enough a single particle density matrix $f_{i,i'}$ may be used; the following operators correspond to it:

$$\hat{f}_{i,i'} = \hat{a}_{i'}^+ \hat{a}_i, \tag{46}$$

where $i$ is a set of quantum numbers characterizing the particle state (e.g. momentum $\mathbf{p}$, spin projection $s$). This description, as we already mentioned above, arises after the chaotization time $\tau_0$, $t \gg \tau_0$. Note, that the quantity $\varepsilon_i$ in (46) is an energy of free particle (or quasi-particle).

It is shown that in case of a single particle density matrix as a sample of reduced description parameters for such system, equations (43) together with (44) – (46) read as

$$\dot{f}_{i,i'} = L^{(0)}_{i,i'}(f) + L^{(1)}_{i,i'}(f) + L^{(2)}_{i,i'}(f),$$

$$L^{(0)}_{i,i'}(f) = i \operatorname{Sp} \rho^{(0)}(f)\left[\hat{H}_0, \hat{a}^+_{i'}\hat{a}_i\right], \quad L^{(1)}_{i,i'}(f) = i \operatorname{Sp} \rho^{(0)}(f)\left[\hat{V}, \hat{a}^+_{i'}\hat{a}_i\right],$$

$$L^{(2)}_{i,i'}(f) = -i \int_{-\infty}^{0} d\tau e^{\eta\tau} \operatorname{Sp} \rho^{(0)}(f)\left[\hat{V}(\tau), \left[\hat{V}, \hat{a}^+_{i'}\hat{a}_i\right] + i \sum_{i_1 i'_1} \hat{a}^+_{i'_1}\hat{a}_{i_1} \frac{\partial L^{(1)}_a(\gamma)}{\partial f_{i_1, i'_1}}\right], \quad (47)$$

$$\hat{V}(\tau) \equiv e^{i\hat{H}_0 \tau} \hat{V} e^{-i\hat{H}_0 \tau},$$

while the statistical operator $\rho^{(0)}(f)$ (see also (32), (33)) is given with

$$\rho^{(0)}(f) = \exp\left\{\Omega(f) - \sum_{ii'} Y_{i,i'}(f)\hat{a}^+_{i'}\hat{a}_i\right\}, \tag{48}$$

in which $\Omega(f)$ and $Y_{i,i'}(f)$ being functionals of the single particle density matrix are to be obtained from the following equations:

$$\operatorname{Sp} \rho^{(0)}(f) = 1, \quad \operatorname{Sp} \rho^{(0)}(f)\hat{a}^+_{i'}\hat{a}_i = f_{i,i'}. \tag{49}$$

After we calculate the commutation relations and traces in (47) taking into account (48), (49) these equations may be reduced to a coupled kind[31]:

$$\frac{\partial f}{\partial t} + i[\varepsilon, f] = L(f), \tag{50}$$

where $\varepsilon_{i,i'}$ is defined by the formula

$$\varepsilon_{i,i'} = \varepsilon_i \delta_{i,i'} + \frac{1}{V} \sum_{i_1 i'_1} \Phi(ii_1; i'_1 i') f_{i'_1, i_1}, \tag{51}$$

and the collision integral $L(f)$ for bosons equals:

$$L_{ii'}(f) = \sum_{i_1 i_2 i_3 i_4} \sum_{i'_1 i'_2 i'_3 i'_4} \Phi(i_1 i_2; i_3 i_4) \Phi(i'_1 i'_2; i'_3 i'_4) \delta_-\left(\varepsilon_{i'_1} + \varepsilon_{i'_2} - \varepsilon_{i'_3} - \varepsilon_{i'_4}\right)$$
$$\times \left\{ f_{i'_4, i_2} f_{i'_3, i_1} \left(\delta_{i_3, i'_1} + f_{i_3, i'_1}\right)\left(\delta_{i, i'_2} + f_{i, i'_2}\right) - f_{i_3, i'_1} f_{i, i'_2}\left(\delta_{i'_1, i'_3} + f_{i'_3, i'_1}\right)\left(\delta_{i'_4, i_2} + f_{i'_4, i_2}\right)\right\} \delta_{i_4 i'} + h.c., \quad (52)$$

where

$$\delta_-(x) \equiv \frac{1}{\pi} \lim_{\eta \to 0} \int_{-\infty}^{0} d\tau e^{i\tau x + \eta \tau}. \quad (53)$$

For fermions, equation (52) has a "minus" instead of a "plus" inside the brackets. We should also note that the left-hand side of equation (50) contains the quantity $\varepsilon_{i,i'}$, which has the amendments to particle free energy $\varepsilon_i$, that are connected with interaction and a single particle density matrix $f_{i,i'}$ (a distribution function, see below). That is why the quantity $\varepsilon_{i,i'}$ also takes into account the effects of and average field. Thus, equation (50) together with the self-consistent field (51) and the collision integral (52), (53) reflects the kinetic equation of the system characterized by the Hamiltonians (44), (45).

To demonstrate this we should pass in equations (50) – (52) (see Ref. 31), to a momentum representation, that is, the quantities characterized by indexes $i$ one should count with a set of momentums $\mathbf{p}$ with a respective enumeration. In this case it is convenient to introduce the Wigner distribution function instead of the single particle density matrix $f_{\mathbf{p},\mathbf{p}'} \equiv Sp \rho^{(0)}(f) \hat{a}^+_{\mathbf{p}'} \hat{a}_{\mathbf{p}}$ (see (48), (49)):

$$f(\mathbf{x},\mathbf{p}) \equiv \sum_{\mathbf{k}} e^{-i\mathbf{k}\mathbf{x}} f_{\mathbf{p}-\frac{\mathbf{k}}{2},\mathbf{p}+\frac{\mathbf{k}}{2}} = \frac{V}{(2\pi)^3} \int d^3k e^{-i\mathbf{k}\mathbf{x}} f_{\mathbf{p}-\frac{\mathbf{k}}{2},\mathbf{p}+\frac{\mathbf{k}}{2}}. \quad (54)$$

Please, pay attention that in a space-homogeneous case the quantity $f_{\mathbf{p}-\frac{\mathbf{k}}{2},\mathbf{p}+\frac{\mathbf{k}}{2}}$ equals to $f_{\mathbf{p}} \delta_{\mathbf{k},0}$. Consequently, $f_{\mathbf{p}-\frac{\mathbf{k}}{2},\mathbf{p}+\frac{\mathbf{k}}{2}}$ should have sharp maximum at $\mathbf{k}=0$. According to equations (50) – (53), we have a following evolution equation for Wigner distribution function $f(\mathbf{x},\mathbf{p})$ (the kinetic equation) in the perturbation theory over small spatial gradients[31]:

$$\frac{\partial f(\mathbf{x},\mathbf{p})}{\partial t}+\frac{\partial \varepsilon(\mathbf{x},\mathbf{p})}{\partial \mathbf{p}}\frac{\partial f(\mathbf{x},\mathbf{p})}{\partial \mathbf{x}}-\frac{\partial \varepsilon(\mathbf{x},\mathbf{p})}{\partial \mathbf{x}}\frac{\partial f(\mathbf{x},\mathbf{p})}{\partial \mathbf{p}}=L(\mathbf{p};f), \tag{54}$$

where particle (or quasi-particle, see Ref. 31) energy $\varepsilon(\mathbf{x},\mathbf{p})$ and the collision integral $L(\mathbf{p};f)$ are defined as follows:

$$\begin{aligned}
\varepsilon(\mathbf{x},\mathbf{p}) &\equiv \sum_{\mathbf{k}} e^{-i\mathbf{k}\mathbf{x}} \varepsilon_{\mathbf{p}-\frac{\mathbf{k}}{2},\mathbf{p}+\frac{\mathbf{k}}{2}} = \frac{V}{(2\pi)^3}\int d^3k\, e^{-i\mathbf{k}\mathbf{x}} \varepsilon_{\mathbf{p}-\frac{\mathbf{k}}{2},\mathbf{p}+\frac{\mathbf{k}}{2}}, \\
L(\mathbf{p};f) &\equiv \frac{\pi}{V^2}\sum_{\mathbf{p}_1\mathbf{p}_2\mathbf{p}_3\mathbf{p}_4} |\Phi(\mathbf{p}_1\mathbf{p}_2;\mathbf{p}_3\mathbf{p}_4)|^2 \delta(\varepsilon(\mathbf{p}_1)+\varepsilon(\mathbf{p}_2)-\varepsilon(\mathbf{p}_3)-\varepsilon(\mathbf{p}_4)) \\
&\quad \times \delta_{\mathbf{p}_4,\mathbf{p}} \{f(\mathbf{p}_1)f(\mathbf{p}_2)[1+f(\mathbf{p}_3)][1+f(\mathbf{p}_4)]-f(\mathbf{p}_3)f(\mathbf{p}_4)[1+f(\mathbf{p}_1)][1+f(\mathbf{p}_2)]\},
\end{aligned} \tag{55}$$

in which $\varepsilon(\mathbf{p})$ is an energy of a free particle or quasi-particle. We note, that a corresponding relation for the collision integral $L(\mathbf{p};f)$ in (55) is correct also for fermions if the sign "plus" in square braces is replaced with a "minus". It is clear, that a kinematic part of equation (54) has the same shape as that of a classical kinetic equation if the energy $\varepsilon(\mathbf{x},\mathbf{p})$ is the particle's Hamiltonian $\varepsilon_{\mathbf{p}}$. However, the collision integral in (55) substantially differs from that of a classical case since in involves the influence of the particles statistics.

The derivation of the kinetic equation (54), (55) within the framework of the reduced description method is given here in detail in order to demonstrate how the suggested procedure may be used to construct the kinetic theory of weakly ionized gases of hydrogen-like atoms.

## V. KINETIC THEORY OF WEAKLY IONIZED GAS OF HYDROGEN-LIKE ATOMS WITHIN THE REDUCED DESCRIPTION METHOD

To construct the stated theory we need to somewhat modify the presented above method mainly taking into account the fact, that in the case of weakly ionized gases of hydrogen-like atoms we deal with a multicomponent system. The core of this description, as we already mentioned, are the system Hamiltonians defined with formulas (5) – (25) (see also Ref. 18). However, here it is worth to remember the following. The Hamiltonian (5) differs from the Hamiltonian (26) or (44), (45) by an additional summand, $\hat{V}(t)$, that describes the interaction of the system's components with external field and irradiation (photons), see (5) – (20). In what follows we neglect this additional summand that allows us to use the results of the previous Section to construct the stated kinetic theory with minor modifications. In fact, this neglecting of $\hat{V}(t)$ is not necessary in view of any principled motivations. Indeed, we could include a single particle photons density matrix $f_{\lambda\mathbf{k},\lambda'\mathbf{k}'}$ into the set of the reduced description parameters having defined it with the formulas (see also (48), (49)):

$$f_{\lambda\mathbf{k},\lambda'\mathbf{k}'} \equiv Sp\rho^{(0)}(f)\hat{f}_{\lambda\mathbf{k},\lambda'\mathbf{k}'}, \quad \hat{f}_{\lambda\mathbf{k},\lambda'\mathbf{k}'} = \hat{C}^+_{\mathbf{k}'\lambda'}\hat{C}_{\mathbf{k}\lambda}. \tag{56}$$

This would allow adding the Hamiltonian $\hat{V}_{int}$ (see (20)) to the set of the interaction Hamiltonians (21) and eventually receiving, following Ref. 31, a kinetic equation for Wigner distribution function of photons. As we show below, this would substantially complicate the calculations and the clarity of results. Taking into account also that the photons have a weak influence on the relaxation processes in a medium, it is possible to therefore neglect the summand $\hat{V}_{int}$ in the Hamiltonian $\hat{V}(t)$. Due to this reason we also exclude from considerations the summand $\sum_{\lambda,\mathbf{k}}\omega(\mathbf{k})\hat{C}^+_\lambda(\mathbf{k})\hat{C}_\lambda(\mathbf{k})$, that defines the kinetic energy of free photons in (6). However, if one is interested in the relaxation processes of photons in a medium, than it is necessary to write the kinetic equation for distribution function of photons (56) and consider both the summand $\sum_{\lambda,\mathbf{k}}\omega(\mathbf{k})\hat{C}^+_\lambda(\mathbf{k})\hat{C}_\lambda(\mathbf{k})$ and the Hamiltonian $\hat{V}_{int}$. These are the summands contained in $\hat{V}_{int}$ that determine the photon subsystem relaxation, see in this regard Ref. 31.

As for the Hamiltonian $\hat{V}_{ext}(t)$ (see (18)) in $\hat{V}(t)$, which is connected with the interaction between the matter and the external electromagnetic field, its influence on the system's evolution in some cases may also be accounted for. In particular, Ref. 31 presents a detailed modification procedure of the reduced description method for the case when an external weak-intensity and slowly-varying force affects the system. However, we are going not to solve this problem here, also, noticing that the consideration of an external electromagnetic field even of low intensity and low frequency extremely complicates both the calculations and the presentation of the results. These two problems (an external field's influence and the system's interaction with photons) are at the time under study by the authors.

When neglecting the summand $\hat{V}(t)$, the Hamiltonian (5) completely coincides with (26) or (44), (45). The difference between the forms of the Hamiltonians (5) and (44) taking into account (6) and (45) is only the multicomponent of the system under consideration in the present paper. However, this circumstance is not an obstacle on the way to consequently construct a kinetic theory of weakly ionized gas of hydrogen-like atoms following the theory expressed in the two Sections above.

A modification of the method is as follows. Provided the system is multicomponent, it is necessary to introduce single particle density matrixes for each component of the system. The whole set of these density matrixes we serve as reduced description parameters for the system at its kinetic stage. Taking into account the above-mentioned neglecting of photons contribution to relaxation processes in the system, we introduce single particle density matrixes $f^{(1)}_{\mathbf{p},\mathbf{p}'}$, $f^{(2)}_{\mathbf{p},\mathbf{p}'}$ for free (non-bound) fermions of the first and the second kind with formulas, see (1), (2), (48), (49) (we remind that the index "1" stands for the electronic subsystem characteristics, while the index "2" denotes the ones of cores):

$$f^{(1)}_{\mathbf{p},\mathbf{p}'} \equiv \mathrm{Sp}\,\rho^{(0)}(f)\hat{f}^{(1)}_{\mathbf{p},\mathbf{p}'}, \quad \hat{f}^{(1)}_{\mathbf{p},\mathbf{p}'} \equiv \hat{a}^+_1(\mathbf{p}')\hat{a}_1(\mathbf{p}),$$
$$f^{(2)}_{\mathbf{p},\mathbf{p}'} \equiv \mathrm{Sp}\,\rho^{(0)}(f)\hat{f}^{(2)}_{\mathbf{p},\mathbf{p}'}, \quad \hat{f}^{(2)}_{\mathbf{p},\mathbf{p}'} \equiv \hat{a}^+_2(\mathbf{p}')\hat{a}_2(\mathbf{p}), \tag{57}$$

and also single particle density matrixes of atoms at different quantum-mechanical states $f^{(0)}_{\alpha\mathbf{p},\beta\mathbf{p}'}$ (see (3)):

$$f^{(0)}_{\alpha\mathbf{p},\beta\mathbf{p}'} \equiv \mathrm{Sp}\,\rho^{(0)}(f)\hat{f}^{(0)}_{\alpha\mathbf{p},\beta\mathbf{p}'}, \quad \hat{f}^{(0)}_{\alpha\mathbf{p},\beta\mathbf{p}'} \equiv \hat{\eta}^+_\beta(\mathbf{p}')\hat{\eta}_\alpha(\mathbf{p}), \quad Sp\,\rho^{(0)}(f) = 1, \tag{58}$$

where the statistical operator $\rho^{(0)}(f)$ according to (33), (48) obeys the relation

$$\rho^{(0)}(f) = \exp\left\{\Omega(f) - \sum_{\mathbf{p}\mathbf{p}'} Y^{(1)}_{\mathbf{p}',\mathbf{p}}(f)\hat{f}^{(1)}_{\mathbf{p},\mathbf{p}'} - \sum_{\mathbf{p}\mathbf{p}'} Y^{(2)}_{\mathbf{p}',\mathbf{p}}(f)\hat{f}^{(2)}_{\mathbf{p},\mathbf{p}'} - \sum_{\mathbf{p}\mathbf{p}'} Y^{(0)}_{\beta\mathbf{p}',\alpha\mathbf{p}}(f)\hat{f}^{(0)}_{\alpha\mathbf{p},\beta\mathbf{p}'}\right\}, \tag{59}$$

in which the thermodynamic potential $\Omega(f)$ and the relation of the quantities $Y^{(1)}_{\mathbf{p}',\mathbf{p}}(f)$, $Y^{(2)}_{\mathbf{p}',\mathbf{p}}(f)$, $Y^{(0)}_{\beta\mathbf{p}',\alpha\mathbf{p}}(f)$ with the introduced single particle density matrixes is defined by expressions (57), (58). Further, for each single particle density matrix (57), (58) one writes down evolution equations following to the procedure given in (47). In present paper we are obtaining the kinetic equations accurate within the first order over weak particles interactions, that goes with an average (or self-consistent) field approximation. Neglecting the second order over the interaction, we avoid the problem of constructing the collision integrals, see (47), (55). In this approximation as in the ones suggested earlier, there is no principled need. Obviously, consulting (47), (5) – (25), the collision integrals may be derived. However, due to the complexity of the final expressions and since the main aim of the paper is by developing the kinetic theory for the medium under study, to obtain dispersion laws for elementary perturbations in it, the problem is to be omitted here. The same problem may be solved starting from the kinetic equations taking into account only average field, see Refs. 31, 36, 37.

Considering the mentioned approximations according to (47) the evolution equations for single particle density matrixes $f^{(1)}_{\mathbf{p},\mathbf{p}'}$, $f^{(2)}_{\mathbf{p},\mathbf{p}'}$ of free fermions of both kinds have the following form (see also (57)):

$$\dot{f}_{\mathbf{p},\mathbf{p}'}^{(1)} = L_{\mathbf{p},\mathbf{p}'}^{(1,0)}(f) + L_{\mathbf{p},\mathbf{p}'}^{(1,1)}(f),$$
$$L_{\mathbf{p},\mathbf{p}'}^{(1,0)}(f) = i\mathrm{Sp}\,\rho^{(0)}(f)\left[\hat{H}_0, \hat{f}_{\mathbf{p},\mathbf{p}'}^{(1)}\right], \quad L_{\mathbf{p},\mathbf{p}'}^{(1,1)}(f) = i\mathrm{Sp}\,\rho^{(0)}(f)\left[\hat{V}, \hat{f}_{\mathbf{p},\mathbf{p}'}^{(1)}\right];$$
$$\dot{f}_{\mathbf{p},\mathbf{p}'}^{(2)} = L_{\mathbf{p},\mathbf{p}'}^{(2,0)}(f) + L_{\mathbf{p},\mathbf{p}'}^{(2,1)}(f),$$
$$L_{\mathbf{p},\mathbf{p}'}^{(2,0)}(f) = i\mathrm{Sp}\,\rho^{(0)}(f)\left[\hat{H}_0, \hat{f}_{\mathbf{p},\mathbf{p}'}^{(2)}\right], \quad L_{\mathbf{p},\mathbf{p}'}^{(1,1)}(f) = i\mathrm{Sp}\,\rho^{(0)}(f)\left[\hat{V}, \hat{f}_{\mathbf{p},\mathbf{p}'}^{(1)}\right],$$
(60)

where the Hamiltonians $\hat{H}_0$ and $\hat{V}$ are defined with relations (6) and (21). The same equation may be written for a single particle density matrix of bound states of these fermions, hydrogen-like atoms at different quantum states (see (58)):

$$\dot{f}_{\alpha\mathbf{p},\beta\mathbf{p}'}^{(0)} = L_{\alpha\mathbf{p},\beta\mathbf{p}'}^{(0,0)}(f) + L_{\alpha\mathbf{p},\beta\mathbf{p}'}^{(0,1)}(f),$$
$$L_{\alpha\mathbf{p},\beta\mathbf{p}'}^{(0,0)}(f) = i\mathrm{Sp}\,\rho^{(0)}(f)\left[\hat{H}_0, \hat{f}_{\alpha\mathbf{p},\beta\mathbf{p}'}^{(0)}\right], \quad L_{\alpha\mathbf{p},\beta\mathbf{p}'}^{(0,1)}(f) = i\mathrm{Sp}\,\rho^{(0)}(f)\left[\hat{V}, \hat{f}_{\mathbf{p},\mathbf{p}'}^{(1)}\right].$$
(61)

Calculating commutation relations and traces in (60), (compare to Ref. 50):

$$\frac{\partial f_{\mathbf{p}_1,\mathbf{p}_2}^{(1)}}{\partial t} + i\left[\varepsilon^{(1)}, f^{(1)}\right]_{\mathbf{p}_1,\mathbf{p}_2} = 0,$$
$$\varepsilon_{\mathbf{p}_1\mathbf{p}_2}^{(1)} \equiv \varepsilon_1(\mathbf{p}_1)\Delta(\mathbf{p}_1-\mathbf{p}_2) - \frac{e}{V}\sum_{\mathbf{p}_1'\mathbf{p}_3'}\Delta(\mathbf{p}_3'-\mathbf{p}_1'+\mathbf{p}_2-\mathbf{p}_1)$$
$$\times\left\{v(\mathbf{p}_3'-\mathbf{p}_1')\left[\sigma_{\alpha\beta}(\mathbf{p}_1'-\mathbf{p}_3')f_{\beta\mathbf{p}_3';\alpha\mathbf{p}_1'}^{(0)} - e\left(f_{\mathbf{p}_3'\mathbf{p}_1'}^{(1)} - f_{\mathbf{p}_3'\mathbf{p}_1'}^{(2)}\right)\right] + ev(\mathbf{p}_1'-\mathbf{p}_2)f_{\mathbf{p}_3'\mathbf{p}_1'}^{(1)}\right\},$$
$$\frac{\partial f_{\mathbf{p}_1\mathbf{p}_2}^{(2)}}{\partial t} + i\left[\varepsilon^{(2)}, f^{(2)}\right]_{\mathbf{p}_1\mathbf{p}_2} = 0,$$
$$\varepsilon_{\mathbf{p}_1\mathbf{p}_2}^{(2)} \equiv \varepsilon_2(\mathbf{p}_1)\Delta(\mathbf{p}_1-\mathbf{p}_2) + \frac{e}{V}\sum_{\mathbf{p}_1'\mathbf{p}_3'}\Delta(\mathbf{p}_3'-\mathbf{p}_1'+\mathbf{p}_2-\mathbf{p}_1)$$
$$\times\left\{v(\mathbf{p}_3'-\mathbf{p}_1')\left[\sigma_{\alpha\beta}(\mathbf{p}_1'-\mathbf{p}_3')f_{\beta\mathbf{p}_3';\alpha\mathbf{p}_1'}^{(0)} - e\left(f_{\mathbf{p}_3'\mathbf{p}_1'}^{(1)} - f_{\mathbf{p}_3'\mathbf{p}_1'}^{(2)}\right)\right] - ev(\mathbf{p}_1'-\mathbf{p}_2)f_{\mathbf{p}_3'\mathbf{p}_1'}^{(2)}\right\}$$
(62)

where $\Delta(\mathbf{p}_1-\mathbf{p}_2)$ is the Kronecker delta, the polarization matrix $\sigma_{\alpha\beta}(\mathbf{p})$ obeys formula (11) and the function $v(\mathbf{p})$ is still defined by (25). We remind that the collision integrals in (61), (62) are absent since we settled to neglect the terms quadratic over the interaction $\hat{V}$ in the perturbation theory, see (21). Similarly, the evolution equation (61) takes the form of

$$\frac{\partial f_{\alpha_1\mathbf{p}_1,\alpha_2\mathbf{p}_2}^{(0)}}{\partial t} = -i\left[\varepsilon^{(0)}, f^{(0)}\right]_{\alpha_1\mathbf{p}_1,\alpha_2\mathbf{p}_2},$$
$$\varepsilon_{\alpha_1\mathbf{p}_1;\alpha_2\mathbf{p}_2}^{(0)} \equiv \varepsilon_{\alpha_1}(\mathbf{p}_1)\delta_{\alpha_1\alpha_2}\Delta(\mathbf{p}_1-\mathbf{p}_2)$$

$$+\frac{e}{V}\sum_{\mathbf{p}_3\mathbf{p}_4}\Delta(\mathbf{p}_1-\mathbf{p}_2-\mathbf{p}_3+\mathbf{p}_4)v(\mathbf{p}_3-\mathbf{p}_4)\sigma_{\alpha_2\alpha_1}(\mathbf{p}_1-\mathbf{p}_2)\left(f^{(2)}_{\mathbf{p}_4\mathbf{p}_3}-f^{(1)}_{\mathbf{p}_4\mathbf{p}_3}\right)$$

$$+\frac{1}{V}\sum_{\mathbf{p}'_1\mathbf{p}'_3}\Delta(\mathbf{p}'_3-\mathbf{p}'_1+\mathbf{p}_2-\mathbf{p}_1)f^{(0)}_{\alpha'_3\mathbf{p}'_3;\alpha'_1\mathbf{p}'_1} \qquad (63)$$

$$\times\left\{v(\mathbf{p}'_3-\mathbf{p}_1)\sigma_{\alpha'_1\alpha_2}(\mathbf{p}'_1-\mathbf{p}_2)\sigma_{\alpha_1\alpha'_3}(\mathbf{p}_1-\mathbf{p}'_3)\right.$$

$$\left.+v(\mathbf{p}'_3-\mathbf{p}'_1)\sigma_{\alpha_1\alpha_2}(\mathbf{p}_1-\mathbf{p}_2)\sigma_{\alpha'_1\alpha'_3}(\mathbf{p}'_1-\mathbf{p}'_3)\right\},$$

where $\varepsilon_\alpha(\mathbf{p})$ is an energy of a hydrogen-like atom in a quantum-mechanical state characterized by the set of indexes $\alpha$, see (6). The operation $\left[\varepsilon^{(1)}, f^{(1)}\right]_{\mathbf{p}_1,\mathbf{p}_2}$ defines the commutation relation for matrixes with indexes $\mathbf{p}_1, \mathbf{p}_2$ in equations (62) or the commutation relation for matrixes with indexes $\alpha_1\mathbf{p}_1, \alpha_2\mathbf{p}_2$ in equation (63). It is convenient to use the Wick's theorem when obtaining the formulas (62), (63) for calculating averages of creation and annihilation operators products.

Further introducing the Wigner distribution functions $f^{(1)}(\mathbf{x},\mathbf{p})$, $f^{(2)}(\mathbf{x},\mathbf{p})$, $f^{(0)}_{\alpha_1,\alpha_2}(\mathbf{x},\mathbf{p})$ as

$$f^{(1)}(\mathbf{x},\mathbf{p}) \equiv \sum_{\mathbf{k}} e^{-i\mathbf{k}\mathbf{x}} f^{(1)}_{\mathbf{p}-\frac{\mathbf{k}}{2},\mathbf{p}+\frac{\mathbf{k}}{2}} = \frac{V}{(2\pi)^3}\int d^3k\, e^{-i\mathbf{k}\mathbf{x}} f^{(1)}_{\mathbf{p}-\frac{\mathbf{k}}{2},\mathbf{p}+\frac{\mathbf{k}}{2}},$$

$$f^{(2)}(\mathbf{x},\mathbf{p}) \equiv \sum_{\mathbf{k}} e^{-i\mathbf{k}\mathbf{x}} f^{(2)}_{\mathbf{p}-\frac{\mathbf{k}}{2},\mathbf{p}+\frac{\mathbf{k}}{2}} = \frac{V}{(2\pi)^3}\int d^3k\, e^{-i\mathbf{k}\mathbf{x}} f^{(2)}_{\mathbf{p}-\frac{\mathbf{k}}{2},\mathbf{p}+\frac{\mathbf{k}}{2}}, \qquad (64)$$

$$f^{(0)}_{\alpha_1,\alpha_2}(\mathbf{x},\mathbf{p}) \equiv \sum_{\mathbf{k}} e^{-i\mathbf{k}\mathbf{x}} f^{(1)}_{\alpha_1\mathbf{p}-\frac{\mathbf{k}}{2},\alpha_2\mathbf{p}+\frac{\mathbf{k}}{2}} = \frac{V}{(2\pi)^3}\int d^3k\, e^{-i\mathbf{k}\mathbf{x}} f^{(1)}_{\alpha_1\mathbf{p}-\frac{\mathbf{k}}{2},\alpha_2\mathbf{p}+\frac{\mathbf{k}}{2}},$$

we may transform equations (62) to have a more "usual" form similar to that of equation (54), where we neglect the collision integral:

$$\frac{\partial f^{(1)}(\mathbf{x},\mathbf{p})}{\partial t}+\frac{\partial \varepsilon^{(1)}(\mathbf{x},\mathbf{p})}{\partial \mathbf{p}}\frac{\partial f^{(1)}(\mathbf{x},\mathbf{p})}{\partial \mathbf{x}}-\frac{\partial \varepsilon^{(1)}(\mathbf{x},\mathbf{p})}{\partial \mathbf{x}}\frac{\partial f^{(1)}(\mathbf{x},\mathbf{p})}{\partial \mathbf{p}}=0,$$

$$\frac{\partial f^{(2)}(\mathbf{x},\mathbf{p})}{\partial t}+\frac{\partial \varepsilon^{(2)}(\mathbf{x},\mathbf{p})}{\partial \mathbf{p}}\frac{\partial f^{(2)}(\mathbf{x},\mathbf{p})}{\partial \mathbf{x}}-\frac{\partial \varepsilon^{(2)}(\mathbf{x},\mathbf{p})}{\partial \mathbf{x}}\frac{\partial f^{(2)}(\mathbf{x},\mathbf{p})}{\partial \mathbf{p}}=0. \qquad (65)$$

Concerning equation (53), it now has a more complex form additionally due to the distribution function $f^{(0)}_{\alpha_1,\alpha_2}(\mathbf{x},\mathbf{p})$ being a single particle density matrix over the indexes $\alpha_1, \alpha_2$:

$$\frac{\partial f_{\alpha_1;\alpha_2}^{(0)}(\mathbf{x},\mathbf{p})}{\partial t}+i\left[\varepsilon^{(0)}(\mathbf{x},\mathbf{p}),f^{(0)}(\mathbf{x},\mathbf{p})\right]_{\alpha_1\alpha_2}$$
$$=-\frac{1}{2}\left\{\frac{\partial \varepsilon^{(0)}(\mathbf{x},\mathbf{p})}{\partial \mathbf{p}},\frac{\partial f^{(0)}(\mathbf{x},\mathbf{p})}{\partial \mathbf{x}}\right\}_{\alpha_1\alpha_2}+\frac{1}{2}\left\{\frac{\partial \varepsilon^{(0)}(\mathbf{x},\mathbf{p})}{\partial \mathbf{x}},\frac{\partial f^{(0)}(\mathbf{x},\mathbf{p})}{\partial \mathbf{p}}\right\}_{\alpha_1\alpha_2} \quad (66)$$

where the operations $[A,B]_{\alpha_1\alpha_2}$, $\{A,B\}_{\alpha_1\alpha_2}$ are the commutation expressions of matrixes with indexes $\alpha_1, \alpha_2$ and the anti-commutation ones, respectively (we remind that a summation is assumed over the repeated indexes):

$$[A,B]_{\alpha_1\alpha_2} \equiv A_{\alpha_1\beta}B_{\beta\alpha_2} - B_{\alpha_1\beta}A_{\beta\alpha_2},$$
$$\{A,B\}_{\alpha_1\alpha_2} \equiv A_{\alpha_1\beta}B_{\beta\alpha_2} + B_{\alpha_1\beta}A_{\beta\alpha_2}. \quad (67)$$

Equations (65), (66) are obtained within the first order of the perturbation theory over small spatial gradients, the same as equation (54) (see in more detail in Ref. 31).

The quantities $\varepsilon^{(1)}(\mathbf{x},\mathbf{p})$, $\varepsilon^{(2)}(\mathbf{x},\mathbf{p})$, $\varepsilon_{\alpha_1\alpha_2}^{(0)}(\mathbf{x},\mathbf{p})$ in equations (65), (66) are defined by analogous to (55) formulas

$$\varepsilon^{(1)}(\mathbf{x},\mathbf{p}) \equiv \sum_{\mathbf{k}} e^{-i\mathbf{k}\mathbf{x}}\varepsilon^{(1)}_{\mathbf{p}-\frac{\mathbf{k}}{2},\mathbf{p}+\frac{\mathbf{k}}{2}} = \frac{V}{(2\pi)^3}\int d^3k\, e^{-i\mathbf{k}\mathbf{x}}\varepsilon^{(1)}_{\mathbf{p}-\frac{\mathbf{k}}{2},\mathbf{p}+\frac{\mathbf{k}}{2}},$$
$$\varepsilon^{(2)}(\mathbf{x},\mathbf{p}) \equiv \sum_{\mathbf{k}} e^{-i\mathbf{k}\mathbf{x}}\varepsilon^{(2)}_{\mathbf{p}-\frac{\mathbf{k}}{2},\mathbf{p}+\frac{\mathbf{k}}{2}} = \frac{V}{(2\pi)^3}\int d^3k\, e^{-i\mathbf{k}\mathbf{x}}\varepsilon^{(2)}_{\mathbf{p}-\frac{\mathbf{k}}{2},\mathbf{p}+\frac{\mathbf{k}}{2}}, \quad (68)$$
$$\varepsilon_{\alpha_1,\alpha_2}^{(0)}(\mathbf{x},\mathbf{p}) \equiv \sum_{\mathbf{k}} e^{-i\mathbf{k}\mathbf{x}}\varepsilon^{(0)}_{\alpha_1\mathbf{p}-\frac{\mathbf{k}}{2},\alpha_2\mathbf{p}+\frac{\mathbf{k}}{2}} = \frac{V}{(2\pi)^3}\int d^3k\, e^{-i\mathbf{k}\mathbf{x}}\varepsilon^{(0)}_{\alpha_1\mathbf{p}-\frac{\mathbf{k}}{2},\alpha_2\mathbf{p}+\frac{\mathbf{k}}{2}}.$$

All the quantities defined with formulas (68), one may express in terms of the Wigner distribution functions introduced above. The functionals $\varepsilon^{(1)}(\mathbf{x},\mathbf{p})$, $\varepsilon^{(2)}(\mathbf{x},\mathbf{p})$ satisfy the following equalities:

$$\varepsilon^{(1)}(\mathbf{x},\mathbf{p}) = \varepsilon_1(\mathbf{p}) - U^{(1)}(\mathbf{x},\mathbf{p};f) - U(\mathbf{x},\mathbf{p};f),$$
$$\varepsilon^{(2)}(\mathbf{x},\mathbf{p}) = \varepsilon_2(\mathbf{p}) - U^{(2)}(\mathbf{x},\mathbf{p};f) + U(\mathbf{x},\mathbf{p};f), \quad (69)$$

where we introduce:

$$U^{(1)}(\mathbf{x},\mathbf{p};f) \equiv \frac{e^2}{V}\sum_{\mathbf{p}'} f^{(1)}(\mathbf{x},\mathbf{p}')\nu(\mathbf{p}'-\mathbf{p}),$$

$$U^{(2)}(\mathbf{x},\mathbf{p};f) \equiv \frac{e^2}{V}\sum_{\mathbf{p}'} f^{(2)}(\mathbf{x},\mathbf{p}')v(\mathbf{p}'-\mathbf{p}),$$

$$U(\mathbf{x},\mathbf{p};f) \equiv \frac{e}{V}\int d\mathbf{x}' \left[ v_{\alpha_1'\alpha_3'}(\mathbf{x}-\mathbf{x}')\sum_{\mathbf{p}_3'} f^{(0)}_{\alpha_3',\alpha_1'}(\mathbf{x}',\mathbf{p}_3') \right.$$
$$\left. -ev(\mathbf{x}-\mathbf{x}')\left(\sum_{\mathbf{p}_3'} f^{(1)}(\mathbf{x}',\mathbf{p}_3') - \sum_{\mathbf{p}_3'} f^{(2)}(\mathbf{x}',\mathbf{p}_3')\right)\right], \quad (70)$$

$$v(\mathbf{x}-\mathbf{x}') \equiv \frac{1}{V}\sum_{\mathbf{k}} e^{i\mathbf{k}(\mathbf{x}-\mathbf{x}')}v(\mathbf{k}),$$

$$v_{\alpha_1'\alpha_3'}(\mathbf{x}-\mathbf{x}') \equiv \frac{1}{V}\sum_{\mathbf{k}} e^{i\mathbf{k}(\mathbf{x}-\mathbf{x}')}v(\mathbf{k})\sigma_{\alpha_1'\alpha_3'}(-\mathbf{k}).$$

For the quantity $\varepsilon^{(0)}_{\alpha_1,\alpha_2}(\mathbf{x},\mathbf{p})$, taking into account (67) that defines the evolution of the atoms distribution function according to equation (66), after some calculations we arrive to

$$\varepsilon^{(0)}_{\alpha_1,\alpha_2}(\mathbf{x},\mathbf{p}) = \varepsilon_{\alpha_1}(\mathbf{p})\delta_{\alpha_1,\alpha_2} + U^{(0)}_{\alpha_1,\alpha_2}(\mathbf{x},\mathbf{p};f) + U_{\alpha_1,\alpha_2}(\mathbf{x},\mathbf{p};f), \quad (71)$$

where we also introduced the following designations:

$$U_{\alpha_1\alpha_2}(\mathbf{x},\mathbf{p};f) \equiv \frac{1}{V^2}\int d\mathbf{x}' \sum_{\mathbf{k}} e^{i\mathbf{k}(\mathbf{x}-\mathbf{x}')}\sigma_{\alpha_1\alpha_2}(\mathbf{k})v(\mathbf{k})$$
$$\times \left[ \sigma_{\alpha_1'\alpha'}(-\mathbf{k})\sum_{\mathbf{p}'} f^{(0)}_{\alpha',\alpha_1'}(\mathbf{x}',\mathbf{p}') - e\left(\sum_{\mathbf{p}'} f^{(1)}(\mathbf{x}',\mathbf{p}') - \sum_{\mathbf{p}'} f^{(2)}(\mathbf{x}',\mathbf{p}')\right)\right], \quad (72)$$

$$U^{(0)}_{\alpha_1\alpha_2}(\mathbf{x},\mathbf{p}) \equiv \frac{1}{V}\sum_{\mathbf{p}'} v(\mathbf{p}'-\mathbf{p})\sigma_{\alpha_1\alpha'}(\mathbf{p}-\mathbf{p}') f^{(0)}_{\alpha',\alpha_1'}(\mathbf{x},\mathbf{p}')\sigma_{\alpha_1'\alpha_2}(\mathbf{p}'-\mathbf{p})$$

The matrix $\sigma_{\alpha_1\alpha_2}(\mathbf{p})$ in (70) – (72) is still given with expression (11). Thus, we have built a system of coupled kinetic equations (65) – (72) for Wigner distribution functions (64) of components of weakly ionized gas of hydrogen-like atoms in a collisionless approximation. We remind, that the main truth condition for a collisionless approximation, and, therefore, for equations (65) – (72), is supplied by the relation (see, e.g., Ref. 31, also Ref. 36):

$$\tau_0 \ll t \ll \tau_r, \quad (73)$$

where $\tau_0$ is the chaotization time mentioned above, $\tau_r$ is the system relaxation time due to collisions between the particles. As we mentioned above, the relaxation time $\tau_r$ is defined by the interaction intensity $\hat{V}$, to be more precise, by the collision integral that is quadratic over an interaction, see (52), (55). The relaxation time tends to infinity,

$\tau_r \to \infty$, at $\hat{V} \to 0$ (in our paper the latter corresponds to neglecting the collision integral). That is, the relaxation time is only large in case of small interaction. Since the chaotization time as we noted, does not depend on the interaction intensity, the condition (73) is applicable for most systems.

Thus, the equations (65), (66) together with (68) – (72) in fact contain the solution of the problem stated in the present paper's title, a development of the kinetic theory of weakly ionized rarified gases of hydrogen-like atoms from the first principles of quantum statistics. We should note, however, that the kinetic equations were derived within the first approximation of the perturbation theory over small interaction, without constructing collision integrals. The latter may be also derived without any principled complexities following the method described in Ref. 31. However, this requires extremely lengthy calculations and a substantial increase of the paper's size. This is the main reason we refused to include such a material the paper. In what follows we demonstrate the capabilities of the theory developed, in context of real applications. In particular, we focus on the derivation of dispersion equation for weakly ionized gas of hydrogen-like atoms.

## VI. EQUILIBRIUM AND LINEARIZED KINETIC EQUATIONS OF WEAKLY IONIZED GAS OF HYDROGEN-LIKE ATOMS

One of the main purposes of the kinetic theory of weakly ionized gas of hydrogen-like atoms is, as stated above, the derivation of the dispersion law for electromagnetic eigenwaves in such system. This dispersion law plays essential role when describing the BEC phenomenon of photons in weakly ionized gas of hydrogen-like atoms, see Refs. 28 - 33. The eigenwaves dispersion law itself, as shown in Ref. 31, may be derived from kinetic equation in a collisionless approximation linearized near the system's equilibrium.

Let us show that the obtained equations (65) – (72) also allow to define the dispersion law for eigenwaves (or the dispersion law of photons in medium) in a weakly ionized gas of hydrogen-like atoms. To do this it is enough to linearized the kinetic equations near a spatially homogeneous state of statistical equilibrium of the system under study,

$$f^{(1)}(\mathbf{x},\mathbf{p}) \approx \overline{f}^{(1)}(\mathbf{p}) + \delta f^{(1)}(\mathbf{x},\mathbf{p}), \quad f^{(2)}(\mathbf{x},\mathbf{p}) \approx \overline{f}^{(2)}(\mathbf{p}) + \delta f^{(2)}(\mathbf{x},\mathbf{p}),$$
$$f^{(0)}_{\alpha_1;\alpha_2}(\mathbf{x},\mathbf{p}) \approx \overline{f}^{(0)}_{\alpha_1;\alpha_2}(\mathbf{p}) + \delta f^{(0)}_{\alpha_1;\alpha_2}(\mathbf{x},\mathbf{p}),$$
(74)

and to find the equations' solutions periodic in space and time. The functions $\delta f^{(1)}(\mathbf{x},\mathbf{p})$, $\delta f^{(2)}(\mathbf{x},\mathbf{p})$, $\delta f^{(0)}_{\alpha_1;\alpha_2}(\mathbf{x},\mathbf{p})$ in (74) are the deviations of Wigner distribution functions $f^{(1)}(\mathbf{x},\mathbf{p})$, $f^{(2)}(\mathbf{x},\mathbf{p})$, $f^{(0)}_{\alpha_1;\alpha_2}(\mathbf{x},\mathbf{p})$

from their equilibrium values $\overline{f}^{(1)}(\mathbf{p})$, $\overline{f}^{(2)}(\mathbf{p})$, $\overline{f}^{(0)}_{\alpha_1;\alpha_2}(\mathbf{p})$. At this we need to take into account a condition of the equilibrium state quasi-neutralness, that is, that the number of oppositely charged fermions is to be the same:

$$\sum_{\mathbf{p}} \overline{f}^{[1]}(\mathbf{p}) = \sum_{\mathbf{p}} \overline{f}^{[2]}(\mathbf{p}). \tag{75}$$

Substituting further the decompositions (74) into (65) – (72), we come to the following system of linearized kinetic motion equations for distribution function $\delta f^{(1)}(\mathbf{x},\mathbf{p})$, $\delta f^{(2)}(\mathbf{x},\mathbf{p})$, $\delta f^{(0)}_{\alpha_1;\alpha_2}(\mathbf{x},\mathbf{p})$:

$$\begin{aligned}
\frac{\partial}{\partial t}\delta f^{(1)}(\mathbf{x},\mathbf{p}) &= -\frac{\partial \overline{\varepsilon}^{(1)}(\mathbf{p})}{\partial \mathbf{p}}\frac{\partial}{\partial \mathbf{x}}\delta f^{(1)}(\mathbf{x},\mathbf{p}) + \frac{\partial \overline{f}^{(1)}(\mathbf{p})}{\partial \mathbf{p}}\frac{\partial}{\partial \mathbf{x}}\delta \varepsilon^{(1)}(\mathbf{x},\mathbf{p}), \\
\frac{\partial}{\partial t}\delta f^{(2)}(\mathbf{x},\mathbf{p}) &= -\frac{\partial \overline{\varepsilon}^{(2)}(\mathbf{p})}{\partial \mathbf{p}}\frac{\partial}{\partial \mathbf{x}}\delta f^{(2)}(\mathbf{x},\mathbf{p}) + \frac{\partial \overline{f}^{(2)}(\mathbf{p})}{\partial \mathbf{p}}\frac{\partial}{\partial \mathbf{x}}\delta \varepsilon^{(2)}(\mathbf{x},\mathbf{p}), \\
\frac{\partial \delta f^{(0)}_{\alpha_1;\alpha_2}(\mathbf{x},\mathbf{p})}{\partial t} &= -i\left[\delta \varepsilon^{(0)}(\mathbf{x},\mathbf{p}), \overline{f}^{(0)}(\mathbf{p})\right]_{\alpha_1\alpha_2} - i\left[\overline{\varepsilon}^{(0)}(\mathbf{p}), \delta f^{(0)}(\mathbf{x},\mathbf{p})\right]_{\alpha_1\alpha_2} \\
&\quad + \frac{1}{2}\left\{\frac{\partial \delta \varepsilon^{(0)}(\mathbf{x},\mathbf{p})}{\partial \mathbf{x}}, \frac{\partial \overline{f}^{(0)}(\mathbf{p})}{\partial \mathbf{p}}\right\}_{\alpha_1\alpha_2} - \frac{1}{2}\left\{\frac{\partial \overline{\varepsilon}^{(0)}(\mathbf{p})}{\partial \mathbf{p}}, \frac{\partial \delta f^{(0)}(\mathbf{x},\mathbf{p})}{\partial \mathbf{x}}\right\}_{\alpha_1\alpha_2}
\end{aligned} \tag{76}$$

where the quantities $\overline{\varepsilon}^{(1)}(\mathbf{p})$, $\overline{\varepsilon}^{(2)}(\mathbf{p})$ together with (75) are defined by the following expressions:

$$\begin{aligned}
\overline{\varepsilon}^{(1)}(\mathbf{p}) &= \varepsilon_1(\mathbf{p}) - U(\overline{f}) - U^{(1)}(\mathbf{p};\overline{f}), \quad \overline{\varepsilon}^{(2)}(\mathbf{p}) = \varepsilon_2(\mathbf{p}) - U(\overline{f}) - U^{(2)}(\mathbf{p};\overline{f}), \\
U^{(1)}(\mathbf{p};\overline{f}) &\equiv \frac{e^2}{V}\sum_{\mathbf{p}'}\overline{f}^{(1)}(\mathbf{p}')v(\mathbf{p}'-\mathbf{p}), \quad U^{(2)}(\mathbf{p};\overline{f}) \equiv \frac{e^2}{V}\sum_{\mathbf{p}'}\overline{f}^{(2)}(\mathbf{p}')v(\mathbf{p}'-\mathbf{p}), \\
U(\overline{f}) &\equiv \frac{e}{V}\sum_{\alpha'\alpha_1'\mathbf{p}'}\overline{f}^{[0]}_{\alpha',\alpha_1'}(\mathbf{p}')\lim_{\mathbf{k}\to 0}\sigma_{\alpha_1'\alpha'}(-\mathbf{k})v(\mathbf{k}),
\end{aligned} \tag{77}$$

and the function $\overline{\varepsilon}^{[0]}_{\alpha_1;\alpha_2}(\mathbf{p})$ together with (75) is given by formulas (see also (6)):

$$\begin{aligned}
\overline{\varepsilon}^{(0)}_{\alpha_1;\alpha_2}(\mathbf{p}) &= \varepsilon_{\alpha_1}(\mathbf{p})\delta_{\alpha_1\alpha_2} + U_{\alpha_1\alpha_2}(\overline{f}) + U^{(0)}_{\alpha_1\alpha_2}(\mathbf{p};\overline{f}), \\
U_{\alpha_1\alpha_2}(\overline{f}) &\equiv \frac{1}{V}\sum_{\alpha'\alpha_1'\mathbf{p}'}\overline{f}^{(0)}_{\alpha',\alpha_1'}(\mathbf{p}')\lim_{\mathbf{k}\to 0}\sigma_{\alpha_1\alpha_2}(\mathbf{k})v(\mathbf{k})\sigma_{\alpha_1'\alpha'}(-\mathbf{k}), \\
U^{(0)}_{\alpha_1\alpha_2}(\mathbf{p};\overline{f}) &\equiv \frac{1}{V}\sum_{\mathbf{p}'}v(\mathbf{p}'-\mathbf{p})\sigma_{\alpha_1\alpha'}(\mathbf{p}-\mathbf{p}')\overline{f}^{(0)}_{\alpha',\alpha_1'}(\mathbf{p}')\sigma_{\alpha_1'\alpha_2}(\mathbf{p}'-\mathbf{p}).
\end{aligned} \tag{78}$$

The deviations $\delta\varepsilon^{(1)}(\mathbf{x},\mathbf{p})$, $\delta\varepsilon^{(2)}(\mathbf{x},\mathbf{p})$, $\delta\varepsilon^{(0)}_{\alpha_1;\alpha_2}(\mathbf{x},\mathbf{p})$ of the quantities $\varepsilon^{(1)}(\mathbf{x},\mathbf{p})$, $\varepsilon^{(2)}(\mathbf{x},\mathbf{p})$, $\varepsilon^{(0)}_{\alpha_1;\alpha_2}(\mathbf{x},\mathbf{p})$ from their equilibrium values (77), (78) can be written as:

$$\delta\varepsilon^{(1)}(\mathbf{x},\mathbf{p}) = -\delta U(\mathbf{x}) - \delta U^{(1)}(\mathbf{x},\mathbf{p}), \quad \delta\varepsilon^{(2)}(\mathbf{x},\mathbf{p}) = \delta U(\mathbf{x}) - \delta U^{(2)}(\mathbf{x},\mathbf{p}),$$
$$\delta\varepsilon^{(0)}_{\alpha_1;\alpha_2}(\mathbf{x},\mathbf{p}) = \delta U_{\alpha_1\alpha_2}(\mathbf{x}) + \delta U^{(0)}_{\alpha_1\alpha_2}(\mathbf{x},\mathbf{p}), \tag{79}$$

where we introduce the following notations (see also (70) – (72)):

$$\delta U^{[1]}(\mathbf{x},\mathbf{p}) \equiv U^{[1]}(\mathbf{x},\mathbf{p};\delta f) = \frac{e^2}{V}\sum_{\mathbf{p}'}\delta f^{[1]}(\mathbf{x},\mathbf{p}')v(\mathbf{p}'-\mathbf{p}),$$

$$\delta U^{[2]}(\mathbf{x},\mathbf{p}) \equiv U^{[2]}(\mathbf{x},\mathbf{p};\delta f) = \frac{e^2}{V}\sum_{\mathbf{p}'}\delta f^{[2]}(\mathbf{x},\mathbf{p}')v(\mathbf{p}'-\mathbf{p}),$$

$$\delta U(\mathbf{x}) \equiv U(\mathbf{x};\delta f)$$
$$= \frac{1}{V^2}\int d\mathbf{x}'\sum_{\mathbf{k}} e^{i\mathbf{k}(\mathbf{x}-\mathbf{x}')}v(\mathbf{k})\left[\sigma_{\alpha'_1\alpha'_3}(-\mathbf{k})\sum_{\mathbf{p}'_3}\delta f^{[0]}_{\alpha'_3,\alpha'_1}(\mathbf{x}',\mathbf{p}'_3)\right.$$
$$\left. -e^2\left(\sum_{\mathbf{p}'_3}\delta f^{[1]}(\mathbf{x}',\mathbf{p}'_3) - \sum_{\mathbf{p}'_3}\delta f^{[2]}(\mathbf{x}',\mathbf{p}'_3)\right)\right],$$

$$\delta U_{\alpha_1\alpha_2}(\mathbf{x}) \equiv U_{\alpha_1\alpha_2}(\mathbf{x};\delta f)$$
$$= \frac{1}{V^2}\int d\mathbf{x}'\sum_{\mathbf{k}} e^{i\mathbf{k}(\mathbf{x}-\mathbf{x}')}\sigma_{\alpha_1\alpha_2}(\mathbf{k})v(\mathbf{k})\left[\sigma_{\alpha'_1\alpha'}(-\mathbf{k})\sum_{\mathbf{p}'}\delta f^{[0]}_{\alpha',\alpha'_1}(\mathbf{x}',\mathbf{p}')\right.$$
$$\left. -e^2\left(\sum_{\mathbf{p}'}\delta f^{[1]}(\mathbf{x}',\mathbf{p}') - \sum_{\mathbf{p}'}\delta f^{[2]}(\mathbf{x}',\mathbf{p}')\right)\right],$$

$$\delta U^{(0)}_{\alpha_1\alpha_2}(\mathbf{x},\mathbf{p}) \equiv U^{(0)}_{\alpha_1\alpha_2}(\mathbf{x},\mathbf{p};\delta f) \tag{80}$$
$$= \frac{1}{V}\sum_{\mathbf{p}'} v(\mathbf{p}'-\mathbf{p})\sigma_{\alpha_1\alpha'}(\mathbf{p}-\mathbf{p}')\delta f^{(0)}_{\alpha',\alpha'_1}(\mathbf{x},\mathbf{p}')\sigma_{\alpha'_1\alpha_2}(\mathbf{p}'-\mathbf{p}).$$

We also need to ensure that the equilibrium distribution functions $\overline{f}^{(1)}(\mathbf{p})$, $\overline{f}^{(2)}(\mathbf{p})$, $\overline{f}^{(0)}_{\alpha_1;\alpha_2}(\mathbf{p})$ also obey the kinetic equations (65), (66). As for the equations (65), the functions $\overline{f}^{(1)}(\mathbf{p})$, $\overline{f}^{(2)}(\mathbf{p})$ obey them due to spatial homogeneousness of the equilibrium system. The requirement that $\overline{f}^{(0)}_{\alpha_1;\alpha_2}(\mathbf{p})$ obeys equation (66), is equivalent to the fulfillment of the equality

$$\left[\overline{\varepsilon}^{(0)}(\mathbf{p}),\overline{f}^{(0)}(\mathbf{p})\right]_{\alpha_1\alpha_2} = \overline{\varepsilon}^{(0)}_{\alpha_1,\beta}(\mathbf{p})\overline{f}^{(0)}_{\beta,\alpha_2}(\mathbf{p}) - \overline{f}^{(0)}_{\alpha_1,\beta}(\mathbf{p})\overline{\varepsilon}^{(0)}_{\beta,\alpha_2}(\mathbf{p}) = 0. \tag{81}$$

where $\bar{\varepsilon}_{\alpha_1,\alpha_2}^{(0)}(\mathbf{p})$ is given with (78). Formulas (76) – (81) show that both the solutions of the linearized kinetic equations (76), and that of equation (81) (and, consequently the dispersion law for eigenwaves itself) should depend largely on Wigner equilibrium distribution functions. To solve equation (81) it is worth reminding that the initial kinetic equations (65) – (72) were obtained in the first-order accuracy over the weak interaction between particles. Due to this reason, the above-mentioned equation (81) to define the equilibrium distribution function should contain only the summands of not higher than first order over the interaction (formally, over the quantity $v(\mathbf{k})$). That is, the solution of this equation we take in the form

$$\bar{f}_{\alpha_1;\alpha_2}^{(0)}(\mathbf{p}) \approx f_{\alpha_1}^{(0)}(\mathbf{p})\delta_{\alpha_1,\alpha_2} + \delta\bar{f}_{\alpha_1;\alpha_2}^{(0)}(\mathbf{p}), \tag{82}$$

expressing (78) for $\bar{\varepsilon}_{\alpha_1,\beta}^{(0)}(\mathbf{p})$ in a more convenient way

$$\bar{\varepsilon}_{\alpha_1,\alpha_2}^{(0)}(\mathbf{p}) \approx \varepsilon_{\alpha_1}(\mathbf{p})\delta_{\alpha_1,\alpha_2} + \delta\bar{\varepsilon}_{\alpha_1,\alpha_2}^{(0)}(\mathbf{p}), \quad \delta\bar{\varepsilon}_{\alpha_1,\alpha_2}^{(0)}(\mathbf{p}) \approx U_{\alpha_1\alpha_2}(\bar{f}) + U_{\alpha_1\alpha_2}^{(0)}(\mathbf{p};\bar{f}), \tag{83}$$

where $\bar{f}_\alpha^{(0)}(\mathbf{p})$ is a distribution function for an ideal Bose-gas of hydrogen-like atoms (the principal approximation over the interaction between the system's particles):

$$f_\alpha^{(0)}(\mathbf{p}) = \left\{e^{\beta(\varepsilon_\alpha(\mathbf{p})-\mu^{(0)})} + 1\right\}^{-1}, \quad \varepsilon_\alpha(\mathbf{p}) = \varepsilon_\alpha + \frac{\mathbf{p}^2}{2M}, \tag{84}$$

and the quantity we want to obtain, $\delta\bar{f}_{\alpha_1;\alpha_2}^{(0)}(\mathbf{p})$, is proportional to the first order of the perturbation theory over the interaction. Since the neglecting of particles interaction complies with the ideal gas approximation, it allows us to choose the function $\bar{f}_{\alpha_1;\alpha_2}^{(0)}(\mathbf{p})$ in the principal approximation in a diagonal representation, see (82). the quantity $\mu^{(0)}$ in (84) is a chemical potential, unknown, and with $\beta$ we designate an inverse temperature, $\beta \equiv 1/T$. Hereinafter we assume that temperature $T$ is measured in energy units. After the relatively simple calculations from (81) together with (82), (83) we get the following expression for $\delta\bar{f}_{\alpha_1;\alpha_2}^{(0)}(\mathbf{p})$:

$$\delta\bar{f}_{\alpha_1,\alpha_2}^{(0)}(\mathbf{p}) = \frac{U_{\alpha_1\alpha_2}^{(0)}(\mathbf{p};f)\left(f_{\alpha_1}^{(0)}(\mathbf{p}) - f_{\alpha_2}^{(0)}(\mathbf{p})\right)}{\varepsilon_{\alpha_1} - \varepsilon_{\alpha_2}}, \tag{85}$$

where the quantity $U_{\alpha_1\alpha_2}^{(0)}(\mathbf{p};\bar{f})$ is obtained from (78) taking into account (82), (83):

$$U^{(0)}_{\alpha_1\alpha_2}(\mathbf{p};\overline{f}) \equiv \frac{1}{V}\sum_{\mathbf{p}'\beta} v(\mathbf{p}'-\mathbf{p})\sigma_{\alpha_1\beta}(\mathbf{p}-\mathbf{p}') f^{(0)}_\beta(\mathbf{p}')\sigma_{\beta\alpha_2}(\mathbf{p}'-\mathbf{p}). \tag{86}$$

In (85) we assumed that $U_{\alpha_1\alpha_2}(f) = 0$,

$$U_{\alpha_1\alpha_2}(\overline{f}) \equiv \frac{1}{V}\sum_{\alpha'\alpha'_1\mathbf{p}'} f^{(0)}_{\alpha'}(\mathbf{p}') \lim_{\mathbf{k}\to 0}(\mathbf{k}\mathbf{d})_{\alpha_1\alpha_2} v(\mathbf{k})(\mathbf{k}\mathbf{d})_{\alpha'\alpha'} = 0. \tag{87}$$

Indeed, at small $\mathbf{k}$, $\mathbf{k}\to 0$ the matrix $\sigma_{\alpha_1\alpha_2}(-\mathbf{k})$, as follows from the determination (11), may be written as

$$\sigma_{\alpha_1\alpha_2}(-\mathbf{k}) \approx i\mathbf{k}\mathbf{d}_{\alpha_1\alpha_2}, \tag{88}$$

where $\mathbf{d}_{\alpha\beta}$ is atom's dipole moment tensor:

$$\mathbf{d}_{\alpha\beta} \equiv \sigma_{\alpha\beta}(\mathbf{k}) \equiv e\int d\mathbf{y}\, \varphi^*_\alpha(\mathbf{y})\mathbf{y}\varphi_\beta(\mathbf{y}), \quad \mathbf{d}_{\alpha\alpha} = 0, \tag{89}$$

and $\varphi_\alpha(\mathbf{y})$ is wave function of hydrogen-like atom in a quantum-mechanical state defined by a set of characteristics $\alpha$. Formula (88) is valid also in a "point" approximation for atom, see Ref. 18. Since the diagonal matrix elements $\mathbf{d}_{\alpha\beta}$ are equal to zero, then also $U_{\alpha_1\alpha_2}(f) = 0$, see (87). We present a one more expression connected with the limits of type (87) at $v(\mathbf{k}) = \frac{4\pi}{\mathbf{k}^2}$, see (25). Ref. 38 shows how to tackle with such limits. Following that receipts one can show that the limit of type (87) should be calculated according to the receipt:

$$\lim_{\mathbf{k}\to 0}(\mathbf{k}\mathbf{d})_{\alpha_1\alpha_2} v(\mathbf{k})(\mathbf{k}\mathbf{d})_{\alpha'\alpha'} = \lim_{\mathbf{k}\to 0}\frac{4\pi}{\mathbf{k}^2}(\mathbf{k}\mathbf{d}_{\alpha_1\alpha_2})(\mathbf{k}\mathbf{d}_{\beta_1\beta_2}) = \frac{4\pi}{3}\mathbf{d}_{\alpha_1\alpha_2}\mathbf{d}_{\beta_1\beta_2}. \tag{90}$$

The expressions (88) – (90) will be helpful for us when conducting further calculations.

We consider now the equilibrium distribution functions of free (non-bound) fermions. They are to have a usual "Fermi" form,

$$\overline{f}^{(1)}(\mathbf{p}) = \left\{e^{\beta(\overline{\varepsilon}^{(1)}(\mathbf{p})-\overline{\mu}^{(1)})}+1\right\}^{-1}, \quad \overline{f}^{(2)}(\mathbf{p}) = \left\{e^{\beta(\overline{\varepsilon}^{(2)}(\mathbf{p})-\overline{\mu}^{(2)})}+1\right\}^{-1}, \tag{91}$$

where the quantities $\bar{\varepsilon}^{(1)}(\mathbf{p})$, $\bar{\varepsilon}^{(2)}(\mathbf{p})$ are defined by formulas (77), $\bar{\mu}^{(1)}$, $\bar{\mu}^{(2)}$ are the chemical potentials of the first and second fermion components, unknown, and $\beta$ is the inverse temperature. Together with (77) the expressions (91) are the self-consistence equations, since the quantities $\bar{\varepsilon}^{(1)}(\mathbf{p})$, $\bar{\varepsilon}^{(2)}(\mathbf{p})$ in their turn, depend on the equilibrium distribution functions $\bar{f}^{(1)}(\mathbf{p})$, $\bar{f}^{(2)}(\mathbf{p})$ and $\bar{f}^{(0)}_{\alpha_1;\alpha_2}(\mathbf{p})$. At this point we should remember that $U(f)$, defined by (77), due to (82), (88), (89) equals to zero. The reason is the same as in the case with $U_{\alpha_1\alpha_2}(f)$, see (87): a diagonal shape of the equilibrium distribution function's principal approximation over the interaction (83) and the equality to zero of the diagonal elements of the dipole moment tensor of atom (89). Thus, the self-consistence equations for equilibrium distribution functions of fermions of both kinds are the equations (82), with the following expressions (see also (77)):

$$\bar{\varepsilon}^{(1)}(\mathbf{p})=\varepsilon_1(\mathbf{p})-\frac{e^2}{V}\sum_{\mathbf{p}'}\bar{f}^{(1)}(\mathbf{p}')\nu(\mathbf{p}'-\mathbf{p}), \quad \bar{\varepsilon}^{(2)}(\mathbf{p})=\varepsilon_2(\mathbf{p})-\frac{e^2}{V}\sum_{\mathbf{p}'}\bar{f}^{(2)}(\mathbf{p}')\nu(\mathbf{p}'-\mathbf{p}). \quad (92)$$

We note that the self-consistence equations similar to (91), (92), raise in the Fermi-liquid theory, in particular, when describing specific phase transitions in it, see, e.g. Ref. 39.

To solve equations (91), (92) (like when tackling with eq. (81)) we also use a perturbation theory over the weak interaction. The principal approximation in this theory is the approximation of an ideal fermion gas, like in the case of constructing an equilibrium distribution function of Bose-atoms, see (84):

$$f^{(1)}(\mathbf{p})=\left\{e^{\beta(\varepsilon_1(\mathbf{p})-\bar{\mu}^{(1)})}+1\right\}^{-1}, \quad f^{(2)}(\mathbf{p})=\left\{e^{\beta(\varepsilon_2(\mathbf{p})-\bar{\mu}^{(2)})}+1\right\}^{-1}, \quad \varepsilon_{1,2}(\mathbf{p})=\frac{\mathbf{p}^2}{2m_{1,2}}. \quad (93)$$

In the present paper, we will not obtain amendments over the interaction to expressions (93) due to the following reason. The linearized equations (76), as it follows from (79), (80), in their right-hand sides contain only the summands proportional to the interaction that is considered to be weak. The initial equations (65), (66) we also obtained within the first order over the weak particles interactions. Consequently, to obtain the quantities $\delta\varepsilon^{(1)}(\mathbf{x},\mathbf{p})$, $\delta\varepsilon^{(2)}(\mathbf{x},\mathbf{p})$, $\delta\varepsilon^{(0)}_{\alpha_1;\alpha_2}(\mathbf{x},\mathbf{p})$, that define the right-hand sides of equations (76), the expressions (84), (93) for the equilibrium distribution functions of an ideal gas, are sufficient. These expressions, hoverer, contain some still unknown quantities, the chemical potentials $\mu^{(1)}$, $\mu^{(2)}$ и $\mu^{(0)}$. The procedure to obtain them is presented in Ref. 23 in detail. To find them, the expressions (84) and (93) should be extended with

$$\mu^{(1)} + \mu^{(2)} = \mu^{(0)},$$
$$\sum_{\mathbf{p}} f^{(1)}(\mathbf{p}) = \sum_{\mathbf{p}} f^{(2)}(\mathbf{p}), \tag{94}$$
$$\sum_{\mathbf{p}} f^{(1)}(\mathbf{p}) + \sum_{\alpha,\mathbf{p}} f_\alpha^{(0)}(\mathbf{p}) = N,$$

where $N$ is the full number of fermions of the first or the second kind in the system. The first equation (94) is the Gibbs phases rule (like in the case of chemical reactions). the second one is equivalent to the quasi-neutralness condition of the system (see (75)). Finally, the third equation (94) reflects the preservation of the full number of fermions of the first or the second kind, taking into account both their free states and them in bound states (hydrogen-like atoms). We here use the solution of equations (93), (94), that describes an equilibrium state of weakly ionized gas far from degeneracy temperatures of all components of the system, see Ref. 23. In this case all the distribution functions have Maxwell form and are defined by the following formulas:

$$f^{(1)}(\mathbf{p}) = \nu^{(1)}(\nu,T)\left(\frac{2\pi}{m_1 T}\right)^{3/2} e^{-\frac{\mathbf{p}^2}{2m_1 T}}, \quad f^{(2)}(\mathbf{p}) = \nu^{(2)}(\nu,T)\left(\frac{2\pi}{T m_2}\right)^{3/2} e^{-\frac{\mathbf{p}^2}{2m_2 T}},$$
$$f_\alpha^{(0)}(\mathbf{p}) = \nu_\alpha^{(0)}(\nu,T)\left(\frac{2\pi}{MT}\right)^{3/2} e^{-\frac{\mathbf{p}^2}{2MT}}, \tag{95}$$

where the fermion components densities $\nu^{(1)}(\nu,T)$, $\nu^{(1)}(\nu,T)$ and the atom density in the state $\alpha$ are given by the following expressions:

$$\nu^{(1)}(\nu,T) = \nu^{(2)}(\nu,T) \approx \sqrt{\nu}\left(\frac{m_1 m_2 T}{2\pi M}\right)^{3/4} e^{\frac{\varepsilon_0}{2T}},$$
$$\nu_0^{(0)}(\nu,T) \approx \nu - \nu^{(1)}(\nu,T), \qquad \varepsilon_\alpha = \varepsilon_0, \tag{96}$$
$$\nu_\alpha^{(0)}(\nu,T) \approx \nu e^{\frac{\varepsilon_0 - \varepsilon_\alpha}{T}}, \qquad \varepsilon_\alpha \neq \varepsilon_0, \quad \nu \equiv \frac{N}{V},$$

where $N$ is the full number of fermions of the first and the second kinds (including the ones in the bound states) and $\varepsilon_0 < 0$ is the ground state energy of the hydrogen-like atom. we remind, that $\varepsilon_\alpha$, the energy of a bound state, is the energy of an atom with a zero momentum, see (84). Consequently, $\varepsilon_\alpha < 0$ for every $\alpha$. Then, at $\varepsilon_\alpha \neq \varepsilon_0$ the inequality $|\varepsilon_\alpha| < |\varepsilon_0|$ always holds. Furthermore, as we stated above, the secondary quantization method itself[18] is applicable in the case when the particles average kinetic energy is small compared to the differences of atomic levels energies, $T/|\varepsilon_\alpha - \varepsilon_\beta| \ll 1$. Here we see that the components densities $\nu^{(1)}(\nu,T)$, $\nu^{(2)}(\nu,T)$ and $\nu_\alpha^{(0)}(\nu,T)$ are

exponentially small over the temperature, whereas $\nu_0^{(0)}(\nu,T)$ differs from $\nu$ with an exponentially small value, see (96). Since the quantities $\nu^{(1)}(\nu,T), \nu^{(2)}(\nu,T)$ are the densities of free (non-bound) charged fermions, then

$$\nu^{(1)}(\nu,T) = \nu^{(2)}(\nu,T) \ll \nu_0^{(0)}(\nu,T) \tag{97}$$

contain the essence of the term "weakly ionized" gas of hydrogen-like atoms. The inequality

$$\nu_\alpha^{(0)}(\nu,T) \ll \nu_0^{(0)}(\nu,T) \tag{98}$$

unfolds the definition of a "weakly excited" gas in which all the components are far from their respective degeneracy temperatures $T_c^{(1)}$, $T_c^{(2)}$, $T_c^{(0)}$:

$$T \gg \max\left\{T_c^{(1)}, T_c^{(2)}, T_c^{(0)}\right\}. \tag{99}$$

Here there is also a need that the main condition of the secondary quantization method is true,

$$T \ll \left|\varepsilon_\alpha - \varepsilon_\beta\right|, \quad \forall \varepsilon_\alpha, \varepsilon_\beta, \tag{100}$$

The synthesis of (99), (100) gives a temperatures interval where the results of our work are valid,

$$\left|\varepsilon_\alpha - \varepsilon_\beta\right| \gg T \gg \max\left\{T_c^{(1)}, T_c^{(2)}, T_c^{(0)}\right\}. \tag{101}$$

Thus, the obtained linearized kinetic equations (76) together with (79), (80), (86) – (90), (95), (96) allow to get a dispersion law for electromagnetic eigenwaves in the system under study (or the photons dispersion law in such medium) only when (101) holds.

## VII. EVOLUTION EQUATIONS OF EIGENWAVES IN WEAKLY IONIZED AND WEAKLY EXCITED GAS OF HYDROGEN-LIKE ATOMS.

In order to obtain the mentioned above dispersion laws it is necessary, as already noted, to solve the linearized equations (76). To do this we closely follow the methods described in Ref. 36, modified them to the case of multicomponent systems. When solving equations (76) it is convenient to use Fourier transformations. Passing to Fourier-images of the coordinate in these equations and according to the expressions

$$F(\mathbf{x},t) = \frac{1}{(2\pi)^4} \int_{-\infty}^{\infty} d\omega \int d\mathbf{q}\, e^{i\mathbf{q}\mathbf{x}-i\omega t} F(\mathbf{q},\omega), \quad F(\mathbf{q},\omega) = \int_{-\infty}^{\infty} d\omega \int d\mathbf{q}\, e^{i\mathbf{q}\mathbf{x}-i\omega t} F(\mathbf{x},t), \tag{102}$$

(where $F(\mathbf{x},t)$ is some arbitrary function of coordinates and time and $F(\mathbf{q},\omega)$ is its Fourier image), we come to the following equations for the Fourier-images of distribution functions $\delta f^{(1)}(\mathbf{k},\omega,\mathbf{p})$, $\delta f^{(2)}(\mathbf{k},\omega,\mathbf{p})$, $\delta f^{(0)}_{\alpha_1,\alpha_2}(\mathbf{k},\omega,\mathbf{p})$:

$$\left(\mathbf{k}\frac{\mathbf{p}}{m_1} - \omega\right)\delta f^{(1)}(\mathbf{k},\omega;\mathbf{p}) = \mathbf{k}\frac{\partial f^{(1)}(\mathbf{p})}{\partial \mathbf{p}} \delta\varepsilon^{(1)}(\mathbf{k},\omega;\mathbf{p}),$$

$$\left(\mathbf{k}\frac{\mathbf{p}}{m_1} - \omega\right)\delta f^{(2)}(\mathbf{k},\omega;\mathbf{p}) = \mathbf{k}\frac{\partial f^{(2)}(\mathbf{p})}{\partial \mathbf{p}} \delta\varepsilon^{(2)}(\mathbf{k},\omega;\mathbf{p}),$$

$$\left(\mathbf{k}\frac{\mathbf{p}}{M} + \varepsilon_{\alpha_1} - \varepsilon_{\alpha_2} - \omega\right)\delta f^{(0)}_{\alpha_1\alpha_2}(\mathbf{k},\omega;\mathbf{p}) \tag{103}$$

$$= \left\{\frac{\mathbf{k}}{2}\frac{\partial f^{(0)}_{\alpha_1}(\mathbf{p})}{\partial \mathbf{p}} + f^{(0)}_{\alpha_1}(\mathbf{p}) + \frac{\mathbf{k}}{2}\frac{\partial f^{(0)}_{\alpha_2}(\mathbf{p})}{\partial \mathbf{p}} - f^{(0)}_{\alpha_2}(\mathbf{p})\right\}\delta\varepsilon^{(0)}_{\alpha_1\alpha_2}(\mathbf{k},\omega;\mathbf{p}),$$

$$M = m_1 + m_2,$$

where the functions $\delta\varepsilon^{(1)}(\mathbf{k},\omega;\mathbf{p})$, $\delta\varepsilon^{(2)}(\mathbf{k},\omega;\mathbf{p})$, $\delta\varepsilon^{(0)}_{\alpha_1\alpha_2}(\mathbf{k},\omega;\mathbf{p})$ according to Eqs. (79), (80), (87), (102) are defined with the following formulae:

$$\delta\varepsilon^{(1)}(\mathbf{k},\omega;\mathbf{p}) \equiv -\frac{e^2}{V}\sum_{\mathbf{p}'}\delta f^{(1)}(\mathbf{k},\omega;\mathbf{p}')v(\mathbf{p}'-\mathbf{p})$$

$$-v(\mathbf{k})\frac{1}{V}\left[\sum_{\alpha'\alpha''\mathbf{p}'}\sigma_{\alpha''\alpha'}(-\mathbf{k})\delta f^{(0)}_{\alpha'\alpha''}(\mathbf{k},\omega;\mathbf{p}')\right.$$

$$\left.-e^2\left(\sum_{\mathbf{p}'}\delta f^{(1)}(\mathbf{k},\omega;,\mathbf{p}') - \sum_{\mathbf{p}'_3}\delta f^{(2)}(\mathbf{k},\omega;,\mathbf{p}'_3)\right)\right],$$

$$\delta\varepsilon^{(2)}(\mathbf{k},\omega;\mathbf{p}) \equiv -\frac{e^2}{V}\sum_{\mathbf{p}'}\delta f^{(2)}(\mathbf{k},\omega;\mathbf{p}')v(\mathbf{p}'-\mathbf{p})$$

$$+v(\mathbf{k})\frac{1}{V}\left[\sum_{\alpha'\alpha''\mathbf{p}'}\sigma_{\alpha''\alpha'}(-\mathbf{k})\delta f^{(0)}_{\alpha'\alpha''}(\mathbf{k},\omega;\mathbf{p}')\right.$$

$$\left.-e^2\left(\sum_{\mathbf{p}'}\delta f^{(1)}(\mathbf{k},\omega;,\mathbf{p}') - \sum_{\mathbf{p}'_3}\delta f^{(2)}(\mathbf{k},\omega;,\mathbf{p}'_3)\right)\right],$$

$$\delta \varepsilon^{(0)}_{\alpha_1\alpha_2}(\mathbf{k},\omega;\mathbf{p}) \equiv$$

$$\equiv \frac{1}{V}\Bigg\{\sigma_{\alpha_1\alpha_2}(\mathbf{k})\nu(\mathbf{k})\Bigg[\sum_{\alpha'\alpha''\mathbf{p}'}\sigma_{\alpha''\alpha'}(-\mathbf{k})\delta f^{(0)}_{\alpha'\alpha''}(\mathbf{k},\omega;\mathbf{p}')$$

$$-e^2\Bigg(\sum_{\mathbf{p}'}\delta f^{(1)}(\mathbf{k},\omega;\mathbf{p}') - \sum_{\mathbf{p}'}\delta f^{(2)}(\mathbf{k},\omega;\mathbf{p}')\Bigg)\Bigg] \quad (104)$$

$$+\frac{1}{V}\sum_{\alpha'\alpha''\mathbf{p}'}\nu(\mathbf{p}'-\mathbf{p})\sigma_{\alpha_1\alpha'}(\mathbf{p}-\mathbf{p}')\delta f^{(0)}_{\alpha'\alpha''}(\mathbf{k},\omega;\mathbf{p}')\sigma_{\alpha''\alpha_2}(\mathbf{p}'-\mathbf{p})\Bigg\}.$$

The most general solution of equations (103) analogous to Ref. 36 reads:

$$\delta f^{(1)}(\mathbf{k},\omega;\mathbf{p}) = -\left(\omega - \mathbf{k}\frac{\mathbf{p}}{m_1} + i0\right)^{-1}\mathbf{k}\frac{\partial f^{(1)}(\mathbf{p})}{\partial \mathbf{p}}\delta\varepsilon^{(1)}(\mathbf{k},\omega;\mathbf{p})$$

$$+\delta A^{(1)}(\mathbf{k},\mathbf{p})\delta\left(\omega - \mathbf{k}\frac{\mathbf{p}}{m_1}\right),$$

$$\delta f^{(2)}(\mathbf{k},\omega;\mathbf{p}) = -\left(\omega - \mathbf{k}\frac{\mathbf{p}}{m_2} + i0\right)^{-1}\mathbf{k}\frac{\partial f^{(2)}(\mathbf{p})}{\partial \mathbf{p}}\delta\varepsilon^{(2)}(\mathbf{k},\omega;\mathbf{p})$$

$$+\delta A^{(2)}(\mathbf{k},\mathbf{p})\delta\left(\omega - \mathbf{k}\frac{\mathbf{p}}{m_2}\right),$$

$$\delta f^{(0)}_{\alpha_1\alpha_2}(\mathbf{k},\omega;\mathbf{p}) = -\left(\omega - \mathbf{k}\frac{\mathbf{p}}{M} - \varepsilon_{\alpha_1} + \varepsilon_{\alpha_2} + i0\right)^{-1}$$

$$\times\left\{\frac{\mathbf{k}}{2}\frac{\partial f^{(0)}_{\alpha_1}(\mathbf{p})}{\partial \mathbf{p}} + f^{(0)}_{\alpha_1}(\mathbf{p}) + \frac{\mathbf{k}}{2}\frac{\partial f^{(0)}_{\alpha_2}(\mathbf{p})}{\partial \mathbf{p}} - f^{(0)}_{\alpha_2}(\mathbf{p})\right\}\delta\varepsilon^{(0)}_{\alpha_1\alpha_2}(\mathbf{k},\omega;\mathbf{p})$$

$$+\delta A^{(0)}_{\alpha_1\alpha_2}(\mathbf{k},\mathbf{p})\delta\left(\omega - \mathbf{k}\frac{\mathbf{p}}{M} - \varepsilon_{\alpha_1} + \varepsilon_{\alpha_2}\right), \quad (105)$$

where the quantities $\delta\varepsilon^{(1)}(\mathbf{k},\omega;\mathbf{p})$, $\delta\varepsilon^{(2)}(\mathbf{k},\omega;\mathbf{p})$, $\delta\varepsilon^{(0)}_{\alpha_1\alpha_2}(\mathbf{k},\omega;\mathbf{p})$ (see (104)) may be defined as follows:

$$\delta\varepsilon^{(1)}(\mathbf{k},\omega;\mathbf{p}) \equiv e^2\left[\nu(\mathbf{k}) - \nu(\mathbf{p})\right]\delta\varphi^{(1)}(\mathbf{k},\omega)$$

$$-\nu(\mathbf{k})\left[e\sum_{\alpha'\alpha''}\sigma_{\alpha''\alpha'}(-\mathbf{k})\delta\varphi^{(0)}_{\alpha'\alpha''}(\mathbf{k},\omega) + e^2\delta\varphi^{(2)}(\mathbf{k},\omega)\right],$$

$$\delta\varepsilon^{(2)}(\mathbf{k},\omega;\mathbf{p}) \equiv e^2\left[\nu(\mathbf{k}) - \nu(\mathbf{p})\right]\delta\varphi^{(2)}(\mathbf{k},\omega)$$

$$+\nu(\mathbf{k})\left[e\sum_{\alpha'\alpha''}\sigma_{\alpha''\alpha'}(-\mathbf{k})\delta\varphi^{(0)}_{\alpha'\alpha''}(\mathbf{k},\omega) - e^2\delta\varphi^{(1)}(\mathbf{k},\omega)\right], \quad (106)$$

$$\delta\varepsilon^{(0)}_{\alpha_1\alpha_2}(\mathbf{k},\omega;\mathbf{p}) \equiv \sum_{\alpha'\alpha''}\delta\varphi^{(0)}_{\alpha'\alpha''}(\mathbf{k},\omega)\left\{\sigma_{\alpha_1\alpha_2}(\mathbf{k})\nu(\mathbf{k})\sigma_{\alpha''\alpha'}(-\mathbf{k})\right.$$

$$\left.+\sigma_{\alpha_1\alpha'}(\mathbf{p})\nu(\mathbf{p})\sigma_{\alpha''\alpha_2}(-\mathbf{p})\right\}$$

$$-\sigma_{\alpha_1\alpha_2}(\mathbf{k})\nu(\mathbf{k})e\left(\delta\varphi^{(1)}(\mathbf{k},\omega) - \delta\varphi^{(2)}(\mathbf{k},\omega)\right).$$

We introduce in the solutions (105) the arbitrary functions $\delta A^{(1)}(\mathbf{k},\mathbf{p})$, $\delta A^{(2)}(\mathbf{k},\mathbf{p})$, $\delta A^{(0)}_{\alpha_1\alpha_2}(\mathbf{k},\mathbf{p})$, which are restricted only with the following: the obtained with their help distribution functions deviations $\delta f^{(1)}(\mathbf{x},\mathbf{p},t)$, $\delta f^{(2)}(\mathbf{x},\mathbf{p},t)$, $\delta f^{(0)}_{\alpha_1,\alpha_2}(\mathbf{x},\mathbf{p},t)$ are to be small in comparison with the corresponding equilibrium distribution functions. Besides, these functions after the operation of complex conjugation according to (102) must fulfill the following:

$$\left(\delta A^{(1)}(\mathbf{k},\mathbf{p})\right)^* = \delta A^{(1)}(-\mathbf{k},\mathbf{p}), \quad \left(\delta A^{(2)}(\mathbf{k},\mathbf{p})\right)^* = \delta A^{(2)}(-\mathbf{k},\mathbf{p}),$$
$$\left(\delta A^{(0)}_{\alpha_1\alpha_2}(\mathbf{k},\mathbf{p})\right)^* = \delta A^{(0)}_{\alpha_2\alpha_1}(-\mathbf{k},\mathbf{p}). \tag{107}$$

In the solutions (105) we introduce also the functions $\delta\varphi^{(1)}(\mathbf{k},\omega)$, $\delta\varphi^{(2)}(\mathbf{k},\omega)$, $\delta\varphi^{(0)}_{\alpha_1\alpha_2}(\mathbf{k},\omega)$, which in turn depend on the linearized distribution functions and, therefore, are subject to the definitions

$$\delta\varphi^{(1)}(\mathbf{k},\omega) \equiv \frac{1}{V}\sum_{\mathbf{p}'}\delta f^{(1)}(\mathbf{k},\omega;\mathbf{p}'), \quad \delta\varphi^{(2)}(\mathbf{k},\omega) \equiv \frac{1}{V}\sum_{\mathbf{p}'}\delta f^{(2)}(\mathbf{k},\omega;\mathbf{p}'),$$
$$\delta\varphi^{(0)}_{\alpha_1\alpha_2}(\mathbf{k},\omega) \equiv \frac{1}{V}\sum_{\mathbf{p}'}\delta f^{(0)}_{\alpha_1\alpha_2}(\mathbf{k},\omega;\mathbf{p}'). \tag{108}$$

From (105) for the quantities (108) we obtain a set of equations

$$\delta\varphi^{(1)}(\mathbf{k},\omega) = \frac{g^{(1,2)}(\mathbf{k},\omega)}{D_1}\left[g^{(2,2)}(\mathbf{k},\omega) - g^{(2,1)}(\mathbf{k},\omega)\right]\sum_{\alpha\alpha''}\sigma_{\alpha''\alpha'}(-\mathbf{k})\delta\varphi^{(0)}_{\alpha'\alpha''}(\mathbf{k},\omega)$$
$$+ \frac{g^{(1,2)}(\mathbf{k},\omega)}{D_1}\delta A^{(2)}(\mathbf{k},\omega) + \frac{g^{(2,2)}(\mathbf{k},\omega)}{D_1}\delta A^{(1)}(\mathbf{k},\omega),$$
$$\delta\varphi^{[2]}(\mathbf{k},\omega) = \frac{g^{(2,1)}(\mathbf{k},\omega)}{D}\left[g^{(1,2)}(\mathbf{k},\omega) - g^{(1,1)}(\mathbf{k},\omega)\right]\sum_{\alpha'\alpha''}\sigma_{\alpha''\alpha'}(-\mathbf{k})\delta\varphi^{(0)}_{\alpha'\alpha''}(\mathbf{k},\omega) \tag{109}$$
$$+ \frac{g^{(1,1)}(\mathbf{k},\omega)}{D_1}\delta A^{(2)}(\mathbf{k},\omega) + \frac{g^{(2,1)}(\mathbf{k},\omega)}{D_1}\delta A^{(1)}(\mathbf{k},\omega),$$
$$\sum_{\alpha'\alpha''}g^{(0,0)}_{\alpha_1\alpha';\alpha_2\alpha''}\delta\varphi^{(0)}_{\alpha'\alpha''}(\mathbf{k},\omega) = g^{(0)}_{\alpha_1\alpha_2}\left[\delta\varphi^{(1)}(\mathbf{k},\omega) - \delta\varphi^{(2)}(\mathbf{k},\omega)\right] + \delta A^{(0)}_{\alpha_1\alpha_2}(\mathbf{k},\omega),$$

in which the coefficients near the unknown variables (108) are defined with

$$g^{(1,1)}(\mathbf{k},\omega) \equiv 1 + \frac{1}{V}\sum_{\mathbf{p}}\left(\omega - \mathbf{k}\frac{\mathbf{p}}{m_1} + i0\right)^{-1}\mathbf{k}\frac{\partial f^{(1)}(\mathbf{p})}{\partial\mathbf{p}}e^2\left[\nu(\mathbf{k}) - \nu(\mathbf{p})\right],$$

$$g^{(1,2)}(\mathbf{k},\omega) \equiv v(\mathbf{k})\frac{e^2}{V}\sum_{\mathbf{p}}\left(\omega-\mathbf{k}\frac{\mathbf{p}}{m_1}+i0\right)^{-1}\mathbf{k}\frac{\partial f^{(1)}(\mathbf{p})}{\partial \mathbf{p}},$$

$$g^{(2,2)}(\mathbf{k},\omega) \equiv 1+\frac{e^2}{V}\sum_{\mathbf{p}}\left(\omega-\mathbf{k}\frac{\mathbf{p}}{m_2}+i0\right)^{-1}\mathbf{k}\frac{\partial f^{(2)}(\mathbf{p})}{\partial \mathbf{p}}\left[v(\mathbf{k})-v(\mathbf{p})\right],$$

$$g^{(2,1)}(\mathbf{k},\omega) \equiv v(\mathbf{k})\frac{e^2}{V}\sum_{\mathbf{p}}\left(\omega-\mathbf{k}\frac{\mathbf{p}}{m_2}+i0\right)^{-1}\mathbf{k}\frac{\partial f^{(2)}(\mathbf{p})}{\partial \mathbf{p}},$$

$$g^{(0,0)}_{\alpha_1\alpha';\alpha_2\alpha''} \equiv \delta_{\alpha_1\alpha'}\delta_{\alpha_2\alpha''}+$$

$$+\frac{1}{V}\sum_{\mathbf{p}}\left(\omega-\mathbf{k}\frac{\mathbf{p}}{M}-\varepsilon_{\alpha_1}+\varepsilon_{\alpha_2}+i0\right)^{-1}\left\{\frac{\mathbf{k}}{2}\frac{\partial f^{(0)}_{\alpha_1}(\mathbf{p})}{\partial \mathbf{p}}+f^{(0)}_{\alpha_1}(\mathbf{p})+\frac{\mathbf{k}}{2}\frac{\partial f^{(0)}_{\alpha_2}(\mathbf{p})}{\partial \mathbf{p}}-f^{(0)}_{\alpha_2}(\mathbf{p})\right\}$$

$$\times\left\{\sigma_{\alpha_1\alpha_2}(\mathbf{k})v(\mathbf{k})\sigma_{\alpha''\alpha'}(-\mathbf{k})+\sigma_{\alpha_1\alpha'}(\mathbf{p})v(\mathbf{p})\sigma_{\alpha''\alpha_2}(-\mathbf{p})\right\},$$

$$g^{(0)}_{\alpha_1\alpha_2} \equiv \sigma_{\alpha_1\alpha_2}(\mathbf{k})v(\mathbf{k})\frac{e}{V}\sum_{\mathbf{p}}\left(\omega-\mathbf{k}\frac{\mathbf{p}}{M}-\varepsilon_{\alpha_1}+\varepsilon_{\alpha_2}+i0\right)^{-1}$$

$$\times\left\{\frac{\mathbf{k}}{2}\frac{\partial f^{(0)}_{\alpha_1}(\mathbf{p})}{\partial \mathbf{p}}+f^{(0)}_{\alpha_1}(\mathbf{p})+\frac{\mathbf{k}}{2}\frac{\partial f^{(0)}_{\alpha_2}(\mathbf{p})}{\partial \mathbf{p}}-f^{(0)}_{\alpha_2}(\mathbf{p})\right\}, \quad (110)$$

and we introduced the following designations (see (107)):

$$D_1 \equiv \left[\varepsilon^{(1,1)}(\mathbf{k},\omega)\varepsilon^{(2,2)}(\mathbf{k},\omega)-\varepsilon^{(1,2)}(\mathbf{k},\omega)\varepsilon^{(2,1)}(\mathbf{k},\omega)\right],$$

$$\delta A^{(1)}(\mathbf{k},\omega) \equiv \frac{1}{V}\sum_{\mathbf{p}}\delta A^{(1)}(\mathbf{k},\mathbf{p})\delta\left(\omega-\mathbf{k}\frac{\mathbf{p}}{m_1}\right),$$

$$\delta A^{(2)}(\mathbf{k},\omega) \equiv \frac{1}{V}\sum_{\mathbf{p}}\delta A^{(2)}(\mathbf{k},\mathbf{p})\delta\left(\omega-\mathbf{k}\frac{\mathbf{p}}{m_2}\right), \quad (111)$$

$$\delta A^{(0)}_{\alpha_1\alpha_2}(\mathbf{k},\omega) \equiv \frac{1}{V}\sum_{\mathbf{p}}\delta A^{(0)}_{\alpha_1\alpha_2}(\mathbf{k},\mathbf{p})\delta\left(\omega-\mathbf{k}\frac{\mathbf{p}}{M}-\varepsilon_{\alpha_1}+\varepsilon_{\alpha_2}\right).$$

The further calculations suppose evaluating the contained in (110) integrals of the type

$$J_1^{(1)} \equiv v(\mathbf{k})\frac{e^2}{V}\sum_{\mathbf{p}}\left(\omega-\mathbf{k}\frac{\mathbf{p}}{m_1}+i0\right)^{-1}\mathbf{k}\frac{\partial f^{(1)}(\mathbf{p})}{\partial \mathbf{p}}$$

$$= e^2 v(\mathbf{k})\frac{1}{(2\pi)^3}\int d^3p\left(\omega-\mathbf{k}\frac{\mathbf{p}}{m_1}+i0\right)^{-1}\mathbf{k}\frac{\partial f^{(1)}(\mathbf{p})}{\partial \mathbf{p}},$$

$$J_2^{(1)} \equiv \frac{e^2}{V}\sum_{\mathbf{p}}v(\mathbf{p})\left(\omega-\mathbf{k}\frac{\mathbf{p}}{m_1}+i0\right)^{-1}\mathbf{k}\frac{\partial f^{(1)}(\mathbf{p})}{\partial \mathbf{p}}$$

$$= e^2\frac{1}{(2\pi)^3}\int d^3p\, v(\mathbf{p})\left(\omega-\mathbf{k}\frac{\mathbf{p}}{m_1}+i0\right)^{-1}\mathbf{k}\frac{\partial f^{(1)}(\mathbf{p})}{\partial \mathbf{p}}, \quad (112)$$

under the condition that the quantity $V(\mathbf{k})$ (as well as $V(\mathbf{p})$) is given by the formula (25), and the expression for $f^{(1)}(\mathbf{p})$ is given by Eq. (93). First we show how to evaluate the integrals similar to $J_1^{(1)}$ (112). We take into account that for fraction $\left(\omega - \mathbf{k}\dfrac{\mathbf{p}}{m_1} + i0\right)^{-1}$ the following is valid:

$$\left(\omega - \mathbf{k}\frac{\mathbf{p}}{m_1} + i0\right)^{-1} = P\left(\omega - \mathbf{k}\frac{\mathbf{p}}{m_1}\right)^{-1} - i\pi\delta\left(\omega - \mathbf{k}\frac{\mathbf{p}}{m_1}\right), \quad (113)$$

where symbol $P$ denotes that in further integrating the principal value of the integral is taken. Taking into account that $\mathbf{k}\dfrac{\partial f^{(1)}(\mathbf{p})}{\partial \mathbf{p}} = -\mathbf{k}\dfrac{\mathbf{p}}{m_1} f^{(1)}(\mathbf{p})$, the integral $J_1$ may be decomposed into the two summands:

$$\begin{aligned}
J_1^{(1)} &= A + iB, \\
A &= -e^2 V(\mathbf{k}) \frac{1}{(2\pi)^3} P\int d\mathbf{p}\left(1 - \frac{\mathbf{kp}}{\omega m_1}\right)^{-1} \frac{\mathbf{kp}}{\omega m_1} f^{(1)}(\mathbf{p}), \\
B &= \pi\omega e^2 V(\mathbf{k}) \frac{1}{(2\pi)^3} \int d\mathbf{p}\,\delta\left(\omega - \frac{\mathbf{kp}}{m_1}\right) f^{(1)}(\mathbf{p}).
\end{aligned} \quad (114)$$

Taking into account that the function $f^{(1)}(\mathbf{p})$ has a sharp maximum at $\mathbf{p} = 0$, in the integral $A$ in Eq. (114) the intra-integral expression may be formally decomposed into the Taylor series over $\mathbf{kp}/\omega m_1$, that allows to calculate this integral in the perturbation theory (see in this relation Refs. 31, 36). The result of these calculations may be presented as:

$$A \approx -\left[\frac{\omega_0^{(1)}(\nu,T)}{\omega}\right]^2 \left\{1 + 3\left[\frac{\omega_0^{(1)}(\nu,T)}{\omega}\right]^2 \left[kr_D^{(1)}(\nu,T)\right]^2\right\}, \quad \left[kr_D^{(1)}(\nu,T)\right]^2 \ll 1, \quad (115)$$

where we introduced

$$\begin{aligned}
\left[\omega_0^{(1,2)}(\nu,T)\right]^2 &\equiv \frac{4\pi e^2 \nu^{(1,2)}(\nu,T)}{m_{1,2}}, \\
\left[r_D^{(1)}(\nu,T)\right]^2 &= \left[r_D^{(2)}(\nu,T)\right]^2 \equiv \left[r_D(\nu,T)\right]^2 \equiv \frac{T}{4\pi e^2 \nu^{(1,2)}(\nu,T)} = \frac{T}{m_{1,2}\left[\omega_0^{(1,2)}(\nu,T)\right]^2}
\end{aligned} \quad (116a)$$

in which the densities of the fermions of the first and the second kinds $v^{(1)}(V,T) = v^{(2)}(V,T)$ are provided with Eq. (96). As we see, from the definitions (116a) it follows that:

$$\left[\frac{\omega_0^{(1,2)}(V,T)}{\omega}\right]^2 \left[kr_D^{(1,2)}(V,T)\right]^2 = \frac{k^2 T}{\omega^2 m_{1,2}} \ll 1. \tag{116b}$$

The inequality (116b), that is the basis for the calculation of the integrals, requires the smallness of the squared heat velocity of non-bound fermions in comparison with the squared phase velocity of the wave. We note, that in plasma physics the frequency $\omega_0^{(1)}(V,T)$, being defined by the first of (116a) formulas, is called a plasma (or Langmuir) frequency. The quantity $r_D^{(1,2)}(V,T) \equiv r_D(V,T)$, given by the second formula in (116a), in plasmas is called a Debye radius of screening. In the formulas (115), (116) these values are introduced for reasons of convenience and to establish some analogies in results of our work with the known results of plasma physics. In this paper, the introduction of them does not have a usual physical sense. In fact, within this approximation (see (95 -. (96)) due to the low density of the charged components of the system the frequency $\omega_0^{(1)}(V,T)$ is vanishingly small, and the quantity $r_D^{(1,2)}(V,T) \equiv r_D(V,T)$ is so large that it cannot be called the screening radius. This also is supported by the quantitative calculations that we supply below. We remind that earlier we agreed that $m_1$ is electron mass, and $m_2$ is the mass of a core of a hydrogen-like atom, so that $m_1 \ll m_2$. Consequently, the frequency $\omega_0^{(2)}(V,T)$ is much smaller than $\omega_0^{(1)}(V,T)$, $\omega_0^{(2)}(V,T) \ll \omega_0^{(1)}(V,T)$, see (116a), that is typical also for a usual hydrogen plasma.

The integral $B$ in (114) may be calculated exactly since the distribution function $f^{(1)}(\mathbf{p})$ depends only on the momentum's absolute value (see in this regard Ref. 36):

$$B = \sqrt{\frac{\pi}{2}} \frac{\omega}{\omega_0^{(1)}(V,T)} \left(kr_D(V,T)\right)^{-3} e^{-\frac{1}{2}\left(\frac{\omega}{\omega_0^{(1)}(V,T)}\right)^2 \frac{1}{(kr_D(V,T))^2}}. \tag{117}$$

Thus, according to (112), (115), (117) the integral $J_1^{(1)}$ may be presented as:

$$J_1^{(1)} \approx -\left[\frac{\omega_0^{(1)}(V,T)}{\omega}\right]^2 \left\{1 + 3\left[\frac{\omega_0^{(1)}(V,T)}{\omega}\right]^2 \left[kr_D(V,T)\right]^2\right\}$$

$$+i\sqrt{\frac{\pi}{2}}\frac{\omega}{\omega_0^{(1)}(v,T)}(kr_D(v,T))^{-3}e^{-\frac{1}{2}\left(\frac{\omega}{\omega_0^{(1)}(v,T)}\right)^2\frac{1}{(kr_D(v,T))^2}}, \quad kr_D(v,T)\ll 1. \tag{118}$$

We note that in entirely ionized plasma expression (118) defines the permittivity and, consequently, the dispersion law of longitudinal and transverse electromagnetic waves, see in this regard, e.g., Ref. 31.

In the similar way we can calculate the integral $J_2^{(1)}$ in (112), taking into account that the quantity $v(\mathbf{p})$ is defined by formula (25). The result is as follows:

$$J_2^{(1)} \approx -\frac{\left[\omega_0^{(1)}(v,T)\right]^4}{3\omega^2 T^2}\left[kr_D(v,T)\right]^2$$
$$+i\sqrt{\frac{\pi}{2}}\frac{\omega\omega_0^{(1)}(v,T)}{2T^2}\left[kr_D(v,T)\right]^{-1}E_1\left(\left(\frac{\omega}{\omega_0^{(1)}(v,T)}\right)^2\frac{1}{2\left[kr_D(v,T)\right]^2}\right), \quad kr_D(v,T)\ll 1, \tag{119}$$

where $E_1(x)$ is the integral exponential function of the first order:

$$E_1(x) \equiv \int_1^\infty \frac{dt}{t}e^{-tx}, \qquad E_1(x) \underset{x\gg 1}{\approx} \frac{e^{-x}}{x}\left(1-\frac{1}{x}+\frac{2}{x^2}-...\right). \tag{120}$$

Considering further asymptotic series (120) of the *exponenta integralis*, expression (119) when $kr_D^{(1)}(v,T)\ll 1$ can be simplified:

$$J_2^{(1)} \approx -\frac{\left[\omega_0^{(1)}(v,T)\right]^4}{3\omega^2 T^2}\left[kr_D(v,T)\right]^2 + i\sqrt{\frac{\pi}{2}}\frac{\left[\omega_0^{(1)}(v,T)\right]^3}{\omega T^2}\left[kr_D(v,T)\right]e^{-\frac{1}{2}\left(\frac{\omega}{\omega_0^{(1)}(v,T)}\right)^2\frac{1}{(kr_D(v,T))^2}}. \tag{121}$$

Similar formulas may be obtained for the integrals with the distribution function $f^{(2)}(\mathbf{p})$, see (110). The latter using (112) – (121) allows to evaluate the expressions for $g^{(1,1)}(\mathbf{k},\omega)$, $g^{(1,2)}(\mathbf{k},\omega)$, $g^{(2,1)}(\mathbf{k},\omega)$ and $g^{(2,2)}(\mathbf{k},\omega)$ in (110):

$$g^{(1,1)}(\mathbf{k},\omega) \approx 1 - \left[\frac{\omega_0^{(1)}(\nu,T)}{\omega}\right]^2 - \left[\frac{\omega_0^{(1)}(\nu,T)}{\omega}\right]^2 \left[kr_D(\nu,T)\right]^2 \left\{3 - \frac{\left[\omega_0^{(1)}(\nu,T)\right]^2}{3T^2}\right\}$$

$$+i\sqrt{\frac{\pi}{2}} \frac{\omega}{\omega_0^{(1)}(\nu,T)} \left(kr_D(\nu,T)\right)^{-3} e^{-\frac{1}{2}\left(\frac{\omega}{\omega_0^{(1)}(\nu,T)}\right)^2 \frac{1}{(kr_D(\nu,T))^2}},$$

$$g^{(2,2)}(\mathbf{k},\omega) \approx 1 - \left[\frac{\omega_0^{(2)}(\nu,T)}{\omega}\right]^2 - \left[\frac{\omega_0^{(2)}(\nu,T)}{\omega}\right]^2 \left[kr_D(\nu,T)\right]^2 \left\{3 - \frac{\left[\omega_0^{(2)}(\nu,T)\right]^2}{3T^2}\right\}$$

$$+i\sqrt{\frac{\pi}{2}} \frac{\omega}{\omega_0^{(2)}(\nu,T)} \left(kr_D(\nu,T)\right)^{-3} e^{-\frac{1}{2}\left(\frac{\omega}{\omega_0^{(2)}(\nu,T)}\right)^2 \frac{1}{(kr_D(\nu,T))^2}},$$

$$g^{(1,2)}(\mathbf{k},\omega) \approx -\left[\frac{\omega_0^{(1)}(\nu,T)}{\omega}\right]^2 \left\{1 + 3\left[\frac{\omega_0^{(1)}(\nu,T)}{\omega}\right]^2 \left[kr_D(\nu,T)\right]^2\right\}$$

$$+i\sqrt{\frac{\pi}{2}} \frac{\omega}{\omega_0^{(1)}(\nu,T)} \left(kr_D(\nu,T)\right)^{-3} e^{-\frac{1}{2}\left(\frac{\omega}{\omega_0^{(1)}(\nu,T)}\right)^2 \frac{1}{(kr_D(\nu,T))^2}},$$

$$g^{(1,2)}(\mathbf{k},\omega) \approx -\left[\frac{\omega_0^{(2)}(\nu,T)}{\omega}\right]^2 \left\{1 + 3\left[\frac{\omega_0^{(2)}(\nu,T)}{\omega}\right]^2 \left[kr_D(\nu,T)\right]^2\right\} \quad (122)$$

$$+i\sqrt{\frac{\pi}{2}} \frac{\omega}{\omega_0^{(2)}(\nu,T)} \left(kr_D(\nu,T)\right)^{-3} e^{-\frac{1}{2}\left(\frac{\omega}{\omega_0^{(2)}(\nu,T)}\right)^2 \frac{1}{(kr_D(\nu,T))^2}}.$$

In these relations the frequency $\omega_0^{(2)}(\nu,T)$ is also defined by the formula (116), if in the expression for $\omega_0^{(1)}(\nu,T)$ the mass $m_1$ is replaced with $m_2$ and assume $\nu^{(1)}(\nu,T) = \nu^{(2)}(\nu,T)$ (see (96)). We also need to use the equality $r_D^{(1)}(\nu,T) = r_D^{(2)}(\nu,T) \equiv r_D(\nu,T)$. We note that in the imaginary part of $g^{(1,1)}(\mathbf{k},\omega)$, $g^{(2,2)}(\mathbf{k},\omega)$ the amendments to unity in braces ($kr_D(\nu,T) \ll 1$) indeed may be not taken into account since the common exponential multiplier is small. Considering (111), the formulas (122) define also the quantity $D_1$.

We show now how we calculate typical integrals that are basics for defining the tensors $g_{\alpha_1\alpha_2}^{(0)}$, $g_{\alpha_1\alpha';\alpha_2\alpha''}^{(0,0)}$ (see (110)). For this, it is sufficient to define a receipt to calculate the integrals of the two types:

$$I_{\alpha_1\alpha_2} \equiv \frac{1}{V} \sum_{\mathbf{p}} \left(\omega - \mathbf{k}\frac{\mathbf{p}}{M} - \varepsilon_{\alpha_1} + \varepsilon_{\alpha_2} + i0\right)^{-1} f_{\alpha_1}^{(0)}(\mathbf{p})$$

$$= \frac{1}{(2\pi)^3} \int d^3p f_{\alpha_1}^{(0)}(\mathbf{p}) \left(\omega - \mathbf{k}\frac{\mathbf{p}}{M} - \varepsilon_{\alpha_1} + \varepsilon_{\alpha_2} + i0\right)^{-1},$$

$$I_{\alpha_1\alpha_2;\alpha'\alpha''} \equiv \frac{1}{V}\sum_{\mathbf{p}}\left(\omega - \mathbf{k}\frac{\mathbf{p}}{M} - \varepsilon_{\alpha_1} + \varepsilon_{\alpha_2} + i0\right)^{-1} f_{\alpha_1}^{(0)}(\mathbf{p})\sigma_{\alpha_1\alpha'}(\mathbf{p})\nu(\mathbf{p})\sigma_{\alpha''\alpha_2}(-\mathbf{p})$$
$$= \frac{1}{(2\pi)^3}\int d^3p \left(\omega - \mathbf{k}\frac{\mathbf{p}}{M} - \varepsilon_{\alpha_1} + \varepsilon_{\alpha_2} + i0\right)^{-1} f_{\alpha_1}^{(0)}(\mathbf{p})\sigma_{\alpha_1\alpha'}(\mathbf{p})\nu(\mathbf{p})\sigma_{\alpha''\alpha_2}(-\mathbf{p}) \quad (123)$$

The first integral in (123), as in earlier cases, can be calculated in the long-wave limit taking into account representations (113). In this case it implies the possibility to decompose the fraction $\left(\omega - \mathbf{k}\frac{\mathbf{p}}{M} - \varepsilon_{\alpha_1} + \varepsilon_{\alpha_2}\right)^{-1}$ into series over small $\frac{\mathbf{kp}}{\omega_{\alpha_1\alpha_2}M}$, where $\omega_{\alpha_1\alpha_2}$ is a frequency detune in relation the resonant one:

$$\omega_{\alpha_1\alpha_2} \equiv \omega - (\varepsilon_{\alpha_1} - \varepsilon_{\alpha_2}), \quad \left|\frac{\mathbf{kp}}{\omega_{\alpha_1\alpha_2}M}\right| \ll 1. \quad (124)$$

Inequality (124) means that an analytical calculation of those integrals is possible away from resonant atomic levels. The result is as follows:

$$I_{\alpha_1\alpha_2} \approx \frac{\nu_{\alpha_1}^{(0)}(\nu,T)}{\omega_{\alpha_1\alpha_2}}\left(1 + \frac{Tk^2}{2M\omega_{\alpha_1\alpha_2}^2}\right) - i\frac{\nu_{\alpha_1}^{(0)}(\nu,T)}{k}\sqrt{\frac{\pi M}{2T}}e^{-\frac{M\omega_{\alpha_1\alpha_2}^2}{2Tk^2}}. \quad (125)$$

Let us show now how the integrals of type $I_{\alpha_1\alpha_2;\alpha'\alpha''}$, see (123) are calculated. For this, in the same long-wave approximation, it is also necessary to use representation (113) and the decomposition over a small parameter. But beforehand it is worth mentioning that the distribution function $f_{\alpha_1}^{(0)}(\mathbf{p})$ has sharp maximum at $\mathbf{p}=0$, and consequently, it is possible to use approximation (90). Thus, the calculations are reduced to already accomplished ones, and the result is as follows:

$$I_{\alpha_1\alpha_2;\alpha'\alpha''} \approx 4\pi \frac{\nu_{\alpha_1}^{(0)}(\nu,T)}{\omega_{\alpha_1\alpha_2}}\left\{\frac{\mathbf{d}_{\alpha_1\alpha'}\mathbf{d}_{\alpha''\alpha_2}}{3} + \frac{Tk^2}{10M\omega_{\alpha_1\alpha_2}^2}\left[\mathbf{d}_{\alpha_1\alpha'}\mathbf{d}_{\alpha''\alpha_2} + 2\frac{(\mathbf{kd}_{\alpha_1\alpha'})(\mathbf{kd}_{\alpha''\alpha_2})}{k^2}\right]\right\}$$
$$-i2\pi^{3/2}\nu_{\alpha_1}^{(0)}(\nu,T)\sqrt{\frac{M}{2T}}\frac{\mathbf{d}_{\alpha_1\alpha'}\mathbf{d}_{\alpha''\alpha_2}}{k}\left\{e^{-\frac{M\omega_{\alpha_1\alpha_2}^2}{2Tk^2}} - \frac{M\omega_{\alpha_1\alpha_2}^2}{2Tk^2}E_1\left(\frac{M\omega_{\alpha_1\alpha_2}^2}{2Tk^2}\right)\right\} \quad (126)$$
$$-i2\pi^{3/2}\nu_{\alpha_1}^{(0)}(\nu,T)\sqrt{\frac{M}{2T}}\frac{(\mathbf{kd}_{\alpha_1\alpha'})(\mathbf{kd}_{\alpha''\alpha_2})}{k^3}\left\{3\frac{M\omega_{\alpha_1\alpha_2}^2}{2Tk^2}E_1\left(\frac{M\omega_{\alpha_1\alpha_2}^2}{2Tk^2}\right) - e^{-\frac{M\omega_{\alpha_1\alpha_2}^2}{2Tk^2}}\right\}$$

If we use the asymptotic decomposition of *exponenta integralis* (120), then in principal approximation over large $M\omega_{\alpha_1\alpha_2}^2/2Tk^2$ expression (126) simplifies significantly:

$$I_{\alpha_1\alpha_2;\alpha'\alpha''} \approx 4\pi \frac{\nu_{\alpha_1}^{(0)}(\nu,T)}{\omega_{\alpha_1\alpha_2}} \left\{ \frac{\mathbf{d}_{\alpha_1\alpha'}\mathbf{d}_{\alpha''\alpha_2}}{3} + \frac{Tk^2}{10M\omega_{\alpha_1\alpha_2}^2}\left[ \mathbf{d}_{\alpha_1\alpha'}\mathbf{d}_{\alpha''\alpha_2} + 2\frac{(\mathbf{kd}_{\alpha_1\alpha'})(\mathbf{kd}_{\alpha''\alpha_2})}{k^2} \right] \right\}$$

$$-i4\pi^{3/2}\nu_{\alpha_1}^{(0)}(\nu,T)\sqrt{\frac{M}{2T}}\frac{(\mathbf{kd}_{\alpha_1\alpha'})(\mathbf{kd}_{\alpha''\alpha_2})}{k^3}e^{-\frac{M\omega_{\alpha_1\alpha_2}^2}{2Tk^2}}, \qquad \frac{Tk^2}{M\omega_{\alpha_1\alpha_2}^2} \ll 1. \tag{127}$$

In the end, after some lengthy calculations according to the receipts (123) – (127) we arrive to the following expressions for the tensors $g_{\alpha_1\alpha_2}^{(0)}$, $g_{\alpha_1\alpha';\alpha_2\alpha''}^{(0,0)}$ (see(110) ):

$$g_{\alpha_1\alpha_2}^{(0)} \approx e\sigma_{\alpha_1\alpha_2}(\mathbf{k})\nu(\mathbf{k})\frac{\left[\nu_{\alpha_1}^{(0)}(\nu,T)-\nu_{\alpha_2}^{(0)}(\nu,T)\right]}{\omega_{\alpha_1\alpha_2}}$$

$$+e\sigma_{\alpha_1\alpha_2}(\mathbf{k})\nu(\mathbf{k})\frac{Tk^2}{2M\omega_{\alpha_1\alpha_2}^3}\left\{\nu_{\alpha_1}^{(0)}(\nu,T)\left(1-\frac{\omega_{\alpha_1\alpha_2}}{T}\right)-\nu_{\alpha_2}^{(0)}(\nu,T)\left(1+\frac{\omega_{\alpha_1\alpha_2}}{T}\right)\right\}$$

$$-ie\sigma_{\alpha_1\alpha_2}(\mathbf{k})\nu(\mathbf{k})\frac{1}{k}\sqrt{\frac{\pi M}{2T}}e^{-\frac{M\omega_{\alpha_1\alpha_2}^2}{2Tk^2}}\left\{\nu_{\alpha_1}^{(0)}(\nu,T)\left(1-\frac{\omega_{\alpha_1\alpha_2}}{2T}\right)-\nu_{\alpha_2}^{(0)}(\nu,T)\left(1+\frac{\omega_{\alpha_1\alpha_2}}{2T}\right)\right\},$$

$$g_{\alpha_1\alpha';\alpha_2\alpha''}^{(0,0)} \approx \delta_{\alpha_1\alpha'}\delta_{\alpha_2\alpha''} + \frac{\nu_{\alpha_1}^{(0)}(\nu,T)-\nu_{\alpha_2}^{(0)}(\nu,T)}{\omega_{\alpha_1\alpha_2}}\left\{\frac{4\pi}{3}\mathbf{d}_{\alpha_1\alpha'}\mathbf{d}_{\alpha''\alpha_2} + \sigma_{\alpha_1\alpha_2}(\mathbf{k})\nu(\mathbf{k})\sigma_{\alpha''\alpha'}(-\mathbf{k})\right\}$$

$$+\frac{Tk^2}{2M\omega_{\alpha_1\alpha_2}^3}\left\{\sigma_{\alpha_1\alpha_2}(\mathbf{k})\nu(\mathbf{k})\sigma_{\alpha''\alpha'}(-\mathbf{k}) + \frac{4\pi}{5}\left[\mathbf{d}_{\alpha_1\alpha'}\mathbf{d}_{\alpha''\alpha_2} + 2\frac{(\mathbf{kd}_{\alpha_1\alpha'})(\mathbf{kd}_{\alpha''\alpha_2})}{k^2}\right]\right\}$$

$$\times\left\{\nu_{\alpha_1}^{(0)}(\nu,T)\left(1-\frac{\omega_{\alpha_1\alpha_2}}{T}\right)-\nu_{\alpha_2}^{(0)}(\nu,T)\left(1+\frac{\omega_{\alpha_1\alpha_2}}{T}\right)\right\}$$

$$-i\left\{\nu_{\alpha_1}^{(0)}(\nu,T)\left(1-\frac{\omega_{\alpha_1\alpha_2}}{2T}\right)-\nu_{\alpha_2}^{(0)}(\nu,T)\left(1+\frac{\omega_{\alpha_1\alpha_2}}{2T}\right)\right\}\frac{1}{k}\sqrt{\frac{\pi M}{2T}}e^{-\frac{M\omega_{\alpha_1\alpha_2}^2}{2Tk^2}}$$

$$\times\left\{\sigma_{\alpha_1\alpha_2}(\mathbf{k})\nu(\mathbf{k})\sigma_{\alpha''\alpha'}(-\mathbf{k}) + 4\pi\frac{(\mathbf{kd}_{\alpha_1\alpha'})(\mathbf{kd}_{\alpha''\alpha_2})}{k^2}\right\}. \tag{128}$$

In the long-wave limit ($\mathbf{k} \to 0$), according to relations (88) – (90), expressions (128) gain even a more simple shape:

$$g_{\alpha_1\alpha';\alpha_2\alpha''}^{(0,0)} \approx \delta_{\alpha_1\alpha'}\delta_{\alpha_2\alpha''} + g_{\alpha_1\alpha_2}d_{\alpha_1\alpha';\alpha_2\alpha''},$$

$$g_{\alpha_1\alpha_2}^{(0)} \approx -3e\frac{i\mathbf{kd}_{\alpha_1\alpha_2}}{k^2}g_{\alpha_1\alpha_2},$$

$$d_{\alpha_1\alpha';\alpha_2\alpha''} \equiv \mathbf{d}_{\alpha_1\alpha'}\mathbf{d}_{\alpha''\alpha_2} + \mathbf{d}_{\alpha_1\alpha_2}\mathbf{d}_{\alpha''\alpha'},$$

$$g_{\alpha_1\alpha_2}(\mathbf{k},\omega) \equiv \frac{4\pi}{3}\frac{v_{\alpha_1}^{(0)}(\nu,T) - v_{\alpha_2}^{(0)}(\nu,T)}{\omega_{\alpha_1\alpha_2}}$$
$$+ \frac{2\pi}{3}\frac{Tk^2}{M\omega_{\alpha_1\alpha_2}^3}\left\{ v_{\alpha_1}^{(0)}(\nu,T)\left(1 - \frac{\omega_{\alpha_1\alpha_2}}{T}\right) - v_{\alpha_2}^{(0)}(\nu,T)\left(1 + \frac{\omega_{\alpha_1\alpha_2}}{T}\right) \right\} \quad (129)$$
$$- i\frac{4\pi}{3}\frac{1}{k}\sqrt{\frac{\pi M}{2T}} e^{-\frac{M\omega_{\alpha_1\alpha_2}^2}{2Tk^2}} \left\{ v_{\alpha_1}^{(0)}(\nu,T)\left(1 - \frac{\omega_{\alpha_1\alpha_2}}{2T}\right) - v_{\alpha_2}^{(0)}(\nu,T)\left(1 + \frac{\omega_{\alpha_1\alpha_2}}{2T}\right) \right\},$$

In principle, expressions (122), (128), (129) allow to solve the system of equations (109) and obtain the values $\delta\varphi^{(1)}(\mathbf{k},\omega)$, $\delta\varphi^{(2)}(\mathbf{k},\omega)$, $\delta\varphi_{\alpha_1\alpha_2}^{(0)}(\mathbf{k},\omega)$, expressing them through the functions $\delta A^{(1)}(\mathbf{k},\omega)$, $\delta A^{(2)}(\mathbf{k},\omega)$, $\delta A_{\alpha_1\alpha_2}^{(0)}(\mathbf{k},\omega)$, см. (111), (105), (106). Thus, the Fourier-images of the distribution functions $\delta f^{(1)}(\mathbf{k},\omega,\mathbf{p})$, $\delta f^{(2)}(\mathbf{k},\omega,\mathbf{p})$, $\delta f_{\alpha_1,\alpha_2}^{(0)}(\mathbf{k},\omega,\mathbf{p})$, can be also calculated, see (105), (106).

Note that in general the system of equations (109) cannot be solved. Firstly, the number of levels of the atomic spectrum of hydrogen atoms is infinite. That is, in general, infinite is also the number of equations in the system (109). Secondly, as is known, with an increase in the bound state level energy and approach it to zero, the atomic levels thicken (see also Ref. 23 in this regard). In other words, the energy difference between adjacent levels is reduced, which will necessarily lead to a breach of the applicability of the theory presented here, see (101). To the solution to this problem, however, contributes the fact that the proposed in this paper theory itself is valid only at relatively low temperatures, see (101), which leads to the expression (98), the evidence of weak excitation of the subsystem of Bose atoms. This allows in theory, to restrict ourselves to the case of the existence of only a few levels of the excited atoms. In this paper we consider the approximation of a two-levels model of atoms, assuming that there is a level of the ground state of the atom and an excited one. In fact, the two-level model of the atom in these temperature conditions (101) may be considered a sufficiently realistic approximation. In fact, as can be seen from (96), (98), the contribution of the excited atoms with energies $\varepsilon_\alpha \neq \varepsilon_0$ in the processes and phenomena in the system are exponentially small (over the temperature) as compared to the contribution of the atoms in the ground state. Contribution of excited atoms can be taken into account in perturbation theory. We, however, in this paper, will not do this, because even a simple model of two-level atom results in extremely cumbersome and voluminous analytical calculations. Also we will not give here these calculations, as they are reduced to a trivial (though laborious) solving the system of equations (109) in the case when the index $\alpha$, characterizing the quantum-mechanical state of the atom, takes only two specific values, one of which corresponds to the ground state of the atom. To be specific, we assume that the physical characteristics of values with $\alpha = 1$ correspond to a ground state of the atom, while the physical characteristics of $\alpha = 2$ correspond to some excited state. We assume also that there exists a dipole transition between these two states with the electric dipole moment $\mathbf{d}_{12}$ (or $\mathbf{d}_{21}$, see (88), (89)). We show only the eventual results of the calculations in the long-wave limit, that is, when using expressions (129). Solving of equations (109) in the case of a double-level atom is given with the expressions:

$$
\begin{aligned}
\delta\varphi_{12}(\mathbf{k},\omega) &= \frac{1}{\varepsilon}\left\{D_1 + g_{21}|\mathbf{d}_{12}|^2 D_2\right\}\delta A_{12}^{(0)}(\mathbf{k},\omega) - \frac{g_{12}}{\varepsilon}\left\{|\mathbf{d}_{12}|^2 D_1 + \mathbf{d}_{12}^2 D_2\right\}\delta A_{21}^{(0)}(\mathbf{k},\omega) \\
&\quad -3i\mathbf{k}\mathbf{d}_{12}\frac{eg_{12}}{\varepsilon k^2}\left\{\left[g^{(1,2)} - g^{(1,1)}\right]\delta A^{(2)}(\mathbf{k},\omega) + \left[g^{(2,2)} - g^{(2,1)}\right]\delta A^{(1)}(\mathbf{k},\omega)\right\}, \\
\delta\varphi_{21}(\mathbf{k},\omega) &= -\frac{g_{21}}{\varepsilon}\left\{|\mathbf{d}_{12}|^2 D_1 + \mathbf{d}_{21}^2 D_2\right\}\delta A_{12}^{(0)}(\mathbf{k},\omega) + \frac{1}{\varepsilon}\left\{D_1 + g_{12}|\mathbf{d}_{12}|^2 D_2\right\}\delta A_{21}^{(0)}(\mathbf{k},\omega) \\
&\quad -3i\mathbf{k}\mathbf{d}_{21}\frac{eg_{21}}{\varepsilon k^2}\left\{\left[g^{(1,2)} - g^{(1,1)}\right]\delta A^{(2)}(\mathbf{k},\omega) + \left[g^{(2,2)} - g^{(2,1)}\right]\delta A^{(1)}(\mathbf{k},\omega)\right\}, \\
\delta\varphi_{11}(\mathbf{k},\omega) &= \delta A_{11}^{(0)}(\mathbf{k},\omega) - g_{11}|\mathbf{d}_{12}|^2 \delta A_{22}^{(0)}(\mathbf{k},\omega), \\
\delta\varphi_{22}(\mathbf{k},\omega) &= \delta A_{22}^{(0)}(\mathbf{k},\omega) - g_{22}|\mathbf{d}_{12}|^2 \delta A_{11}^{(0)}(\mathbf{k},\omega), \\
\delta\varphi^{(1)}(\mathbf{k},\omega) &= i\frac{D_{1,2}}{e\varepsilon}\left\{\mathbf{k}\mathbf{d}_{21}\delta A_{12}^{(0)}(\mathbf{k},\omega) + \mathbf{k}\mathbf{d}_{12}\delta A_{21}^{(0)}(\mathbf{k},\omega)\right\} \\
&\quad +\frac{1}{\varepsilon}|\mathbf{d}_{12}|^2(g_{12}+g_{21})\left[g^{(2,2)} - g^{(2,1)}\right]\delta A^{(1)}(\mathbf{k},\omega) + \frac{1}{\varepsilon}\left\{g^{(2,2)}\delta A^{(1)}(\mathbf{k},\omega) + g^{(1,2)}\delta A^{(2)}(\mathbf{k},\omega)\right\}, \\
\delta\varphi^{(2)}(\mathbf{k},\omega) &= \frac{D_{2,1}}{e\varepsilon}\left\{\mathbf{k}\mathbf{d}_{12}\delta A_{21}^{(0)}(\mathbf{k},\omega) + \mathbf{k}\mathbf{d}_{21}\delta A_{12}^{(0)}(\mathbf{k},\omega)\right\} \\
&\quad +\frac{1}{\varepsilon}|\mathbf{d}_{12}|^2(g_{12}+g_{21})\left[g^{(1,1)} - g^{(1,2)}\right]\delta A^{(2)}(\mathbf{k},\omega) + \frac{1}{\varepsilon}\left\{g^{(2,1)}\delta A^{(1)}(\mathbf{k},\omega) + g^{(1,1)}\delta A^{(2)}(\mathbf{k},\omega)\right\},
\end{aligned} \quad (130)
$$

in which the quantity $D_1$ is defined with formulas (110), (111), (122) and besides, we introduce the following notations:

$$
\begin{gathered}
D_{1,2} \equiv g^{(1,2)}\left(g^{(2,2)} - g^{(2,1)}\right), \qquad D_{2,1} \equiv g^{(2,1)}\left(g^{(1,2)} - g^{(1,1)}\right), \\
D_2 \equiv \left(g^{(2,2)} - g^{(2,1)}\right)\left(g^{(1,1)} - g^{(1,2)}\right), \qquad \varepsilon \equiv D_1 + |\mathbf{d}_{12}|^2(g_{12}+g_{21})D_2.
\end{gathered} \quad (131)
$$

When obtaining equations (130) we considered also the following circumstance. All the equations, including (109), were obtained within the first-order approximation over the interaction, that is, in the case of interaction between the bound states, up to the quantities proportional to $(\mathbf{d}_{12})^2$, $(\mathbf{d}_{21})^2$. That is why, in further calculations we need to take into account the summands not larger than the first order over the quadrate of the dipole moment. Such a requirement will allow us to clarify the concept of weak interaction between the particles, which was the basis of perturbation theory used in our work. Indeed, let us compare the summands $(g_{12}+g_{21})|\mathbf{d}_{12}|^2$ with neglected ones as $g_{12}g_{21}|\mathbf{d}_{12}|^4$, see (129), by using the inequality:

$$|g_{12}+g_{21}||\mathbf{d}_{12}|^2 \gg |g_{12}g_{21}||\mathbf{d}_{12}|^4.$$

By simple calculations we see that this relation is equivalent to the strong inequality:

$$|\varepsilon_1-\varepsilon_2| \gg |v_1^{(0)}(v,T)-v_2^{(0)}(v,T)||\mathbf{d}_{12}|^2, \qquad (132)$$

that specifies the term "weak interaction" and in accordance with (96) – (101) seems to be quite realistic, even for a sufficiently dense gases (see also specific evaluations below).

## VIII. LAWS OF DISPERSION OF ELEMENTARY EXCITATIONS IN WEAKLY IONIZED AND WEAKLY EXCITED GAS OF HYDROGEN-LIKE ATOMS.

The quantity $\varepsilon(\mathbf{k},\omega)$ in expressions (130), defined by the formula (131), is the dielectric constant of the system, see in this regard, e.g., Refs. 31, 39. The dielectric constant, as is known, in turn determines the dispersion equations for electromagnetic waves, which can propagate in a medium:

$$\varepsilon(\mathbf{k},\omega)=0, \quad \frac{\omega^2}{c^2}\varepsilon(\mathbf{k},\omega)=k^2. \qquad (133)$$

Equations (133) have reduced shape when $\mu(\mathbf{k},\omega)\approx 1$, where $\mu(\mathbf{k},\omega)$ is permeability of the medium. This value may also be calculated within our proposed theory. However, we shall not give here the corresponding calculations, referring to the fact that the magnetic permeability of the medium differs little from unity for most real gases. Recall that the first of the equations (133) or (134) determines the dispersion of the longitudinal electromagnetic waves, while the second that of the transverse ones in the medium (see for example Ref. 31.).

The dielectric constant of the system $\varepsilon(\mathbf{k},\omega)$ can be represented as:

$$\varepsilon(k,\omega)=\varepsilon_1(k,\omega)+i\varepsilon_2(k,\omega), \qquad (134)$$

where the real part $\varepsilon_1(k,\omega)$ of the quantity $\varepsilon(k,\omega)$ should determine the proper dispersion of electromagnetic waves, and the imaginary part $\varepsilon_2(k,\omega)$ is responsible for the stability of the electromagnetic waves in a medium that is, determines their attenuation or growth. Before we write out the explicit form of the quantities $\varepsilon_1(k,\omega)$, $\varepsilon_2(k,\omega)$, we remind that all functions $g(\mathbf{k},\omega)$, that according to (131) define the dielectric constant $\varepsilon(k,\omega)$, obtained correct up to first order over the small parameters $[kr_D(\nu,T)]^2 = Tk^2/4\pi e^2 \nu^{(1,2)}(\nu,T) \ll 1$, $Tk^2/M[\omega-(\varepsilon_1-\varepsilon_2)]^2 \ll 1$, see (115), (116), (124), (127) – (129) (we remind that due to a quasi-neutrality of the system $\nu^{(1)}(\nu,T) = \nu^{(2)}(\nu,T)$). At the same time, as easy to ensure, the imaginary parts of $g(\mathbf{k},\omega)$ over the same parameters are exponentially small. In consequence, also the quantities $\varepsilon_1(k,\omega)$ and $\varepsilon_2(k,\omega)$ can be calculated only with the same precision of the used perturbation theory. Taking this into account, the expression for the real part $\varepsilon_1(k,\omega)$ of the quantity $\varepsilon(k,\omega)$ may be represented as

$$\varepsilon_1(k,\omega) = 1 - \left[\frac{\omega_0^{(1)}(\nu,T)}{\omega}\right]^2 - \left[\frac{\omega_0^{(2)}(\nu,T)}{\omega}\right]^2 - \frac{8\pi}{3}\frac{|\mathbf{d}_{12}|^2\left[\nu_1^{(0)}(\nu,T)-\nu_2^{(0)}(\nu,T)\right]|\varepsilon_1-\varepsilon_2|}{\omega^2-(\varepsilon_1-\varepsilon_2)^2}$$

$$+\frac{\omega^2}{3T^2}[kr_D(\nu,T)]^2\left\{\left(\frac{\omega_0^{(1)}(\nu,T)}{\omega}\right)^4\left[1-\left(\frac{\omega_0^{(2)}(\nu,T)}{\omega}\right)^2\right]\right.$$

$$+\left(\frac{\omega_0^{(2)}(\nu,T)}{\omega}\right)^4\left[1-\left(\frac{\omega_0^{(1)}(\nu,T)}{\omega}\right)^2\right]\right\}$$

$$-3[kr_D(\nu,T)]^2\left\{\left[\frac{\omega_0^{(1)}(\nu,T)}{\omega}\right]^4+\left[\frac{\omega_0^{(2)}(\nu,T)}{\omega}\right]^4\right.$$

$$+\frac{8\pi}{9}\frac{\omega^2|\mathbf{d}_{12}|^2}{3T^2}\frac{\left[\nu_1^{(0)}(\nu,T)-\nu_2^{(0)}(\nu,T)\right]|\varepsilon_1-\varepsilon_2|}{\omega^2-(\varepsilon_1-\varepsilon_2)^2}\right\}$$

$$+|\mathbf{d}_{12}|^2\frac{2\pi}{3}\left\{\frac{Tk^2}{M\omega_{12}^3}\left[\nu_1^{(0)}(\nu,T)\left(1-\frac{\omega_{12}}{T}\right)-\nu_2^{(0)}(\nu,T)\left(1+\frac{\omega_{12}}{T}\right)\right]\right.$$

$$\left.+\frac{Tk^2}{M\omega_{21}^3}\left[\nu_2^{(0)}(\nu,T)\left(1-\frac{\omega_{21}}{T}\right)-\nu_1^{(0)}(\nu,T)\left(1+\frac{\omega_{21}}{T}\right)\right]\right\}. \qquad (135)$$

Where we used a notation $r_D(\nu,T) \equiv r_D^{(1)}(\nu,T) = r_D^{(2)}(\nu,T)$ (see (116)). The imaginary part of the dielectric constant $\varepsilon_2(k,\omega)$ now is given by

$$\varepsilon_2(k,\omega) = \sqrt{\frac{\pi}{2}} (kr_D(\nu,T))^{-3} \left\{ \frac{\omega}{\omega_0^{(1)}(\nu,T)} e^{-\frac{1}{2}\left(\frac{\omega}{\omega_0^{(1)}(\nu,T)}\right)^2 \frac{1}{(kr_D(\nu,T))^2}} \right.$$

$$\left. + \frac{\omega}{\omega_0^{(2)}(\nu,T)} e^{-\frac{1}{2}\left(\frac{\omega}{\omega_0^{(1)}(\nu,T)}\right)^2 \frac{1}{(kr_D(\nu,T))^2}} \right\} \quad (136)$$

$$-|\mathbf{d}_{12}|^2 \frac{4\pi}{3} \frac{1}{k} \sqrt{\frac{\pi M}{2T}} \left\{ e^{-\frac{M\omega_{12}^2}{2Tk^2}} \left[ \nu_1^{(0)}(\nu,T)\left(1-\frac{\omega_{12}}{2T}\right) - \nu_2^{(0)}(\nu,T)\left(1+\frac{\omega_{12}}{2T}\right) \right] \right.$$

$$\left. + e^{-\frac{M\omega_{21}^2}{2Tk^2}} \left[ \nu_2^{(0)}(\nu,T)\left(1-\frac{\omega_{21}}{2T}\right) - \nu_1^{(0)}(\nu,T)\left(1+\frac{\omega_{21}}{2T}\right) \right] \right\}.$$

Formulas (135), (136) allow further significant simplifications for two reasons. We have previously agreed to name electrons as the fermions of kind "1", and the cores of hydrogen atoms as the fermions of kind "2". But then the mass ratio of fermions $m_1/m_2$ needs to be regarded as a small value. Then, we have (see (116)) a widely used in plasmas theory relation

$$\frac{\omega_0^{(1)}(\nu,T)}{\omega_0^{(2)}(\nu,T)} = \sqrt{\frac{m_2}{m_1}} \gg 1, \qquad \left[\omega_0^{(1,2)}(\nu,T)\right]^2 \equiv \frac{4\pi e^2 \nu^{(1,2)}(\nu,T)}{m_{1,2}}, \qquad \nu^{(1)}(\nu,T) = \nu^{(2)}(\nu,T) \quad (137)$$

That allows us not to take into account the frequency $\omega_0^{(2)}(\nu,T)$ in equations (135), (136). The second reason of simplifications occurs if one of the atomic levels is the level of the ground state of the atom (to be specific in this article it is believed that this level is indicated by subscript "1"). Then, in accordance with (96) – (98) a strict inequality is valid:

$$\nu_1^{(0)}(\nu,T) \gg \nu_2^{(0)}(\nu,T), \quad (138)$$

due to which additional significant simplifications in (135), (136) are possible:

$$\varepsilon_1(k,\omega) = \varepsilon_1^0(k,\omega) + \varepsilon_1^1(k,\omega),$$

$$\varepsilon_1^0(k,\omega) \equiv 1 - \left[\frac{\omega_0^{(1)}(\nu,T)}{\omega}\right]^2 - \frac{8\pi}{3} \frac{|\mathbf{d}_{12}|^2 \nu_1^{(1)}(\nu,T)|\varepsilon_1 - \varepsilon_2|}{\omega^2 - (\varepsilon_1 - \varepsilon_2)^2},$$

$$\varepsilon_1^1(k,\omega) \equiv \left(\frac{\omega^2}{3T^2} - 3\right)\left[kr_D(\nu,T)\right]^2 \left(\frac{\omega_0^{(1)}(\nu,T)}{\omega}\right)^4$$
$$-\frac{8\pi}{3}\frac{\omega^2 |\mathbf{d}_{12}|^2}{3T^2}\frac{\nu_1^{(0)}(\nu,T)|\varepsilon_1-\varepsilon_2|}{\omega^2-(\varepsilon_1-\varepsilon_2)^2}\left[kr_D(\nu,T)\right]^2$$
$$+|\mathbf{d}_{12}|^2 \frac{2\pi}{3}\nu_1^{(0)}(\nu,T)\left\{\frac{Tk^2}{M\omega_{12}^3}\left(1-\frac{\omega_{12}}{T}\right) - \frac{Tk^2}{M\omega_{21}^3}\left(1+\frac{\omega_{21}}{T}\right)\right\}, \quad (139)$$

$$\varepsilon_2(k,\omega) \approx \sqrt{\frac{\pi}{2}}(kr_D(\nu,T))^{-3}\frac{\omega}{\omega_0^{(1)}(\nu,T)}e^{-\frac{1}{2}\left(\frac{\omega}{\omega_0^{(1)}(\nu,T)}\right)^2 \frac{1}{(kr_D(\nu,T))^2}}$$
$$-|\mathbf{d}_{12}|^2 \nu_1^{(0)}(\nu,T)\frac{4\pi}{3}\frac{1}{k}\sqrt{\frac{\pi M}{2T}}\left\{e^{-\frac{M\omega_{12}^2}{2Tk^2}}\left(1-\frac{\omega_{12}}{2T}\right) - e^{-\frac{M\omega_{21}^2}{2Tk^2}}\left(1+\frac{\omega_{21}}{2T}\right)\right\}.$$

In (139) in expression for $\varepsilon_1(k,\omega)$ we marked terms of the zero and first-order perturbation theory over the small quantities $\left[kr_D(\nu,T)\right]^2 = Tk^2/m_1\left[\omega_0^{(1)}(\nu,T)\right]^2 \ll 1$, $Tk^2/M\left[\omega-(\varepsilon_1-\varepsilon_2)\right]^2 \ll 1$, see the above. Formulas (137) – (139) will be the starting point for calculating the dispersion laws of electromagnetic waves in a weakly ionized and weakly excited gas of hydrogen-like atoms.

Let us first obtain the dispersion law $\omega_l(k)$ of longitudinal waves in the medium, using the first of equations (133). As known, the existence of an imaginary part in the dielectric constant $\varepsilon(k,\omega)$ means the dissipation (or pumping) of energy of the waves, the dispersion law of which according to (133) should be defined with an equation (see also Refs. 31, 39):

$$\varepsilon(k,\omega_l(k) - i\gamma_l(k)) = 0, \quad (140)$$

while the decrement $\gamma_l(k)$ (or increment!) of the wave is to be defined by the imaginary part $\varepsilon_2(k,\omega)$ of the quantity $\varepsilon(k,\omega)$. That is why, the weakly-decaying (or weakly-growing) oscillations in the system

$$|\omega_l(k)| \gg |\gamma_l(k)| \quad (141)$$

may exist only under condition (see (134), (139)):

$$|\varepsilon_1(k,\omega)| \gg |\varepsilon_2(k,\omega)| \quad (142)$$

Thus, according to (140) – (142) the longitudinal oscillations frequency $\omega_l(k)$ is to be obtained through the relation

$$\varepsilon_1(k, \omega_l^0 + \omega_l^1(k)) = 0,$$
$$\omega_l(k) = \omega_l^0 + \omega_l^1(k), \quad |\omega_l^0| \gg |\omega_l^1(k)|. \tag{143}$$

within the mentioned perturbation theory over small parameters $[kr_D(v,T)]^2 \ll 1$, $Tk^2/M[\omega-(\varepsilon_1-\varepsilon_2)]^2 \ll 1$. The upper index in (143), as in (139), counts the order of such perturbation theory. Now the frequency $\omega_l^0(k)$, the correction to it $\omega_l^1(k)$ and the decrement or increment $\gamma_l(k)$ are given with:

$$\varepsilon_1^0(k, \omega_l^0) = 0,$$
$$\omega_l^1(k) = -\left\{\frac{\partial \varepsilon_1^0(k,\omega)}{\partial \omega}\right\}_{\omega=\omega_l^0(k)}^{-1} \varepsilon_1^1(k, \omega_l^0(k)), \tag{144}$$
$$\gamma_l(k) = \left\{\frac{\partial \varepsilon_1(k,\omega)}{\partial \omega}\right\}_{\omega=\omega_l^0(k)}^{-1} \varepsilon_2(k, \omega_l^0(k)),$$

where all the quantities are taken from formulas (139). Equation (144) defines two branches of oscillations $(\omega_l^0)_-$ и $(\omega_l^0)_+$, which approximately are represented as

$$\left(\omega_l^0\right)_-^2 \approx \left(\omega_0^{(1)}(v,T)\right)^2 \left\{1 - \frac{8\pi}{3}\frac{|\mathbf{d}_{12}|^2 v_1^{(0)}(v,T)}{|\varepsilon_1-\varepsilon_2|} - \frac{\left(\omega_0^{(1)}(v,T)\right)^2}{|\varepsilon_1-\varepsilon_2|^2}\right\},$$
$$\left(\omega_l^0\right)_+^2 \approx (\varepsilon_1-\varepsilon_2)^2 + \frac{8\pi}{3}|\mathbf{d}_{12}|^2 v_1^{(0)}(v,T)|\varepsilon_1-\varepsilon_2|\left\{1+\frac{\left(\omega_0^{(1)}(v,T)\right)^2}{(\varepsilon_1-\varepsilon_2)^2}\right\} - \frac{\left(\omega_0^{(1)}(v,T)\right)^4}{(\varepsilon_1-\varepsilon_2)^2} \tag{145}$$

The latter formulas are obtained within an assumption of the strong hierarchy:

$$\left[\omega_0^{(1)}(v,T)\right]^2 \ll \frac{8\pi}{3}|\mathbf{d}_{12}|^2 v_1^{(0)}(v,T)|\varepsilon_1-\varepsilon_2| \ll (\varepsilon_1-\varepsilon_2)^2, \tag{146a}$$

or (see (95) – (97))

$$\frac{4\pi e^2}{m_1}\sqrt{\nu}\left(\frac{m_1 m_2 T}{2\pi M}\right)^{3/4} e^{\frac{\varepsilon_1}{2T}} \ll \frac{8\pi}{3}\nu|\mathbf{d}_{12}|^2 |\varepsilon_1 - \varepsilon_2| \ll (\varepsilon_1 - \varepsilon_2)^2. \tag{146b}$$

As we see, due to (132) and the existence of an exponential multiplier in (146b), such inequalities should be valid for a number of levels with energy $\varepsilon_2$, connected with the ground state energy $\varepsilon_1$ with a transition with a dipole moment $\mathbf{d}_{12}$. This is confirmed by numerical estimates for a specific system under specific physical conditions, given below. For this reason, the expression in brackets in both formulas (145) accurately enough may be replaced by a unity:

$$\left(\omega_l^0\right)_-^2 \approx \left(\omega_0^{(1)}(\nu,T)\right)^2, \quad \left(\omega_l^0\right)_+^2 \approx (\varepsilon_1 - \varepsilon_2)^2 + \frac{8\pi}{3}|\mathbf{d}_{12}|^2 \nu_1^{(0)}(\nu,T)|\varepsilon_1 - \varepsilon_2|.. \tag{147}$$

This is not done just because as a result of such operation, the terms governing the mutual influence of selected branches of the oscillations will be lost. The second term on the right side of the expression for $\left(\omega_l^0\right)_+^2$ holds in (147) since we need to satisfy the condition of calculation of the integrals (124). Thus, in the principal approximation frequency $\left(\omega_l^0\right)_-$ coincides with the plasma (or Langmuir) frequency. The dispersion of this oscillations branch is defined by the correction $\left[\omega_l^1(k)\right]_-$, and its decay is defined by a decrement $\left[\gamma_l(k)\right]_-$, that according to (144) are given by formulas:

$$\begin{aligned}
\left[\omega_l^1(k)\right]_- &\approx \frac{3}{2}\omega_0^{(1)}(\nu,T)\left[kr_D(\nu,T)\right]^2, \\
\left[\gamma_l(k)\right]_- &\approx \sqrt{\frac{\pi}{8}}\frac{\omega_0^{(1)}(\nu,T)}{\left(kr_D(\nu,T)\right)^3} e^{-\frac{1}{(kr_D(\nu,T))^2}}, \quad \left[kr_D(\nu,T)\right]^2 \ll 1,
\end{aligned} \tag{148}$$

that also coincide with the corresponding longitudinal waves dispersion laws in plasma and the Landau decrement in it (see in this regard, e.g., Ref. 31).

The frequency $\left(\omega_l^0\right)_+$ is a frequency of longitudinal polarization oscillations in a weakly ionized hydrogen-like gas with dispersion $\left[\omega_l^1(k)\right]_+$ and the decrement $\left[\gamma_l(k)\right]_+$, the explicit form of which can be found using (144) – (146):

$$\left[\omega_l^1(k)\right]_+ \approx -\frac{4\pi}{9}|\mathbf{d}_{12}|^2 \nu_1^{(0)}(\nu,T)\frac{(\varepsilon_1 - \varepsilon_2)^2}{T^2}\left(\frac{\omega_0^{(1)}(\nu,T)}{\varepsilon_1 - \varepsilon_2}\right)^4 \left[kr_D(\nu,T)\right]^2,$$

$$\left[\gamma_l(k)\right]_{+} \approx \frac{4\pi}{3}|\mathbf{d}_{12}|^2 v_1^{(0)}(\nu,T)\sqrt{\frac{\pi}{2}}\left(kr_D(\nu,T)\right)^{-3}\frac{|\varepsilon_1-\varepsilon_2|}{\omega_0^{(1)}(\nu,T)}e^{-\frac{m_1|\varepsilon_1-\varepsilon_2|^2}{2Tk^2}},\quad \left[kr_D(\nu,T)\right]^2 \ll 1,. \tag{149}$$

Now we turn to the second of the equations (133), that is the dispersion equation for the transverse electromagnetic waves in a weakly ionized weakly excited hydrogen-like gas. The main approximation for the frequency of the transverse oscillations $\omega_t^0(k)$, and the correction to it $\omega_t^1(k)$ and the decrement $\gamma_t(k)$ analogically to (140) – (144) should be calculated from

$$\begin{aligned}&\left\{\omega_t^0(k)+\omega_t^1(k)-i\gamma_t(k)\right\}^2 \varepsilon\left(k,\omega_t^0(k)+\omega_t^1(k)-i\gamma_t(k)\right)=c^2k^2,\\ &\left|\omega_t^0(k)\right| \gg \left|\omega_t^1(k)\right|,\quad \left|\omega_t^0(k)\right| \gg \left|\gamma_t(k)\right|.\end{aligned} \tag{150}$$

where the quantity $\varepsilon(k,\omega)$ is still given by equations (134) – (135). Within the framework of the used perturbation theory over the small parameters $\left[kr_D(\nu,T)\right]^2 \ll 1$, $Tk^2/M\left[\omega-(\varepsilon_1-\varepsilon_2)\right]^2 \ll 1$ (see above), equation (150) is equivalent to the system of equations:

$$\begin{aligned}&\left[\omega_t^0(k)\right]^2 \varepsilon_1^0\left(k,\omega_t^0(k)\right)-c^2k^2 = 0,\\ &\omega_t^0(k)\varepsilon_1^1\left(k,\omega_t^0(k)\right)+\left[2\varepsilon_1^0\left(k,\omega_t^0(k)\right)+\omega_t^0(k)\left(\frac{\partial \varepsilon_1^0(k,\omega)}{\partial \omega}\right)_{\omega=\omega_t^0(k)}\right]\omega_t^1(k)=0,\\ &\omega_t^0(k)\varepsilon_2\left(k,\omega_t^0(k)\right)-\gamma_t(k)\left[2\varepsilon_1^0\left(k,\omega_t^0(k)\right)+\omega_t^0(k)\left(\frac{\partial \varepsilon_1^0(k,\omega)}{\partial \omega}\right)_{\omega=\omega_t^0(k)}\right]=0,\end{aligned} \tag{151}$$

in which all the quantities are given with expressions (139). As before, see (144), (145), (147) the solution of the first of equations (151) defines two branches of electromagnetic waves (transverse):

$$\begin{aligned}\left(\omega_t^2(k)\right)_{+} &\approx \left[\left(\omega_l^0\right)_{+}^2+c^2k^2\right]\left\{1-\frac{\left[\left(\omega_l^0\right)_{-}^2+c^2k^2\right](\varepsilon_1-\varepsilon_2)^2}{\left[\left(\omega_l^0\right)_{+}^2+c^2k^2\right]^2}\right\},\\ \left(\omega_t^2(k)\right)_{-} &\approx \left[\left(\omega_l^0\right)_{-}^2+c^2k^2\right]\frac{(\varepsilon_1-\varepsilon_2)^2}{\left[\left(\omega_l^0\right)_{+}^2+c^2k^2\right]}.\end{aligned} \tag{152}$$

obtained assuming $\left(\omega_l^0\right)_+^2 \gg \left(\omega_l^0\right)_-^2$, which holds due to the strict hierarchy (146). The quantities $\left(\omega_l^0\right)_\pm^2$ in (152) are defined by expressions (147). The summand with $c^2 k^2$ is preserved in these formulas as the addendum to the largest value $(\varepsilon_1 - \varepsilon_2)^2$ in expressions (146) since it may be comparable with $(\varepsilon_1 - \varepsilon_2)^2$ or even larger than the latter depending on the value of the wave vector $k$.

It is easy to see, that the laws of dispersion of transverse electromagnetic waves in general is quite complicated. The same observation can be made with respect to the decrement $\left[\gamma_t(k, \omega_0)\right]_\pm$ and corrections $\left[\omega_t^1(k)\right]_\pm$ to the frequencies $\left[\omega_t(k)\right]_\pm$:

$$\left[\omega_t^1(k)\right]_\pm = -\left.\frac{\omega_0^3\left[\omega_0^2 - (\varepsilon_1 - \varepsilon_2)^2\right]\varepsilon_1^1(k, \omega_0)}{2\omega_0^2\left[\omega_0^2 + c^2 k^2\right] - \left[\left(\omega_0^{(1)}\right)^2 + c^2 k^2\right](\varepsilon_1 - \varepsilon_2)^2}\right|_{\omega_0 = \left[\omega_t(k)\right]_\pm},$$

$$\left[\gamma_t(k)\right]_\pm = \left.\frac{\omega_0^3\left[\omega_0^2 - (\varepsilon_1 - \varepsilon_2)^2\right]\varepsilon_2(k, \omega_0)}{2\omega_0^2\left[\omega_0^2 + c^2 k^2\right] - \left[\left(\omega_0^{(1)}\right)^2 + c^2 k^2\right](\varepsilon_1 - \varepsilon_2)^2}\right|_{\omega_0 = \left[\omega_t(k)\right]_\pm}.$$ (153)

where all the values determined by the expressions (139), (152), (145).

In range of low wave vector values $k$, $c^2 k^2 \ll (\varepsilon_1 - \varepsilon_2)^2$, lateral oscillation frequencies are given by (see. also in this regard, e.g., Ref. 31):

$$\left(\omega_t^2(k)\right)_\pm \approx \left(\omega_l^0\right)_\pm^2 + c^2 k^2, \quad c^2 k^2 \ll (\varepsilon_1 - \varepsilon_2)^2,$$ (153)

in which the summands $\left(\omega_l^0\right)_\pm^2$ are defined by (147). We note, that in the small wave vectors domain the dispersion law (153) coincides with the dispersion laws for transverse electromagnetic waves in plasma, see, e.g., Ref. 31. The decay coefficients of excitations in this wave vectors domain coincide with those for longitudinal waves, which are defined with expressions (148), (149). In the large wave vectors domain, $c^2 k^2 \gg \left(\omega_l^0\right)_+^2$, the dispersion laws of transverse waves read (inequalities (146) are still valid):

$$\left(\omega_t^2(k)\right)_+ \approx \left(\omega_l^0\right)_+^2 + c^2 k^2 \approx (\varepsilon_1 - \varepsilon_2)^2 + c^2 k^2 \approx c^2 k^2,$$

$$\left(\omega_t^2(k)\right)_- \approx \left(\omega_l^0\right)_+^2 \left[1 - \frac{\left(\omega_l^0\right)_+^2}{c^2 k^2}\right] \approx \left(\varepsilon_1 - \varepsilon_2\right)^2 \left[1 - \frac{\left(\omega_l^0\right)_+^2}{c^2 k^2}\right] \approx \left(\varepsilon_1 - \varepsilon_2\right)^2, \quad (154)$$

$$c^2 k^2 \gg \left(\omega_l^0\right)_+^2.$$

Decrements in this region of the wave vectors are given by:

$$\gamma_-(k) \approx \frac{2\pi}{3} \sqrt{\frac{\pi M}{2T}} |\mathbf{d}_{12}|^2 \nu_1^{(0)}(\nu,T) \frac{1}{k} \frac{|\varepsilon_1 - \varepsilon_2|^3}{c^2 k^2}, \quad \gamma_+(k) \approx \sqrt{\frac{\pi}{8}} ck \left(kr_D(\nu,T)\right)^{-3} \frac{ck}{\left(\omega_l^0\right)_-} e^{-\frac{m_1 c^2}{2T}}, \quad (155)$$

where the characteristic frequency $\omega_0^{(1)}(\nu,T)$ is still defined with the formula (137). It is easy to see the character of the dependencies of the attenuation coefficients of the wave vector in this case varies considerably. Namely, the exponential dependence of the excitation attenuation coefficients, characteristic of the small wave vectors (see. (148) (149)), in the region of large wave vectors is replaced by a power dependence, expressed in formulas (155). In the domain where $c^2 k^2 \sim \left(\omega_l^0\right)_+^2$ the studied branches of the spectrum of oscillations overlap. Thus, you may notice that the behavior of the dispersion laws of elementary excitations in non-degenerate low-temperature dilute gases of hydrogen-like atoms is completely analogous to the behavior of polariton oscillations branches that occur in ionic crystals, see, e.g., Ref. 40.

To conclude this section for an example of the system we present some quantitative estimates to justify the fact that in this article come together the concept of low-temperature approach, "non-degeneracy" gas of hydrogen-like atoms, as well as the approximation of the weak interaction. We recall that such a synthesis of two seemingly contradictory requirements is mathematically expressed in the inequalities (99) - (101). The concept of the weak interaction is specified by (132) and in accordance with (96) - (101) seems to be quite realistic, even for sufficiently dense gases. As mentioned in the introduction to this paper, in both experimental and theoretical studies, related to the Bose-Einstein condensate, the gases of atoms of vapor of alkali metals play a special role[9-23, 26-30]. In the papers devoted to a slow-down and stop of light in BEC, the experimental and theoretical studies are tied to the so-called $D_2$-line of $^{23}$Na, the characteristics of which are now well known[41]. For this reason, the realism of the approximations used in the present study we demonstrate on the example of a gas consisting of atoms $^{23}$Na at temperatures close to room, $T \sim 300 K$ (or in energy units $T \sim 4 \cdot 10^{-14} erg$). We also assume that the energy structure of the atoms of the gas is characterized by only two levels - the level of the ground state and one excited level (justification of the two-level approximation see before the formulas (130).). We will consider the excited state coinciding with $D_2$- line of $^{23}$Na, the position of which

$\varepsilon_2 - \varepsilon_1$ in respect to the ionization energy $\varepsilon_1$, together with the value of the corresponding atomic dipole moment $d_{12}$, the mass of natrium atom $M$ and its radius $R_{Na}$ may be taken from Ref. 41:

$$\varepsilon_2 - \varepsilon_1 \approx 2.104 \cdot eV \approx 3{,}371 \times 10^{-12}\,\text{erg},$$
$$-\varepsilon_1 = 5.139\,076\,50(28)\ eV \approx 8 \times 10^{-12}\,\text{erg},$$
$$d_{12} \approx 6.4 \cdot 10^{-18}\,SGSE \cdot cm, \quad (156)$$
$$M \approx 0.381\,754\,035 \cdot 10^{-22}\,g,$$
$$R_{Na} \approx 1{,}9 \times 10^{-10}\,cm.$$

These values (156) allow to estimate the densities $\nu^{(1,2)}(\nu,T)$ of free fermion components and the densities of atoms in the two assumed energetic states $\nu_1^{(0)}(\nu,T) \gg \nu_2^{(0)}(\nu,T)$ (см. (96) –(98), (138)):

$$\nu^{(1)}(\nu,T) = \nu^{(2)}(\nu,T) \sim 10^{-44}\sqrt{\nu},$$
$$\nu_1^{(0)}(\nu,T) \approx \nu, \quad \nu_2^{(0)}(\nu,T) \sim \nu^{(1)}(\nu,T). \quad (157)$$

We remind, that $\nu$ is the density of the total number of fermions, of one sort, both free and bound states forming - atoms. Inequality (132), which guarantees the validity of the approximation of the weak interaction in the framework of the evaluations is as follows:

$$\left|\nu_1^{(0)}(\nu,T) - \nu_2^{(0)}(\nu,T)\right|10^{-24} \sim 10^{-24}\nu \ll 1. \quad (158)$$

In the formulas (157), (158) we do not write out the dimensions of decimal power functions. It is readily seen due to (157), (158) in (156) the criteria (99) - (101) and (146) of gas non-degeneracy at low temperatures in a large range of densities $\nu$ are fulfilled.

Formulas (157), (158) allow to estimate the frequencies $\omega_0^{(1)}(\nu,T)$ so the values of plasma frequencies of longitudinal waves (see (116a), (145), (147), (148)) are:

$$\omega_0^{(1)}(\nu,T) \sim 10^{-17}\nu^{1/4}\ (\text{Hz}). \quad (159)$$

We see, that at gas densities $\nu$ the frequency $\omega_0^{(1)}(\nu,T)$ is negligibly small. In fact, if a system carries out one oscillation per year, its frequency is about $0.3 \cdot 10^{-7}\,Hz$. The summand $\left[\omega_l^1(k)\right]_- \ll \omega_0^{(1)}(\nu,T)$, that is responsible for the dispersion of plasma oscillations, as well as the damping factor $\left[\gamma_l(k)\right]_- \ll \omega_0^{(1)}(\nu,T)$, are defined by a dimensionless factor $\left[kr_D(\nu,T)\right] \ll 1$, see (148). The estimates show that the inequality $\left[kr_D(\nu,T)\right] \ll 1$ can be true in this case only for waves with wave vector $k \ll 10^{-24}\nu^{1/4} cm^{-1}$. The latter inequality shows an extremely small dispersion of plasma waves. The damping coefficient in this case coincides with the Landau damping coefficient of the plasma waves. Thus, the estimates show that the presence of longitudinal plasma waves in the system under study in conditions (156) can be ignored.

Estimates within the specified values (156) of characteristics of the other longitudinal oscillation frequencies in the system, $\left[\omega_l^1(k)\right]_+$ (see (149)), also evidence of its extremely small dispersion and a small damping coefficient. Indeed, its exponential index is $-10^{64}\nu^{-1/2}\left(kr_D(\nu,T)\right)^{-2}$. Thus, in this approximation the system has a virtually undamped longitudinal wave with a frequency $\left[\omega_0^2(k)\right]_+ \approx \left(\omega_l^0\right)_+^2 \approx (\varepsilon_1 - \varepsilon_2)^2 + \frac{8\pi}{3}|\mathbf{d}_{12}|^2 \nu_1^{(0)}(\nu,T)|\varepsilon_1 - \varepsilon_2|$ and a small dispersion ($k \ll 10^{-9}\nu^{-1/4} cm^{-1}$). The value of this frequency in the weak interaction is practically independent of the density of particles in the system and is determined by the difference between energy levels $\varepsilon_1 - \varepsilon_2$:

$$\left[\omega_0(k)\right]_+ \approx \left(\omega_l^0\right)_+ \approx \frac{|\varepsilon_1 - \varepsilon_2|}{\hbar} Hz \sim 10^{15} Hz = 1 PHz. \quad (160)$$

These estimates (157) - (160) characterize the values of the corresponding quantities in the expressions for the dispersion laws of the transverse waves in the studied systems, which, as noted above, demonstrate polariton behavior.

We also note in concluding this section that, as already noted above, in a Bose-Einstein condensation of photons the frequency of photons $\omega_c$ value plays an important role at zero wave vector value (the cut-off spectrum frequency). This cut-off frequency determines the effective photon mass $m^*$, which differs from zero only in medium[24, 25, 28, 30]:

$$m^* \equiv \frac{\hbar \omega_c}{c^2}.$$

In the cases described above, as is easy to see, for some branches of the oscillations the cut-off frequency coincides with the Langmuir frequency $\omega_0^{(1)}(\nu,T)$, for other branches with polarization oscillations frequency

$$\left(\omega_l^0\right)_+ \approx \varepsilon_1 - \varepsilon_2 + \frac{4\pi}{3}|\mathbf{d}_{12}|^2 \nu_1^{(0)}(\nu, T).$$ Correspondingly, in the first case the effective photons mass is defined by the Langmuir frequency:

$$m^* = \frac{\hbar \omega_0^{(1)}(\nu, T)}{c^2} \approx 0.5 \cdot 10^{-61}\,(g)$$

(in the calculations in the latter formula the value of the fermion density $\nu$ (see (159)) is taken $\nu \sim 10^{16}\,cm^{-3}$ for convenience), while in another case, the effective mass of the photons has the following estimate:

$$m^* = \frac{\hbar\left(\omega_l^0\right)_+}{c^2} \approx \frac{|\varepsilon_1 - \varepsilon_2|}{c^2} \sim 10^{-33}\,(g).$$

These are the values of the effective mass of photons like in the latter case, that were used in the description of experimental results[24, 25] in some theoretical studies (see., e.g., Refs. 28, 30). We emphasize, however, that in describing properties of a Bose condensate of photons it is important to know the dispersion relation for photons in all areas of the wave vector of values, not only the cut-off frequency. This all gives us the theory set out in this paper. The latter fact, as noted in the introduction to this paper, is another confirmation of the importance of construction of the kinetic theory of dilute weakly ionized gas of hydrogen-like atoms from the first principles of quantum statistical mechanics.

**CONCLUSION**

Thus, in this paper we propose a microscopic approach to the consistent construction of the kinetic theory of dilute weakly ionized gas of hydrogen-like atoms. The approach is based on the statements of the second quantization method in the presence of bound states of particles. It is assumed that a bound state (hydrogen-like alkali metal atom) is formed by two charged fermions of different kinds, the valence electron and the core. The basis of the development of kinetic equations is the method of reduced description of relaxation processes, suggested by N.N. Bogolyubov and generalized to the case of quantum systems in the works by S.V. Peletminskii with students. Within the developed approach we received linked kinetic equations for the Wigner distribution functions of free fermions of both kinds and their bound states - of hydrogen- like atoms in the first-order perturbation theory over the weak interaction. These equations are used to study the spectra of elementary excitations in the system. For this purpose, the kinetic equations are linearized near spatially - homogeneous state of equilibrium, characterized by the fact that all components of the system are far from degeneration and described by the Maxwell distribution functions. It is shown that the conditions of low-temperature approximation, non-degeneracy of gas and the approximation of weak interaction are realistic and can be fulfilled in a wide range of temperatures and the densities of the studied system. We obtained dispersion equations for determining the frequencies and wave damping coefficients in dilute weakly ionized gas of hydrogen-like atoms near the described equilibrium state. At the same time the approximation of a two-level atom was used. It was shown that in the system there are longitudinal waves of polarization of matter and transverse waves with the behavior characteristic for plasmon polaritons. The expressions for the dependence of the frequency and damping coefficients (Landau damping!) on wave vector for all branches of the oscillations detected were obtained. Numerical evaluation of characteristics of the elementary perturbations in the system on an example of a weakly ionized dilute gas of $^{23}$Na atoms was carried out. The possibility of using the results of the theory developed to describe the properties of a Bose condensate of photons in the diluted weakly ionized gas of hydrogen-like atoms was noted.

We emphasize, however, that all of this work, including the content of section 6 up to (94), are sufficiently general in the range of criteria used by the formulation of the second quantization method and perturbation theory over the interaction between the structural units of subsystems. After that, the theory introduces first the assumption of an equilibrium state of the system in which all components of its far from areas of quantum degeneracy. Further simplification of the description is to use a two-level atom approximation. According to this the theory can be further extended to the case of other equilibrium states, taking into account the possibility of degeneration of any of the components. In addition, it is desirable to study the effect on the behavior of the wave properties of the system of additional levels of energy structure of the atom, i.e., a departure from the model of two-level atom. In addition, the completion of the kinetic description of the diluted weakly ionized gas of hydrogen-like atoms requires the knowledge of the collision integrals of the coupled system of kinetic equations. To do this it is necessary to draw a second-order perturbation theory over the weak interaction. A separate study is also required for relaxation questions of a photon component subsystems, for which it is obligatory to obtain a kinetic equation associated with the already obtained kinetic equations. Note that all the above summary of the theory requires a large amount of very cumbersome calculations and a significant increase in the volume of the article. For this reason, the authors found it necessary to make all such issues out of the scope of this work.